\documentclass[%
 reprint,
 amsmath,amssymb,
 aps,
floatfix,
]{revtex4-2}

\usepackage{multirow}
\usepackage{graphicx}% Include figure files
\usepackage{dcolumn}% Align table columns on decimal point
\usepackage{bm}% bold math
\usepackage{hyperref}% add hypertext capabilities
\usepackage{dsfont}
\usepackage{subfigure}
\usepackage{comment}
\usepackage{color}
\usepackage[normalem]{ulem}
\usepackage{soul}
\usepackage{float}
\begin{document}

\preprint{APS/123-QED}

\title{Binary black hole mergers from merged stars in the Galactic field}

\author{Jakob Stegmann$^{1}$}
 \email{StegmannJ@cardiff.ac.uk}
\author{Fabio Antonini$^{1}$}%
\author{Fabian R.~N. Schneider$^{2,3}$}%
\author{Vaibhav Tiwari$^{1}$}%
\author{Debatri Chattopadhyay$^{1}$}%
\affiliation{%
 $^{1}$Gravity Exploration Institute, School of Physics and Astronomy, Cardiff University, Cardiff, CF24 3AA, United Kingdom\\
 $^{2}$Heidelberger Institut für Theoretische Studien, Schloß-Wolfsbrunnenweg 35, 69118 Heidelberg, Germany\\
 $^{3}$Astronomisches Rechen-Institut, Zentrum für Astronomie der Universität Heidelberg, Mönchhofstraße 12-14, 69120 Heidelberg, Germany
}%

\date{\today}

\begin{abstract}
The majority of massive stars are found in close binaries which: (i) are prone to merge and (ii) are accompanied by another distant tertiary star (triples). Here, we study the evolution of the stellar postmerger binaries composed of the merger product and the tertiary companion. We find that postmerger binaries originating from compact stellar triples with outer semimajor axes $a_{\rm out,init}\lesssim10^1$~--~$10^2\,\rm AU$ provide a new way to form binary black hole mergers in the galactic field. By means of a population synthesis, we estimate their contribution to the total black hole merger rate to be $\mathcal{R}(z=0)=0.3$~--~$25.2\,\rm Gpc^{-3}\,yr^{-1}$. Merging binary black holes that form from stellar postmerger binaries have exceptionally low mass ratios. We identify a critical mass ratio $q\simeq0.5$ below which they dominate the total black hole merger rate in the field. We show that after including their additional contribution, the mass ratio distribution of binary black hole mergers in the galactic field scenario is in better agreement with that inferred from gravitational wave detections.
\end{abstract}

%\keywords{Suggested keywords}%Use showkeys class option if keyword
                              %display desired
\maketitle

\section{Introduction}
Since the first direct detection of gravitational waves from a coalescing binary black hole (BBH) \citep[][]{2016PhRvL.116f1102A} the observational sample grew to nearly seventy of these events \citep[][GWTC-3]{2021arXiv211103634T}. It is an open astrophysical question how these BBHs were formed in the first place.
Several formation channels have been proposed including scenarios in which the mergers were driven by some dynamical interaction within a dense stellar environment, e.g., the dense cores of globular clusters \citep{2016PhRvD..93h4029R,2017MNRAS.469.4665P,2018ApJ...866L...5R,2020PhRvD.102l3016A}, massive young clusters \citep[][]{2010MNRAS.402..371B,2014MNRAS.441.3703Z,2019MNRAS.487.2947D,2021ApJ...913L..29F}, and galactic nuclei \citep[][]{2016ApJ...831..187A,2017ApJ...846..146P,2019ApJ...881L..13H,2020ApJ...894...15B,2021ApJ...917...76W}. More exotic formation channels, e.g., favor a primordial origin \citep[][]{2018PDU....22..137C,2020arXiv200706481C,2020JCAP...06..044D,2021JCAP...03..078B}.

One of the most popular channels considers the formation of BBHs from massive binary stars in the galactic field \citep[][]{2012ApJ...759...52D,2016Natur.534..512B,2018MNRAS.480.2011G,2021A&A...651A.100O,2021MNRAS.508.5028B,2021arXiv211205763B}. It is well known that these are prone to undergo some close interaction \citep[][]{2012Sci...337..444S}, e.g., an episode of stable mass transfer or common-envelope (CE) evolution that could shrink their orbit \citep[][]{1967AcA....17..355P,1992ApJ...391..246P,2012ApJ...759...52D,2016Natur.534..512B,2018MNRAS.480.2011G,2021A&A...651A.100O,2013ApJ...764..166D,2016A&A...588A..10R,2021PhRvD.103f3007S,2021MNRAS.507.5013M,2021A&A...651A.100O}. If a bound BBH is subsequently formed it may be close enough to merge within a Hubble time due to gravitational wave radiation.

In this work, we take into account the high degree of multiplicity observed for massive stars in the Galactic field. It is found that roughly $\simeq50\%$ and $70\%$ of early B-type and O-type binary stars, respectively, are accompanied by one or more distant stellar companions, while $\gtrsim 80\%$ of black hole progenitors are in triples or higher multiplicity systems \citep[][]{2017ApJS..230...15M,2022arXiv220305036B}. Previous works show that the gravitational perturbation from a bound hierarchical tertiary companion could drive close interactions between the inner binary stars \citep[][]{2021arXiv211210786S,2020A&A...640A..16T,2019ApJ...872..119H} and promote a merger after the stars formed compact objects \citep[][]{2017ApJ...836...39S,2017ApJ...841...77A,2018MNRAS.480L..58A,2018ApJ...863...68L,2018ApJ...863....7R,2020MNRAS.493.3920F,2021arXiv210501671M}. Moreover, triples and higher order systems could lead to hierarchical compact object mergers in which a compact object merger remnant coalesces with another compact object that originated from a distant progenitor star \citep[][]{2017ApJ...840L..24F,2021ApJ...907L..19V,2021NatAs...5..749G}.

We consider a scenario in which a stellar merger occurs in the inner binary of a massive stellar triple. A stellar merger is a frequent outcome of close interactions of massive binary stars \citep[][]{2012Sci...337..444S,2021arXiv211210786S,2021MNRAS.507.5013M}, which would prevent the formation of a BBH merger in an isolated binary. Here, we investigate whether the subsequent evolution of the postmerger star and a tertiary companion could form a BBH merger. For this purpose, we employ a binary stellar evolution code which we adjust to triple evolution. We use this code to simulate a population of hierarchical triple stars in the galactic field, i.e., without any environmental perturbation.

Throughout this work, $G$ and $Z_\odot=0.02$ refer to the gravitational constant and solar metallicity, respectively. Colored versions of the figures are available in the online journal.

\section{Methods}

\subsection{Stellar evolution}\label{sec:Stellar evolution}
In this work, we are interested in the formation of BBH mergers from the evolution of stars in the field of galaxies.
%stellar evolution of hierarchical triples and, as a reference, isolated binaries which lead to BBH mergers.
We consider and compare the two following populations.
\begin{itemize}
\item Triple population: Starting from a hierarchical triple population, stable BBHs are formed which subsequently merge within a Hubble time due to the emission of gravitational waves. These BBHs might form in two different ways as following.
\begin{itemize}
 \item Inner binary channel: The two stars in the inner binary form a stable BBH.
 \item Outer binary channel: The two stars in the inner binary merge and the postmerger star and tertiary companion subsequently form a stable BBH.
\end{itemize}
\item Isolated binary population: Starting from an isolated binary population the binary stars form stable BBHs which subsequently merge within a Hubble time due to the emission of gravitational waves. This is a standard population model used in the literature for which the effect of a tertiary companion is not considered \citep[][]{1967AcA....17..355P,1992ApJ...391..246P,2012ApJ...759...52D,2016Natur.534..512B,2018MNRAS.480.2011G,2021A&A...651A.100O,2013ApJ...764..166D,2016A&A...588A..10R,2021PhRvD.103f3007S,2021MNRAS.507.5013M,2021A&A...651A.100O}. 
\end{itemize}
Throughout this work, \textit{stellar merger} refers to the merger of two stars (or one star and a compact object), whereas \textit{BBH merger} to the merger of a BBH promoted by gravitational wave radiation. The formation of BBH mergers via the outer binary channel is the novel mechanism  investigated in this work.

To follow the stellar evolution with time $t$, we use the code {\tt MOBSE} \citep[][]{2018MNRAS.474.2959G,2018MNRAS.480.2011G,2019MNRAS.482.2234G,2020ApJ...891..141G} which is an update of the widely-adopted binary stellar evolution code {\tt BSE} \citep[][]{2002MNRAS.329..897H}. {\tt BSE/MOBSE} models all relevant evolutionary steps of stellar binaries including mass transfer episodes, CE evolution, and tidal interaction. {\tt MOBSE} improves {\tt BSE} by including up-to-date metal-dependent stellar wind prescriptions, fallback kicks imparted to the remnants of supernovae (SNe) explosions, and (pulsational) pair instability supernovae \citep[][]{2018MNRAS.474.2959G,2018MNRAS.480.2011G,2019MNRAS.482.2234G,2020ApJ...891..141G}. For our purpose, {\tt MOBSE} provides an adequate tool to simulate isolated binaries, starting on the zero-age-main-sequence.

\begin{figure*}
    \centering
     \begin{subfigure}
         \centering
         \includegraphics[height=120pt]{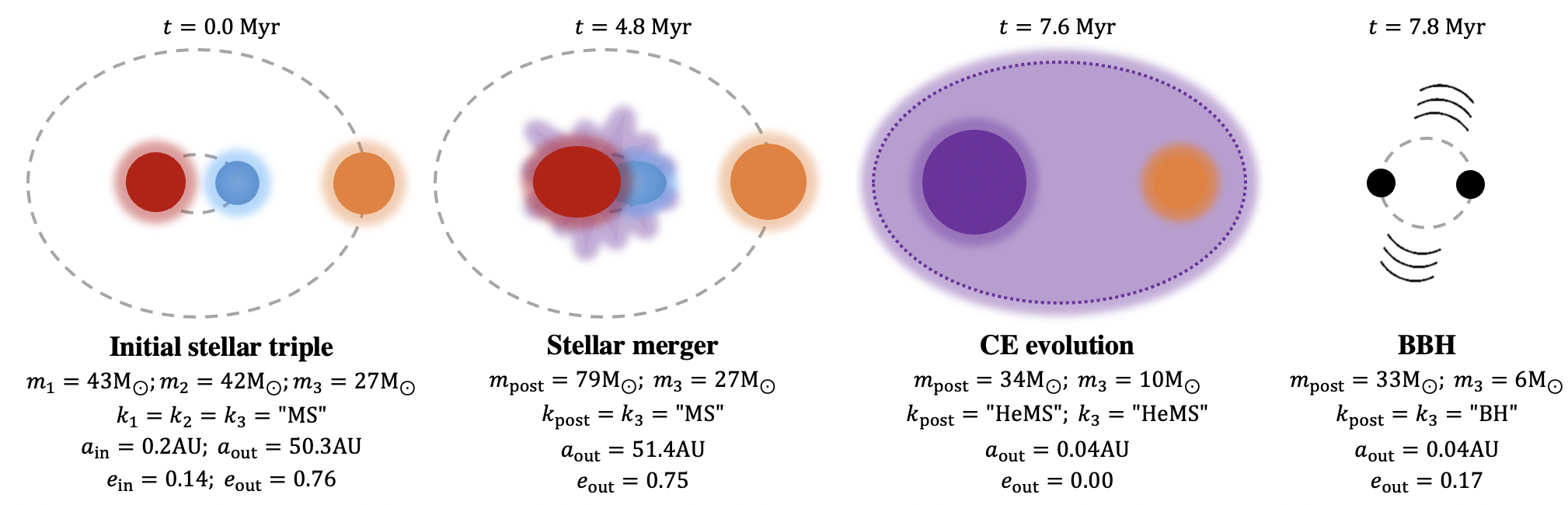}
     \end{subfigure}\\
     \vspace{12pt}
     \begin{subfigure}
         \centering
         \includegraphics[height=120pt]{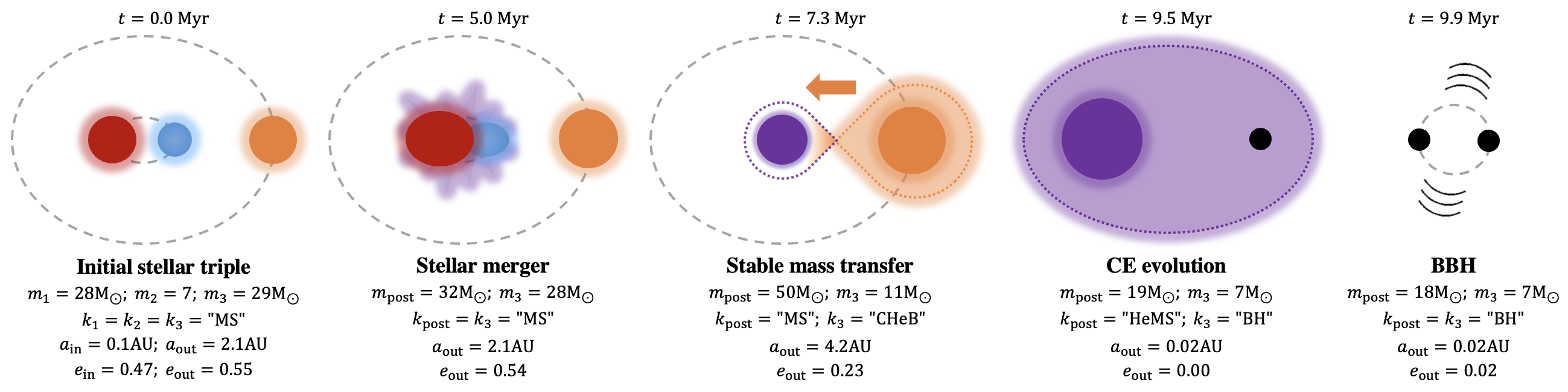}
     \end{subfigure}
        \caption{Two examples of low mass ratio BBH mergers due to the outer binary channel in the {\tt Rapid1} model at metallicity $Z\simeq1\times10^{-4}$ (upper panel) and $5\times10^{-4}$ (lower panel). The final BBHs merge after $t_{\rm GW}\simeq172\,\rm Myr$ and $29\,\rm Myr$, respectively. The type of the stars is described by $k_{1,2,3,{\rm post}}$, where CHeB and HeMS refers to a Core Helium Burning and Naked Helium MS star, respectively. The provided parameter values refer to the end of each event, e.g., to the time at which the two stars fully merge, the stable mass transfer peters out, or the CE is successfully ejected leaving behind a close binary. Distances and eccentricities are not drawn to scale.}
        \label{fig:example}
\end{figure*}

Applied to triples, we use {\tt MOBSE} to evolve the inner binary and tertiary companion as two dynamically independent entities as done in previous triple studies \citep[][]{2017ApJ...841...77A,2018MNRAS.480L..58A,2018ApJ...863....7R,2021arXiv210501671M}. We consider hierarchical stellar triples in which a close inner binary with semimajor axis $a_{\rm in}$ and eccentricity $e_{\rm in}$ is orbited by a distant tertiary star with semimajor axis $a_{\rm out}\gg a_{\rm in}$ and eccentricity $e_{\rm out}$. We refer to the masses of the inner binary stars and the tertiary companion as $m_{1,2}$ and $m_3$, respectively (see Figure~\ref{fig:example}).
\begin{comment}
\begin{figure*}
 \includegraphics[width=2\columnwidth]{Figures/example.png}
 \caption{Example of a BBH merger due to the outer binary channel in the {\tt Tout97} model at metallicity $Z=4\times10^{-4}$. The final BBH merges after  $t_{\rm GW}\simeq30.7\,\rm Myr$. The type of the stars is described by $k_{1,2,3,{\rm post}}$, where CHeB and HeMS refers to a Core Helium Burning and Naked Helium MS star, respectively. Distances and eccentricities are not drawn to scale.}
 \label{fig:example}
\end{figure*}
\end{comment}

In reality, the distant tertiary companion could perturb the dynamics of the inner binary through the Lidov-Kozai mechanism, leading to large amplitude oscillations of $e_{\rm in}(t)$ \citep[][]{1962P&SS....9..719L,1962AJ.....67..591K}. Previous studies show that this could alter the evolution of the inner binary stars by inducing eccentric mass transfer \citep[][]{2021arXiv211210786S,2020A&A...640A..16T,2019ApJ...872..119H}. Moreover, the long-term interaction with the companion might drive the inner binary to a merger after a BBH is formed \citep[][]{2017ApJ...836...39S,2017ApJ...841...77A,2018MNRAS.480L..58A,2018ApJ...863...68L,2018ApJ...863....7R,2019ApJ...881...41L,2020MNRAS.493.3920F,2021arXiv210501671M}. In order to investigate the effect of dynamics on the outer binary channel, we include one computationally expensive model ({\tt 3BodyDynamics}) in which we reevolve all systems which above lead to a BBH merger using the secular three-body integrator, {\tt TSE}, presented by \citet[][]{2021arXiv211210786S}. This code allows to simultaneously evolve the stellar physics and the secular dynamical equations due to the Lidov-Kozai effect, tides, and post-Newtonian corrections. Only reevolving the systems which above lead to a BBH merger potentially ignores systems in which a stellar merger is solely driven by the dynamical effect of tertiary \citep[][]{2022A&A...661A..61T}, i.e., it would not occur if the inner binary was in isolation. Hence, the resulting rates of BBH mergers in the {\tt 3BodyDynamics} must be treated as lower limits. As a result, we will show below that the tertiary effect on the distribution of these BBH mergers is small. Thus, in all the other models we neglect dynamics, which allows us to efficiently explore the relevant parameter space.

\begin{comment}
Firstly, we expect that Lidov-Kozai oscillations would only increase the number of stellar mergers since they periodically cause a close encounter of the inner binary stars. Thus, not including three-body dynamics yields a lower limit of the number of BBH mergers formed via the outer binary channel.

Secondly, by evaluating relevant timescales \citep[][]{2016ARA&A..54..441N} we find that in the majority of triples which form BBHs via the outer binary channel ($\simeq75\,\%$) the Lidov-Kozai effect is quenched by the fast \citet{1916AbhKP1916..189S} precession in the inner binary.

Thirdly, we will see that BBH mergers via the outer binary channel originate from exceptionally compact stellar triples with initial outer semimajor axes $a_{\rm out,init}\lesssim10^1$~--~$10^2\,\rm AU$. Observations for solar-type stars suggest that the inner and outer orbits of these kind of triples tend to be coplanar \citep[][]{2016MNRAS.455.4136B,2017ApJ...844..103T}, for which the tertiary perturbation is suppressed at leading order \citep[][]{2016ARA&A..54..441N}. If future observations show that this trend persists in massive stellar triples, three-body dynamics can be neglected for the compact triples of interest.
\end{comment}

During the inner binary evolution $a_{\rm in}$ changes due to tides, SN kicks, gravitational wave emission, stellar winds or during an episode of mass transfer. These processes are self-consistently treated by {\tt MOBSE} (and likewise for the isolated binary population). For the outer binary evolution we proceed as \citet[][Section~3]{2018ApJ...863....7R} and expand $a_{\rm out}(t)$ according to the fractional mass loss from the system, e.g., due to winds and during a mass transfer episode in the inner binary.

At any point in time we check whether the triples become dynamically unstable \citep[]{2001MNRAS.321..398M},
or if the tertiary companions fills their Roche lobe \citep[][]{1983ApJ...268..368E}. In either events we stop the evolution of the systems because their subsequent evolution is uncertain. Recent studies propose that additional stellar mergers could be triggered by a mass-transfer phase initiated by Roche lobe filling tertiary companions \citep[][]{2021MNRAS.500.1921G,2022arXiv220305357L}.

During the stellar evolution the inner binary stars might merge. This happens when the two stars undergo a CE in which their inspiralling cores coalesce before the envelope could be ejected (see below), or when two stars of similar compactness, e.g., two main-sequence (MS) stars, collide. Our description of stellar mergers follows closely that of \citet[][]{Glebbeek} and \citet[][]{2002MNRAS.329..897H}. Most relevant to our work are MS-MS stellar mergers (see Section~\ref{sec:results}). This kind of merger yields another MS star which is rejuvenated. That is, the additional hydrogen fuel delays the time at which the postmerger star leaves its MS. In general, the rejuvenation process can be described as
\begin{equation}\label{eq:Tout}
    \tau_{\rm post}=\frac{1}{\alpha q_{c,\rm post}}\frac{1}{1-\phi}\frac{q_{c,1}m_1\tau_1+q_{c,2}m_2\tau_2}{m_{\rm post}},
\end{equation}
where $\alpha$ parameterizes the amount of mixing and $\tau_{1,2,\rm post}\in[0,1]$ is the fractional timescale of the primary, secondary, and postmerger star on their MS, respectively \citep[][]{Glebbeek}. $\phi$ is the fractional mass loss during the merger (see below) and $q_{c,1,2,\rm post}$ the effective core mass fraction defined as the fraction of hydrogen that is burned during the MS of the primary, secondary, and postmerger star, respectively. Stellar observables like the radius and luminosity substantially increase only toward the end of the MS evolution, $\tau_{\rm post}\rightarrow1.0$ \citep[][]{2000MNRAS.315..543H}. Rejuvenation is equivalent to $\tau_{\rm post}<{\rm max}(\tau_1,\tau_2)$, i.e., the postmerger star appears younger than the most evolved inner binary star did (or both).

Here, we adopt the mass-dependent approximation for the effective core mass fractions of \citet[][]{Glebbeek} and a mixing parameter of $\alpha=1.14$ \citep[][]{2016MNRAS.457.2355S}. As a default, {\tt BSE/MOBSE} follows \citet[][]{1997MNRAS.291..732T} using $\phi=0$, $q_{c,1}=q_{c,2}=q_{c,\rm post}=0.1$, and $\alpha=10$ \citep[][]{Glebbeek}. This is based on the supposition that the merging stars fully mix and that the end of the MS is reached when 10 per cent of the total hydrogen fuel has been burnt \citep[][]{1997MNRAS.291..732T}. Thus, this prescription is likely to overestimate rejuvenation since it is expected that the MS stars do not fully mix and that less core hydrogen is replenished \citep[][]{Glebbeek,2013ApJ...764..166D}. We explore the original prescription in one additional model ({\tt Tout97}).

The mass $m_{\rm post}$ of the postmerger star is uncertain. Here, we follow \citet[][]{10.1093/mnras/stt1268} and assume that during a MS-MS merger the system suffers a fractional mass-loss 
\begin{equation}
    \phi=\frac{m_{12}-m_{\rm post}}{m_{12}}=0.3\frac{q_{\rm in}}{(1+q_{\rm in})^2},
\end{equation}
where, $m_{12}=m_1+m_2$ and $q_{\rm in}={\rm min}(m_1,m_2)/{\rm max}(m_1,m_2)$.

Immediately after the merger, it is expected that the resulting star undergoes a bloated phase where its radius expands before it contracts to its equilibrium state on the (thermal) Kelvin-Helmholtz timescale \citep[][]{2007ApJ...668..435S,2021MNRAS.503.4276H}. If it is sufficiently close, the postmerger orbit may be partially or entirely enclosed by the outermost parts of the bloated postmerger star. This situation may give rise to interesting transient phenomena as the tertiary companion plunges into the bloated envelope \citep[][]{2016MNRAS.456.3401P,2021MNRAS.503.4276H}. In general, it is not expected that the postmerger orbit gets significantly perturbed by the interaction between companion and envelope since the bloated phase is brief and only a small fraction of the stellar mass undergoes a large expansion \citep[][]{2020MNRAS.495.2796S}. Nevertheless, we also include a  conservative model ({\tt DiscardBloated}) in which we discard any system whose outer periapsis $a_{\rm out}(1-e_{\rm out})$ at the time of the merger is smaller than the radius of the bloated star. Unfortunately, it is not well understood by how much the radius of the bloated postmerger star expands. Here, we use the Hayashi limit as a theoretical upper bound to the radius, $R_{\rm post}=\sqrt{L_{\rm post}/4\pi\sigma T_{\rm post}^4}$, where $\sigma$ is the Stefan-Boltzmann constant and we assume a surface temperature limit of $T_{\rm post}=10^{3.7}\,\rm K$. For the luminosity $L$ we use the Eddington limit $L_{\rm Edd}/L_\odot\simeq3.2\times10^4m_{\rm post}/M_\odot$ beyond which the star would leave hydrostatic equilibrium and its hydrogen envelope would suffer intense radiation-driven winds. Computed in this way the radius serves as strict upper limit to that of the bloated star.

After a stellar merger we use {\tt MOBSE} to continue the integration of the outer orbit composed of the postmerger star and the tertiary companion. If the system survives the subsequent stellar evolution to form a stable BBH, we then calculate the merger timescale induced by gravitational wave emission \citep[][]{1964PhRv..136.1224P}. We add this timescale to the time elapsed until BBH formation to obtain the delay time between star formation and BBH merger. An analogue expression holds for BBHs formed in inner and isolated binaries, respectively. An example evolution of two BBHs formed via the outer binary channel is sketched in Figure~\ref{fig:example}.

In our models, most BBHs merge only if the orbits of the isolated, inner, or postmerger binaries, respectively, significantly inspiraled during a CE evolution. Otherwise, the progenitor stars are too far apart for the resulting BBH to merge within a Hubble time. 

A CE evolution occurs when there is a collision involving a giant star with a dense core or if there is a mass transfer phase from a giant on a dynamical timescale. In either cases the giant's envelope engulfs the orbit of the binary companion and the giant's core. The orbit suffers a friction-driven decay within the envelope which heats up the latter. As a result, the companion and core either coalesce within the envelope or the latter is ejected leaving behind a tight binary which could subsequently form a (merging) BBH. Thus, the efficiency at which orbital energy of the inspiralling cores is transferred to the envelope significantly affects the number of surviving binaries but is very uncertain \citep[][]{2013A&ARv..21...59I}. Here, we adopt the standard $\alpha_{\rm CE}$-formalism \citep[][]{2002MNRAS.329..897H} where the efficiency is described by the free parameter $\alpha_{\rm CE}$.

In this work, we study a set of plausible values for $\alpha_{\rm CE}$ which are summarized in Table~\ref{tab:parameters}. Additionally, we explore different BH formation mechanisms following the ``rapid" and ``delayed" SN model from \citet[][]{10.1093/mnras/stv1161} and assume different treatments of rejuvenation and the bloated stars. We also include three-body dynamics in one of our models as described above.
\begin{table*}
	\centering
	\caption{Model parameters. In the model {\tt 3BodyDynamics}, only systems are reevolved that lead to BBH mergers in the default model {\tt Rapid1} via the outer binary channel.}	
	\label{tab:parameters}
	\begingroup
	\renewcommand*{\arraystretch}{1.2}
	\begin{tabular}{c|c|c|c|c|c}
	\hline
	\hline
	Model name & SN prescription & $\alpha_{\rm CE}$ & Rejuvenation & Three-body dynamics & Bloated star encloses tertiary \\
    \hline
    %{\tt Rapid0.5} & Rapid & 0.5 & \citet[][]{1997MNRAS.291..732T} & \\
    {\tt Rapid1} (default) & Rapid \citep[][]{10.1093/mnras/stv1161} & 1.0 & \citet[][]{Glebbeek} & Neglected & Continued\\
    {\tt Rapid3} & Rapid \citep[][]{10.1093/mnras/stv1161} & 3.0 & \citet[][]{Glebbeek} & Neglected & Continued\\
    {\tt Rapid5} & Rapid \citep[][]{10.1093/mnras/stv1161} & 5.0 & \citet[][]{Glebbeek} & Neglected & Continued\\
    {\tt Delayed} & Delayed \citep[][]{10.1093/mnras/stv1161} & 1.0 & \citet[][]{Glebbeek} & Neglected & Continued\\
    {\tt DiscardBloated} & Rapid \citep[][]{10.1093/mnras/stv1161} & 1.0 & \citet[][]{Glebbeek} & Neglected & Discarded\\
    {\tt Tout97} & Rapid \citep[][]{10.1093/mnras/stv1161} & 1.0 & \citet[][]{1997MNRAS.291..732T} & Neglected & Continued\\
    {\tt 3BodyDynamics} & Rapid \citep[][]{10.1093/mnras/stv1161} & 1.0 & \citet[][]{Glebbeek} & Included & Continued \\
    \hline
    \hline
	\end{tabular}
	\endgroup
\end{table*}

\subsection{Initial conditions}\label{sec:Initial conditions}
In order to set up the initial conditions of the triple parameters on the zero-age-main-sequence, we adopt simple probability density functions motivated by observations. We sample the primary mass $m_1$ from a standard \citet[][]{2001MNRAS.322..231K} mass function 
%\begin{equation}
    $N(m_1)\propto m_1^{-2.3},$
%\end{equation} 
between $5$ and $150\,\rm M_\odot$. For the % density functions of the 
mass ratio $q_{\rm in}=m_2/m_1$, eccentricity $e_{\rm in}$, and orbital period $P_{\rm in}=2\pi a_{\rm in}^{3/2}/G^{1/2}(m_1+m_2)^{1/2}$ we adopt simple power-law fits to the observational data of Galactic binaries with O-type primaries \citep[][]{2012Sci...337..444S}: $N(q_{\rm in})\propto q_{\rm in}^{-0.1}$ between $0.1$ and $1.0$, $N(e_{\rm in})\propto e_{\rm in}^{-0.45}$ between $0.0$ and $0.9$,
and $N\left(\log (P_{\rm in}/{\rm days})\right)\propto \left(\log (P_{\rm in}/{\rm days})\right)^{-0.55}$
between $0.15$ and $5.5$. We only keep systems in which also $m_2\ge5\,\rm M_\odot$ since otherwise in neither of the investigated populations BBHs could be formed.

To any inner binary we repetitively propose tertiary companions until they meet the stability criterion \citep[]{2001MNRAS.321..398M}, %Eq.~\eqref{eq:stability}
beyond which the triple would become chaotic and our adopted method breaks down. To this end, we sample the tertiary mass $m_3$ from a uniform distribution between $8\,\rm M_\odot$ and $m_1+m_2$, the outer eccentricity from a thermal distribution $N(e_{\rm out})\propto e_{\rm out}$ between $0.0$ and $0.9$, and the outer semimajor axis from a log-uniform distribution between $a_{\rm in}$ and $10^4\,\rm AU$.

For our model with three-body dynamics included, we assume random orientations of the inner and outer orbit.

Our agnostic choices for the parameter distributions of the outer binary reflects the poor statistics with which tertiary companions to massive binary stars have been observed so far \citep[][]{2017ApJS..230...15M}. In turn, keeping the sample of inner binaries while discarding tertiary companions which would lead to instability ensures that the observed distribution of close (inner) binaries is recovered \citep[][]{2021arXiv211210786S}.

The massive stellar progenitors of BHs are consistent with a near hundred percent fraction of triples and higher order systems \citep[][]{2017ApJS..230...15M}. 
Nevertheless, most previous work focused on the evolution of isolated binaries \citep[e.g.,][]{1967AcA....17..355P,1992ApJ...391..246P,2012ApJ...759...52D,2016Natur.534..512B,2018MNRAS.480.2011G,2021A&A...651A.100O,2013ApJ...764..166D,2016A&A...588A..10R,2021PhRvD.103f3007S,2021MNRAS.507.5013M,2021A&A...651A.100O}.
Thus, as a reference, we also study the evolution of an isolated binary population without tertiary companions. Comparing it to the triple population allows us to discern the impact of the simplified assumption made in the literature. For our isolated binary population we simply use the same distributions of masses and orbital parameters as for the inner binaries of the triple population. Since the orbital distributions of the latter were taken from observational surveys of binaries, their initial conditions are consistent with observations.

\section{Results}\label{sec:results}
\begin{figure}
 \includegraphics[width=\columnwidth]{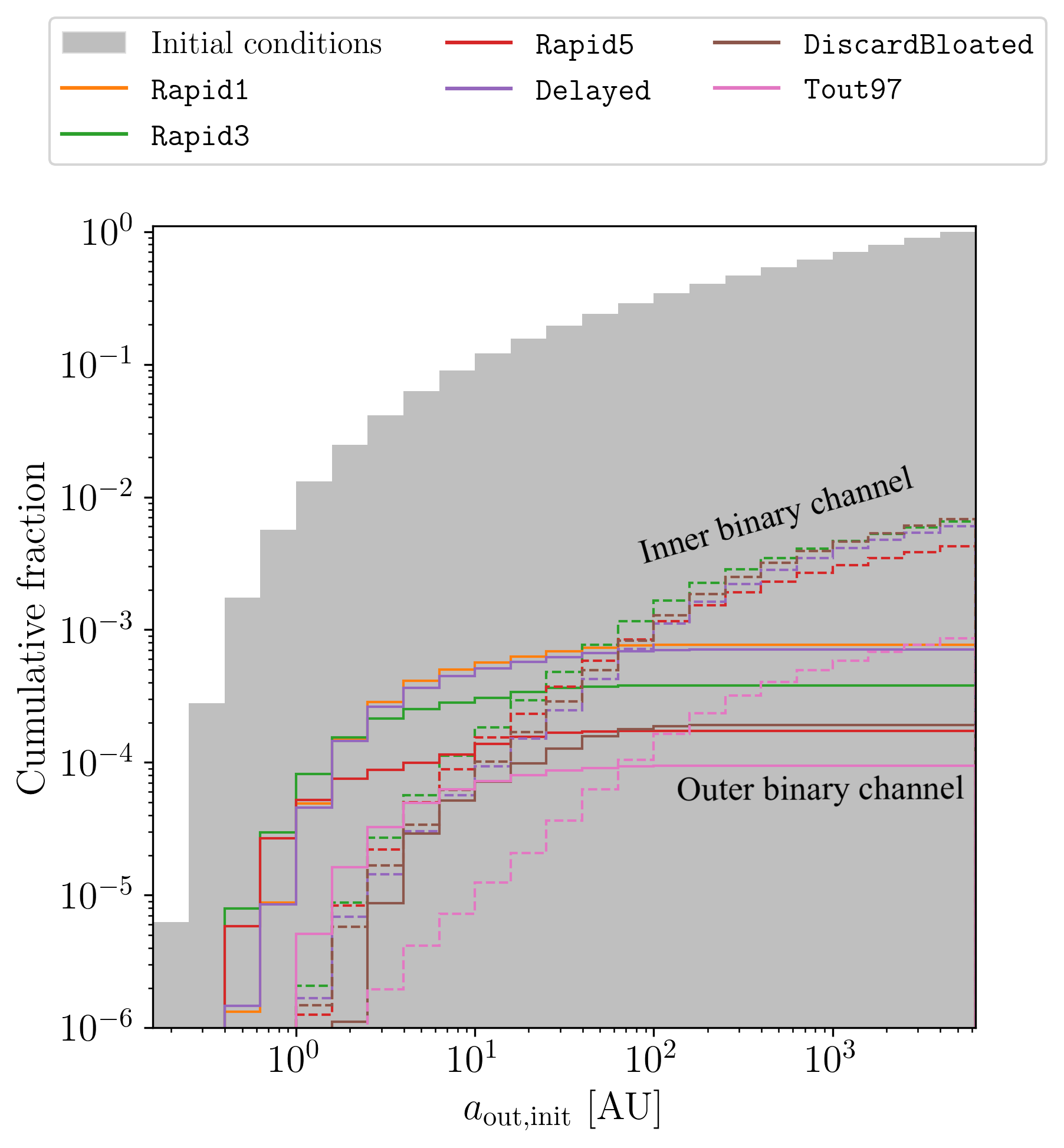}
 \caption{Cumulative distribution of BBH mergers from triple population as a function of the initial outer semimajor axis $a_{\rm out,init}$. Solid lines correspond to mergers via the outer binary channel and dashed lines to the inner binary channel. The outer binary channel dominates the formation of BBH mergers from compact triples with $a_{\rm out,init}\lesssim10^1$~--~$10^2\,\rm AU$, where the precise value depends on the assumed model. The metallicity is sampled log-uniformly.}
 \label{fig:aout}
\end{figure}
A key parameter that determines the relative efficiency of both triple channels is their initial outer semimajor axis $a_{\rm out,init}$. In Figure~\ref{fig:aout}, we show the cumulative fraction of BBH mergers formed in either channel as a function of $a_{\rm out,init}$. In any of our models, we find the outer binary channel to be the dominant way of forming BBH mergers from triples with $a_{\rm out,init}\lesssim10^1$~--~$10^2\,\rm AU$. This is simply because the inner orbits of these compact triples must be even closer in order to ensure dynamical stability and hierarchy. This makes the inner binary stars prone to undergo a stellar merger. Above $a_{\rm out,init}\gtrsim10^1$~--~$10^2\,\rm AU$, the inner binary channel dominates. In these systems, a postmerger star and tertiary companion are too far apart from each other to undergo a mass transfer episode which is necessary to shrink their orbit. Hence, these systems are unable to form BBHs that end up close enough to merge within a Hubble time. Thus, in the entire triple population the inner BBH mergers outweigh those formed via the outer binary channel by a factor of $\sim\mathcal{O}(10)$.

\begin{figure}
 \includegraphics[width=\columnwidth]{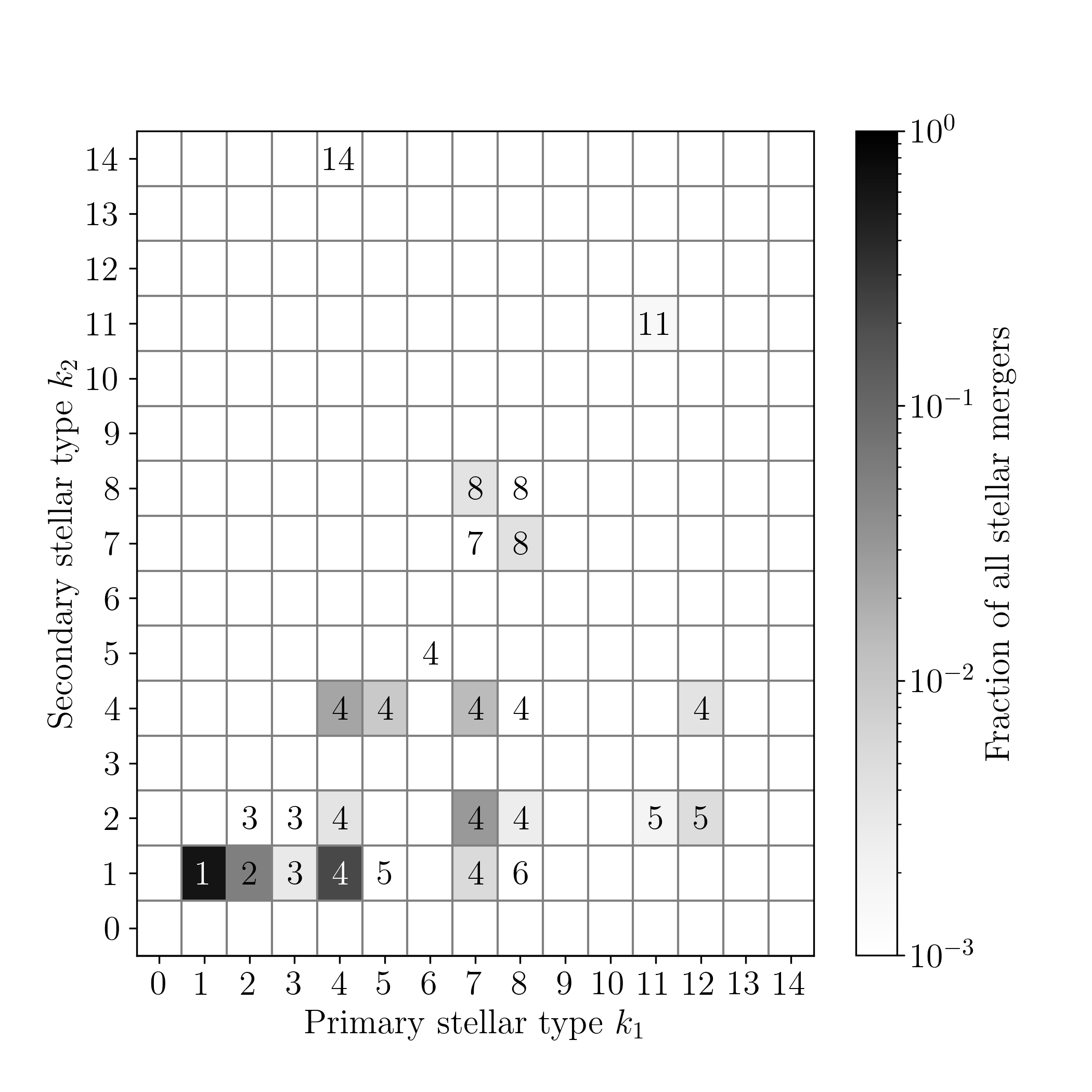}
 \caption{Collision matrix of stellar mergers in the {\tt Rapid1} model. The grayscale correspond to the fraction of each merger type normalized to one. The axes and integers in each cell indicate the stellar type of the merging stars and the postmerger star, respectively (only cells with nonzero fractions are described). The integers are defined as in \citet[][Section~1]{2002MNRAS.329..897H}, e.g., $k_{1,2}=1$ for MS stars.}
 \label{fig:collision-matrix}
\end{figure}
As a typical example we show in Figure~\ref{fig:collision-matrix} the distribution of stellar types which merge in the {\tt Rapid1} model. In any model, the majority (ranging from $\simeq 60\,\%$ to $80\,\%$ in the {\tt Rapid1} and {\tt Rapid5} model, respectively) of stars already merge on their MS yielding another star on the MS. The merger of these stars is a direct collision. By the time of the stellar merger, most ($\simeq 80\,\%$ to $90\,\%$) of the tertiary stars are still on their MS as well. Hence, these merger occur at a relatively early evolutionary stage of all three stars, typically after a few $\rm Myr$.

Previously it has been suggested that stellar mergers of an evolved star with a carbon-oxygen core and a MS star could produce a postmerger star that circumvents a pair-instability SN \citep[][]{2019MNRAS.487.2947D,2022arXiv220403493B}. Thus, it has been suggested that GW190521-like events \citep[][]{2020PhRvL.125j1102A} with the primary BH mass being in the upper mass gap are possibly formed in young stellar clusters \citep[][]{2019MNRAS.487.2947D,2020MNRAS.497.1043D,2020MNRAS.498..495D,2020ApJ...903...45K}. We note that our MS-MS mergers are not expected to produce a star which could populate the upper mass gap.

In both populations and channels, the fraction of systems which lead to BBH mergers is higher at low metallicity. We find that a fraction $\sim\mathcal{O}(10^{-2})$ of the isolated binaries and inner binaries, respectively, evolve to merging BBHs if $Z\lesssim\mathcal{O}(10^{-3})$. Above $Z\gtrsim\mathcal{O}(10^{-3})$ the fraction sharply drops to fractions $\sim\mathcal{O}(10^{-5})$ at solar metallicity. Similarly, the number of BBH mergers via the outer binary channel falls from $\sim\mathcal{O}(10^{-3})$ at low metallicities to $\sim\mathcal{O}(10^{-5})$ at solar metallicity.

\begin{figure}
 \includegraphics[width=\columnwidth]{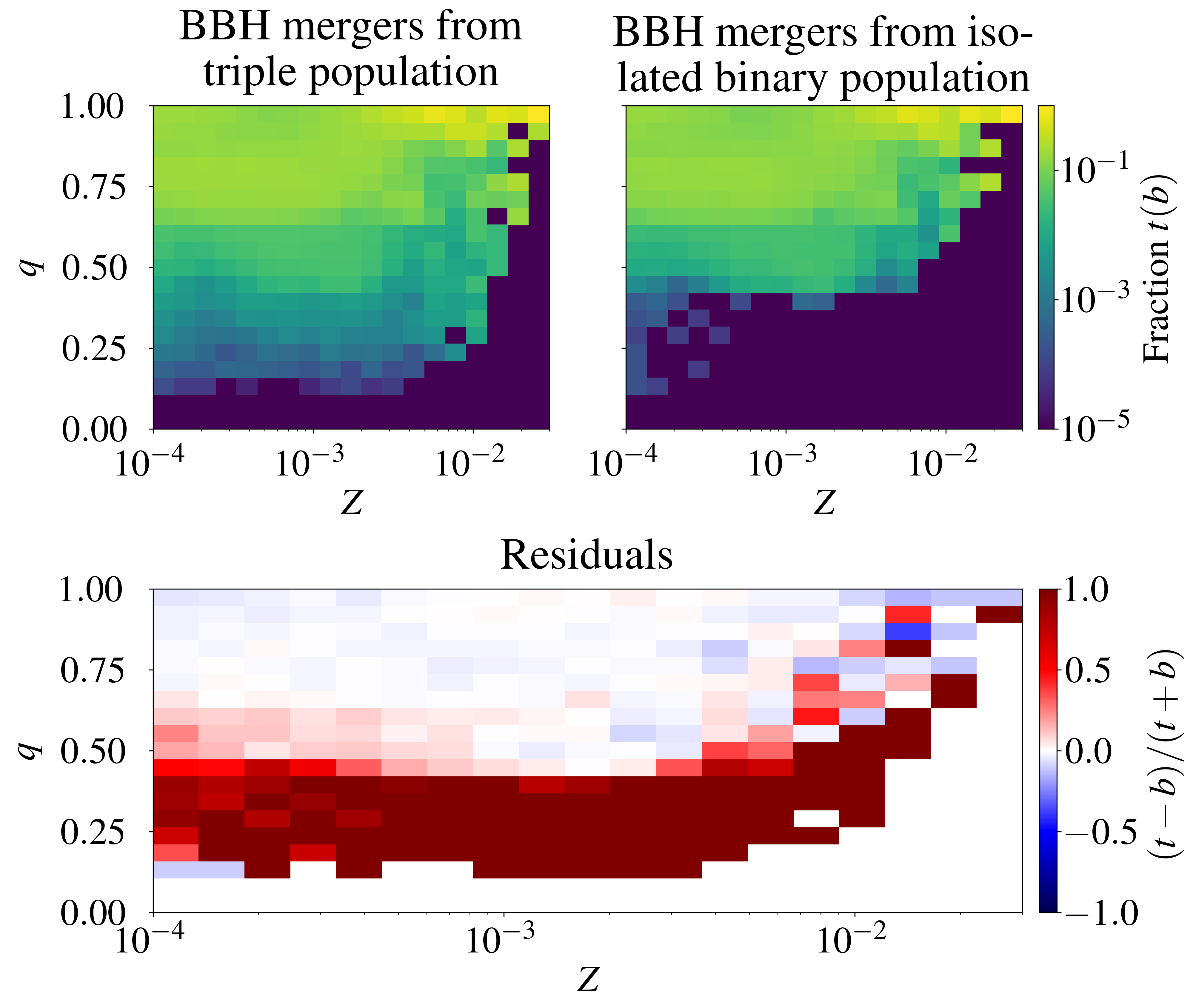}
 \caption{BBH merger distribution as a function of metallicity in the {\tt Rapid1} model. The upper panels show for each metallicity the mass ratio distribution from triples and binaries, respectively. The lower panel shows the relative residuals between both populations. For any bin, they are defined as $(t-b)/(t+b)$ where $t$ corresponds to the bin value in the triple population and $b$ to that of the binary population. Thus, colored areas indicate a large difference between the two populations.}
 \label{fig:metallicity}
\end{figure}

As shown in Figure~\ref{fig:metallicity}, BBH mergers from the triple and isolated binary population are predominantly formed with equal masses ($q\simeq1$). This results from the CE evolution of the progenitor stars which precedes most of our BBH mergers. In the $\alpha_{\rm CE}$-formalism, low mass ratio stellar binaries are more susceptible to merge within a CE rather than forming a close binary which could eventually lead to a BBH merger.

Yet, the BBH mergers formed from the triple population tend to have lower mass ratios than those from the isolated binary population. Although  in both populations, we find BBH mergers with mass ratios as low as $q\simeq0.1$, BBH mergers with $q\lesssim0.4$~--~$0.6$ are much more frequently formed from triples because of the outer binary channel. In this channel, we identify two ways that facilitate the formation of low mass ratio BBH mergers which are shown in Figure~\ref{fig:example}.

%\begin{figure}
% \includegraphics[width=\columnwidth]{Figures/delay-time.png}
% \caption{Delay time distribution of BBHs as a function of mass ratio in the {\tt Rapid1} model. The time $t-t_b$ refers to the interval between zero-age-main-sequence of the progenitor stars and merger of the resulting black holes.}
% \label{fig:delay-time}
%\end{figure}

First, a stellar merger simply produces a more massive star. Thus, there are systems in which the postmerger star is much more massive than the tertiary companion. Typically, in these systems the mass of the postmerger star is larger than $m_{\rm post}\gtrsim60\,\rm M_\odot$ whereas that of the tertiary companion is $m_3\simeq20$~--~$30\,\rm M_\odot$. While details depend on metallicity and the SN prescription \citep[][]{10.1093/mnras/stv1161,2012ApJ...749...91F}, there is the general tendency that more massive stars also form more massive BHs. As exemplified in the upper panel of Figure~\ref{fig:example}, a low mass ratio BBH merger is formed after the massive postmerger star initiated a CE evolution during which the orbit of both stellar cores efficiently decays. This evolutionary pathway is responsible for the formation of the lowest BBH mass ratios down to $q\simeq0.1$.

Second, as shown above, the very large majority of BBH mergers in the outer binary channel are preceded by a MS-MS stellar merger. The resulting MS star is rejuvenated. Thus, there are systems in which the postmerger star is more massive but less evolved than the tertiary companion star. Consequently, the latter fills its Roche lobe first and initiates a stable mass transfer phase onto the postmerger star as exemplified in the lower panel of Figure~\ref{fig:example}. This further increases the imbalance of the stellar progenitor masses and the resulting BH masses. In this way, low mass ratio BBH mergers with $q\gtrsim0.3$ can be formed. It requires the mass of the postmerger star to be only a few solar masses $\lesssim10\,\rm M_\odot$ larger than that of the tertiary progenitor star.

BBH mergers from an isolated binary population evolve differently. Typically, the progenitor of the primary BH was the donor star of the first stable mass transfer \citep[e.g.,][Figure 1]{2016Natur.534..512B} which reduces the imbalance of the resulting BH masses. We note that previous models of isolated binary populations can produce a larger number of low mass ratio BBH mergers only under certain assumptions \citep{2017NatCo...814906S,2020ApJ...901L..39O,2020A&A...636A.104B}. As discussed by \citet{2020A&A...636A.104B}, the resulting mass ratios of BBH mergers crucially depend on the fraction $f_a\in[0,1]$ of transferred mass that is accreted by the progenitor of the secondary BH. Lower values of $f_a$ tend to produce lower mass secondary stars/BHs and therefore lead to smaller mass ratios of BBH mergers. For example, \citet[][]{2022arXiv220302515Z} explore different values of $f_a$ and find that the bulk of their BBH population has $q\gtrsim0.2$ for $f_a=0$ (fully nonconservative) whereas  $q\gtrsim0.4$ for $f_a=1$ (fully conservative). Unfortunately, $f_a$ is poorly constrained and any value between zero and one seems possible \citep{2001A&A...369..939W,2007A&A...467.1181D,2014MNRAS.441.1166E,2015ApJ...805...20S,2016ApJ...833..108S,2017A&A...599A..84M,2018MNRAS.481.5660D,2020A&A...636A.104B}.

In the remainder of this work, we investigate the BBH merger rates from both populations and compare them to the one inferred from
gravitational wave detections. 

In the local Universe the total BBH merger rate inferred from GWTC-3 is $\mathcal{R}(z=0)=16.7^{+16.5}_{-8.7}\,\rm Gpc^{-3}\,yr^{-1}$ ($90\,\%$ C.L.) by using the ``flexible mixture model" \citep{2021CQGra..38o5007T, 2021ApJ...913L..19T, 2021arXiv211113991T}.
To calculate the BBH merger rate from our models, we consider the rate and metallicity at which the stellar progenitor systems are formed throughout cosmic history and convolve it with the delay time distribution. Thus, the rate at a given redshift $z$ takes into account all systems that were formed at some earlier redshift $z_b>z$ and whose delay time matches the cosmic time elapsed between $z_b$ and $z$. We use the metallicity-dependent cosmic star formation rate of \citet[][]{2017ApJ...840...39M} and adopt concordant $\Lambda$CDM cosmology. Details of our calculations are provided in the Appendix~\ref{sec:Merger rate density}. 

In Table~\ref{tab:rates}, we report the resulting total rates for in our models at $z=0$. Across all models, we find the median contribution of the outer binary channel to the total BBH merger rate to be in the range $\mathcal{R}(z=0)=0.3$~--~$25.2\,\rm Gpc^{-3}\,yr^{-1}$ which amounts to a typical fraction $\sim\mathcal{O}(0.01$~--~$0.1)$ of the BBH merger rate from the triple population. In any model, the total rates from the triple and isolated binary population are in the same order of magnitude as the inferred. Thus, although our medians tend to overpredict the inferred local BBH merger rate by a factor of two to three they are in good agreement compared to formation channels that were previously proposed \citep[][]{2021arXiv210714239M}.
\begin{table*}
	\centering
	\caption{BBH merger rates $\mathcal{R}(z=0)$ in $\rm Gpc^{-3}\,yr^{-1}$. For the total rate of the triple population, the contributions from both channels have to be added. The numbers in the table indicate the median rates and $5$~--~$95\,\%$ credible intervals, respectively.}	
	\label{tab:rates}
	\begingroup
	\renewcommand*{\arraystretch}{1.2}
	\begin{tabular}{c|cc|c|c}
	\hline
	\hline
	\multirow{2}{*}{Model name} & \multicolumn{2}{c|}{Triple population} & \multirow{2}{*}{Isolated binary population} & \multirow{2}{*}{GWTC-3} \\
	& Inner binary channel & Outer binary channel & &\\
    \hline
$\tt Rapid1$ & $42.6 ^{+ 27.4 }_{- 18.4 }$ & $12.5 ^{+ 12.7 }_{- 6.1 }$ & $46.1 ^{+ 32.1 }_{- 17.6 }$
& \multirow{ 6 }{*}{$ 16.7 ^{+ 16.5 }_{- 8.7 }$}
\\
$\tt Rapid3$ & $48.3 ^{+ 33.0 }_{- 17.8 }$ & $1.6 ^{+ 2.4 }_{- 0.8 }$ & $37.2 ^{+ 31.9 }_{- 13.4 }$
&
\\
$\tt Rapid5$ & $16.5 ^{+ 13.9 }_{- 4.3 }$ & $0.7 ^{+ 1.8 }_{- 0.4 }$ & $16.9 ^{+ 14.8 }_{- 7.5 }$
&
\\
$\tt Delayed$ & $21.7 ^{+ 15.2 }_{- 8.0 }$ & $5.3 ^{+ 14.6 }_{- 3.2 }$ & $22.1 ^{+ 26.7 }_{- 9.2 }$
&
\\
$\tt DiscardBloated$ & $42.6 ^{+ 31.7 }_{- 14.6 }$ & $5.7 ^{+ 9.5 }_{- 2.8 }$ & $46.1 ^{+ 32.1 }_{- 17.6 }$
&
\\
$\tt Tout97$ & $44.5 ^{+ 19.1 }_{- 12.8 }$ & $2.8 ^{+ 5.6 }_{- 1.5 }$ & $46.1 ^{+ 32.1 }_{- 17.6 }$
&
\\
    \hline
    \hline
	\end{tabular}
	\endgroup
\end{table*}
\begin{figure*}
 \includegraphics[width=2\columnwidth]{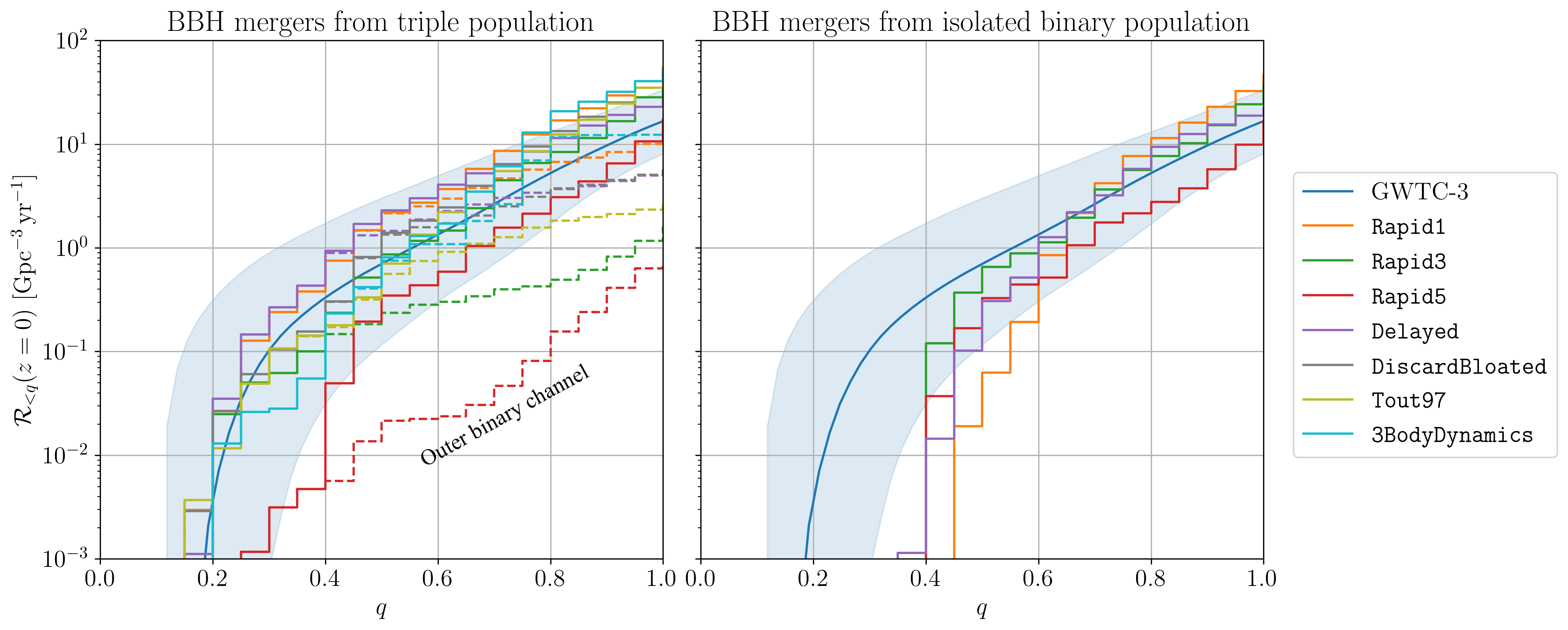}
 \caption{Cumulative mass ratio distribution of BBH mergers in the local Universe ($z=0$). In the left panel, we show the BBH mergers from the triple population and in the right panel from the isolated binary population (solid lines). In the former, we highlight the contribution of BBH mergers via the outer binary channel (dashed lines) which start to dominate below $q\lesssim0.4$~--~$0.6$. All lines indicate median values. The blue envelope shows the $5$~--~$95\,\%$ credible interval of GWTC-3.}
 \label{fig:q-rate}
\end{figure*}

Despite only adding a subdominant fraction to the total merger rate, the BBH mergers formed via the outer binary channel significantly affect the mass ratio distribution of the entire population. In Figure~\ref{fig:q-rate}, we plot the cumulative rate 
\begin{equation}
    \mathcal{R}_{<q}(z=0)=\int_0^q\frac{{\rm d}\mathcal{R}(z=0)}{{\rm d}q'}\,{\rm d}q'
\end{equation}
of BBH mergers as a function of their mass ratio $q$. Neither the inner binary channel in triples nor isolated binaries are found to reproduce well the mass ratio distribution inferred by the gravitational wave detections. While they agree well with the cumulative rate above $q\simeq0.4$~--~$0.6$ they fail at lower mass ratios. The inferred mass ratio distribution can be better recovered after including the contribution from the outer binary channel (dashed lines). For the aforementioned reasons, it is found to be more efficient in producing low mass ratio BBH mergers leading to a less steeply decreasing distribution toward low values of $q$. Below the threshold mass ratio $q\simeq0.4$~--~$0.6$, BBH mergers from the outer binary channel dominate the inner binary channel. The precise threshold value depends on the assumed model. 

In summary, the stellar mergers and the subsequent coevolution with the tertiary companions inflate the number of heavy primary stars which generally lead to heavier BHs. Consequently, more low mass ratio BBH mergers are produced in the triple population. ized{Thus, exceptional BBH merger events like GW190412 with reported component masses $m_1=30.1^{+4.6}_{-5.3}\,\rm M_\odot$ and $m_2=8.3^{+1.6}_{-0.9}\,\rm M_\odot$ \citep[][]{2020PhRvD.102d3015A} are a possible outcome of the outer binary channel.}
Meanwhile, the marginalized mass distribution of BBH mergers is not significantly altered. We find the chirp mass distribution everywhere to be dominated by the high mass ratio BBH mergers formed via the inner binary channel (see Figure~\ref{fig:chirp} in Appendix~\ref{sec:chirp}). Likewise, we do not expect the outer binary channel to leave a distinctive imprint on the eccentricity distribution. Since a CE evolution precedes most BBH mergers in both channels, we expect high eccentricities to be suppressed. A thorough investigation of the spins of merging BHs is beyond the scope of this work. Yet, the outer binary channel might explain the observed correlation between low mass ratios and higher effective spin parameters \citep[][]{1981A&A...102...17P,2021arXiv211103634T}. It is expected that even little mass accretion efficiently spins up the postmerger star during the stable mass transfer episode \citep[][]{2021ApJ...923..277R}. Previous magneto-hydrodynamical simulations suggest that the postmerger star is strongly magnetized \citep[][]{2019Natur.574..211S,2020MNRAS.495.2796S}. A strong core-envelope coupling by the magnetic fields would ensure that the core also spins up due to the mass transfer phase \citep[][]{2019ApJ...881L...1F}. As a result, low mass ratio BBH mergers with highly spinning primary BHs may be formed.

\section*{Acknowledgements}
The authors thank Mark Gieles, Stephen Fairhurst, Ilya Mandel, Simon Stevenson, Aleksandra Olejak, Alejandro Vigna-G\'{o}mez, and the anonymous referee for helpful discussions and comments. FA acknowledges support from a Rutherford fellowship (ST/P00492X/1) of the Science and Technology Facilities Council. This work has received funding from the European Research Council (ERC) under the European Union’s Horizon 2020 research and innovation programme (Grant agreement No.\ 945806) and is supported by the Deutsche Forschungsgemeinschaft (DFG, German Research Foundation) under Germany’s Excellence Strategy EXC 2181/1-390900948 (the Heidelberg STRUCTURES Excellence Cluster).

\section*{Data Availability}
The data underlying this article will be shared on reasonable request to the authors.

\bibliography{ms}% Produces the bibliography via BibTeX.

%apsrev4-2.bst 2019-01-14 (MD) hand-edited version of apsrev4-1.bst
%Control: key (0)
%Control: author (8) initials jnrlst
%Control: editor formatted (1) identically to author
%Control: production of article title (0) allowed
%Control: page (0) single
%Control: year (1) truncated
%Control: production of eprint (0) enabled
\begin{thebibliography}{102}%
\makeatletter
\providecommand \@ifxundefined [1]{%
 \@ifx{#1\undefined}
}%
\providecommand \@ifnum [1]{%
 \ifnum #1\expandafter \@firstoftwo
 \else \expandafter \@secondoftwo
 \fi
}%
\providecommand \@ifx [1]{%
 \ifx #1\expandafter \@firstoftwo
 \else \expandafter \@secondoftwo
 \fi
}%
\providecommand \natexlab [1]{#1}%
\providecommand \enquote  [1]{``#1''}%
\providecommand \bibnamefont  [1]{#1}%
\providecommand \bibfnamefont [1]{#1}%
\providecommand \citenamefont [1]{#1}%
\providecommand \href@noop [0]{\@secondoftwo}%
\providecommand \href [0]{\begingroup \@sanitize@url \@href}%
\providecommand \@href[1]{\@@startlink{#1}\@@href}%
\providecommand \@@href[1]{\endgroup#1\@@endlink}%
\providecommand \@sanitize@url [0]{\catcode `\\12\catcode `\$12\catcode
  `\&12\catcode `\#12\catcode `\^12\catcode `\_12\catcode `\%12\relax}%
\providecommand \@@startlink[1]{}%
\providecommand \@@endlink[0]{}%
\providecommand \url  [0]{\begingroup\@sanitize@url \@url }%
\providecommand \@url [1]{\endgroup\@href {#1}{\urlprefix }}%
\providecommand \urlprefix  [0]{URL }%
\providecommand \Eprint [0]{\href }%
\providecommand \doibase [0]{https://doi.org/}%
\providecommand \selectlanguage [0]{\@gobble}%
\providecommand \bibinfo  [0]{\@secondoftwo}%
\providecommand \bibfield  [0]{\@secondoftwo}%
\providecommand \translation [1]{[#1]}%
\providecommand \BibitemOpen [0]{}%
\providecommand \bibitemStop [0]{}%
\providecommand \bibitemNoStop [0]{.\EOS\space}%
\providecommand \EOS [0]{\spacefactor3000\relax}%
\providecommand \BibitemShut  [1]{\csname bibitem#1\endcsname}%
\let\auto@bib@innerbib\@empty
%</preamble>
\bibitem [{\citenamefont {{Abbott, B.~P. et al.}}(2016)}]{2016PhRvL.116f1102A}%
  \BibitemOpen
  \bibfield  {author} {\bibinfo {author} {\bibnamefont {{Abbott, B.~P. et
  al.}}},\ }\bibfield  {title} {\bibinfo {title} {{Observation of Gravitational
  Waves from a Binary Black Hole Merger}},\ }\href
  {https://doi.org/10.1103/PhysRevLett.116.061102} {\bibfield  {journal}
  {\bibinfo  {journal} {PRL}\ }\textbf {\bibinfo {volume} {116}},\ \bibinfo
  {eid} {061102} (\bibinfo {year} {2016})},\ \Eprint
  {https://arxiv.org/abs/1602.03837} {arXiv:1602.03837 [gr-qc]} \BibitemShut
  {NoStop}%
\bibitem [{\citenamefont {{The LIGO Scientific Collaboration}}\ \emph
  {et~al.}(2021)\citenamefont {{The LIGO Scientific Collaboration}},
  \citenamefont {{The Virgo Collaboration}},\ and\ \citenamefont {{The KAGRA
  Scientific Collaboration}}}]{2021arXiv211103634T}%
  \BibitemOpen
  \bibfield  {author} {\bibinfo {author} {\bibnamefont {{The LIGO Scientific
  Collaboration}}}, \bibinfo {author} {\bibnamefont {{The Virgo
  Collaboration}}},\ and\ \bibinfo {author} {\bibnamefont {{The KAGRA
  Scientific Collaboration}}},\ }\bibfield  {title} {\bibinfo {title} {{The
  population of merging compact binaries inferred using gravitational waves
  through GWTC-3}},\ }\href@noop {} {\bibfield  {journal} {\bibinfo  {journal}
  {arXiv e-prints}\ ,\ \bibinfo {eid} {arXiv:2111.03634}} (\bibinfo {year}
  {2021})},\ \Eprint {https://arxiv.org/abs/2111.03634} {arXiv:2111.03634
  [astro-ph.HE]} \BibitemShut {NoStop}%
\bibitem [{\citenamefont {{Rodriguez}}\ \emph {et~al.}(2016)\citenamefont
  {{Rodriguez}}, \citenamefont {{Chatterjee}},\ and\ \citenamefont
  {{Rasio}}}]{2016PhRvD..93h4029R}%
  \BibitemOpen
  \bibfield  {author} {\bibinfo {author} {\bibfnamefont {C.~L.}\ \bibnamefont
  {{Rodriguez}}}, \bibinfo {author} {\bibfnamefont {S.}~\bibnamefont
  {{Chatterjee}}},\ and\ \bibinfo {author} {\bibfnamefont {F.~A.}\ \bibnamefont
  {{Rasio}}},\ }\bibfield  {title} {\bibinfo {title} {{Binary black hole
  mergers from globular clusters: Masses, merger rates, and the impact of
  stellar evolution}},\ }\href {https://doi.org/10.1103/PhysRevD.93.084029}
  {\bibfield  {journal} {\bibinfo  {journal} {PRD}\ }\textbf {\bibinfo {volume}
  {93}},\ \bibinfo {eid} {084029} (\bibinfo {year} {2016})},\ \Eprint
  {https://arxiv.org/abs/1602.02444} {arXiv:1602.02444 [astro-ph.HE]}
  \BibitemShut {NoStop}%
\bibitem [{\citenamefont {{Park}}\ \emph {et~al.}(2017)\citenamefont {{Park}},
  \citenamefont {{Kim}}, \citenamefont {{Lee}}, \citenamefont {{Bae}},\ and\
  \citenamefont {{Belczynski}}}]{2017MNRAS.469.4665P}%
  \BibitemOpen
  \bibfield  {author} {\bibinfo {author} {\bibfnamefont {D.}~\bibnamefont
  {{Park}}}, \bibinfo {author} {\bibfnamefont {C.}~\bibnamefont {{Kim}}},
  \bibinfo {author} {\bibfnamefont {H.~M.}\ \bibnamefont {{Lee}}}, \bibinfo
  {author} {\bibfnamefont {Y.-B.}\ \bibnamefont {{Bae}}},\ and\ \bibinfo
  {author} {\bibfnamefont {K.}~\bibnamefont {{Belczynski}}},\ }\bibfield
  {title} {\bibinfo {title} {{Black hole binaries dynamically formed in
  globular clusters}},\ }\href {https://doi.org/10.1093/mnras/stx1015}
  {\bibfield  {journal} {\bibinfo  {journal} {MNRAS}\ }\textbf {\bibinfo
  {volume} {469}},\ \bibinfo {pages} {4665} (\bibinfo {year} {2017})},\ \Eprint
  {https://arxiv.org/abs/1703.01568} {arXiv:1703.01568 [astro-ph.HE]}
  \BibitemShut {NoStop}%
\bibitem [{\citenamefont {{Rodriguez}}\ and\ \citenamefont
  {{Loeb}}(2018)}]{2018ApJ...866L...5R}%
  \BibitemOpen
  \bibfield  {author} {\bibinfo {author} {\bibfnamefont {C.~L.}\ \bibnamefont
  {{Rodriguez}}}\ and\ \bibinfo {author} {\bibfnamefont {A.}~\bibnamefont
  {{Loeb}}},\ }\bibfield  {title} {\bibinfo {title} {{Redshift Evolution of the
  Black Hole Merger Rate from Globular Clusters}},\ }\href
  {https://doi.org/10.3847/2041-8213/aae377} {\bibfield  {journal} {\bibinfo
  {journal} {APJl}\ }\textbf {\bibinfo {volume} {866}},\ \bibinfo {eid} {L5}
  (\bibinfo {year} {2018})},\ \Eprint {https://arxiv.org/abs/1809.01152}
  {arXiv:1809.01152 [astro-ph.HE]} \BibitemShut {NoStop}%
\bibitem [{\citenamefont {{Antonini}}\ and\ \citenamefont
  {{Gieles}}(2020)}]{2020PhRvD.102l3016A}%
  \BibitemOpen
  \bibfield  {author} {\bibinfo {author} {\bibfnamefont {F.}~\bibnamefont
  {{Antonini}}}\ and\ \bibinfo {author} {\bibfnamefont {M.}~\bibnamefont
  {{Gieles}}},\ }\bibfield  {title} {\bibinfo {title} {{Merger rate of black
  hole binaries from globular clusters: Theoretical error bars and comparison
  to gravitational wave data from GWTC-2}},\ }\href
  {https://doi.org/10.1103/PhysRevD.102.123016} {\bibfield  {journal} {\bibinfo
   {journal} {PRD}\ }\textbf {\bibinfo {volume} {102}},\ \bibinfo {eid}
  {123016} (\bibinfo {year} {2020})},\ \Eprint
  {https://arxiv.org/abs/2009.01861} {arXiv:2009.01861 [astro-ph.HE]}
  \BibitemShut {NoStop}%
\bibitem [{\citenamefont {{Banerjee}}\ \emph {et~al.}(2010)\citenamefont
  {{Banerjee}}, \citenamefont {{Baumgardt}},\ and\ \citenamefont
  {{Kroupa}}}]{2010MNRAS.402..371B}%
  \BibitemOpen
  \bibfield  {author} {\bibinfo {author} {\bibfnamefont {S.}~\bibnamefont
  {{Banerjee}}}, \bibinfo {author} {\bibfnamefont {H.}~\bibnamefont
  {{Baumgardt}}},\ and\ \bibinfo {author} {\bibfnamefont {P.}~\bibnamefont
  {{Kroupa}}},\ }\bibfield  {title} {\bibinfo {title} {{Stellar-mass black
  holes in star clusters: implications for gravitational wave radiation}},\
  }\href {https://doi.org/10.1111/j.1365-2966.2009.15880.x} {\bibfield
  {journal} {\bibinfo  {journal} {MNRAS}\ }\textbf {\bibinfo {volume} {402}},\
  \bibinfo {pages} {371} (\bibinfo {year} {2010})},\ \Eprint
  {https://arxiv.org/abs/0910.3954} {arXiv:0910.3954 [astro-ph.SR]}
  \BibitemShut {NoStop}%
\bibitem [{\citenamefont {{Ziosi}}\ \emph {et~al.}(2014)\citenamefont
  {{Ziosi}}, \citenamefont {{Mapelli}}, \citenamefont {{Branchesi}},\ and\
  \citenamefont {{Tormen}}}]{2014MNRAS.441.3703Z}%
  \BibitemOpen
  \bibfield  {author} {\bibinfo {author} {\bibfnamefont {B.~M.}\ \bibnamefont
  {{Ziosi}}}, \bibinfo {author} {\bibfnamefont {M.}~\bibnamefont {{Mapelli}}},
  \bibinfo {author} {\bibfnamefont {M.}~\bibnamefont {{Branchesi}}},\ and\
  \bibinfo {author} {\bibfnamefont {G.}~\bibnamefont {{Tormen}}},\ }\bibfield
  {title} {\bibinfo {title} {{Dynamics of stellar black holes in young star
  clusters with different metallicities - II. Black hole-black hole
  binaries}},\ }\href {https://doi.org/10.1093/mnras/stu824} {\bibfield
  {journal} {\bibinfo  {journal} {MNRAS}\ }\textbf {\bibinfo {volume} {441}},\
  \bibinfo {pages} {3703} (\bibinfo {year} {2014})},\ \Eprint
  {https://arxiv.org/abs/1404.7147} {arXiv:1404.7147 [astro-ph.GA]}
  \BibitemShut {NoStop}%
\bibitem [{\citenamefont {{Di Carlo}}\ \emph {et~al.}(2019)\citenamefont {{Di
  Carlo}}, \citenamefont {{Giacobbo}}, \citenamefont {{Mapelli}}, \citenamefont
  {{Pasquato}}, \citenamefont {{Spera}}, \citenamefont {{Wang}},\ and\
  \citenamefont {{Haardt}}}]{2019MNRAS.487.2947D}%
  \BibitemOpen
  \bibfield  {author} {\bibinfo {author} {\bibfnamefont {U.~N.}\ \bibnamefont
  {{Di Carlo}}}, \bibinfo {author} {\bibfnamefont {N.}~\bibnamefont
  {{Giacobbo}}}, \bibinfo {author} {\bibfnamefont {M.}~\bibnamefont
  {{Mapelli}}}, \bibinfo {author} {\bibfnamefont {M.}~\bibnamefont
  {{Pasquato}}}, \bibinfo {author} {\bibfnamefont {M.}~\bibnamefont {{Spera}}},
  \bibinfo {author} {\bibfnamefont {L.}~\bibnamefont {{Wang}}},\ and\ \bibinfo
  {author} {\bibfnamefont {F.}~\bibnamefont {{Haardt}}},\ }\bibfield  {title}
  {\bibinfo {title} {{Merging black holes in young star clusters}},\ }\href
  {https://doi.org/10.1093/mnras/stz1453} {\bibfield  {journal} {\bibinfo
  {journal} {MNRAS}\ }\textbf {\bibinfo {volume} {487}},\ \bibinfo {pages}
  {2947} (\bibinfo {year} {2019})},\ \Eprint {https://arxiv.org/abs/1901.00863}
  {arXiv:1901.00863 [astro-ph.HE]} \BibitemShut {NoStop}%
\bibitem [{\citenamefont {{Fragione}}\ and\ \citenamefont
  {{Banerjee}}(2021)}]{2021ApJ...913L..29F}%
  \BibitemOpen
  \bibfield  {author} {\bibinfo {author} {\bibfnamefont {G.}~\bibnamefont
  {{Fragione}}}\ and\ \bibinfo {author} {\bibfnamefont {S.}~\bibnamefont
  {{Banerjee}}},\ }\bibfield  {title} {\bibinfo {title} {{Binary Black Hole
  Mergers from Young Massive and Open Clusters: Comparison to GWTC-2
  Gravitational Wave Data}},\ }\href {https://doi.org/10.3847/2041-8213/ac00a7}
  {\bibfield  {journal} {\bibinfo  {journal} {APJl}\ }\textbf {\bibinfo
  {volume} {913}},\ \bibinfo {eid} {L29} (\bibinfo {year} {2021})},\ \Eprint
  {https://arxiv.org/abs/2103.10447} {arXiv:2103.10447 [astro-ph.HE]}
  \BibitemShut {NoStop}%
\bibitem [{\citenamefont {{Antonini}}\ and\ \citenamefont
  {{Rasio}}(2016)}]{2016ApJ...831..187A}%
  \BibitemOpen
  \bibfield  {author} {\bibinfo {author} {\bibfnamefont {F.}~\bibnamefont
  {{Antonini}}}\ and\ \bibinfo {author} {\bibfnamefont {F.~A.}\ \bibnamefont
  {{Rasio}}},\ }\bibfield  {title} {\bibinfo {title} {{Merging Black Hole
  Binaries in Galactic Nuclei: Implications for Advanced-LIGO Detections}},\
  }\href {https://doi.org/10.3847/0004-637X/831/2/187} {\bibfield  {journal}
  {\bibinfo  {journal} {APJ}\ }\textbf {\bibinfo {volume} {831}},\ \bibinfo
  {eid} {187} (\bibinfo {year} {2016})},\ \Eprint
  {https://arxiv.org/abs/1606.04889} {arXiv:1606.04889 [astro-ph.HE]}
  \BibitemShut {NoStop}%
\bibitem [{\citenamefont {{Petrovich}}\ and\ \citenamefont
  {{Antonini}}(2017)}]{2017ApJ...846..146P}%
  \BibitemOpen
  \bibfield  {author} {\bibinfo {author} {\bibfnamefont {C.}~\bibnamefont
  {{Petrovich}}}\ and\ \bibinfo {author} {\bibfnamefont {F.}~\bibnamefont
  {{Antonini}}},\ }\bibfield  {title} {\bibinfo {title} {{Greatly Enhanced
  Merger Rates of Compact-object Binaries in Non-spherical Nuclear Star
  Clusters}},\ }\href {https://doi.org/10.3847/1538-4357/aa8628} {\bibfield
  {journal} {\bibinfo  {journal} {APJ}\ }\textbf {\bibinfo {volume} {846}},\
  \bibinfo {eid} {146} (\bibinfo {year} {2017})},\ \Eprint
  {https://arxiv.org/abs/1705.05848} {arXiv:1705.05848 [astro-ph.HE]}
  \BibitemShut {NoStop}%
\bibitem [{\citenamefont {{Hamilton}}\ and\ \citenamefont
  {{Rafikov}}(2019)}]{2019ApJ...881L..13H}%
  \BibitemOpen
  \bibfield  {author} {\bibinfo {author} {\bibfnamefont {C.}~\bibnamefont
  {{Hamilton}}}\ and\ \bibinfo {author} {\bibfnamefont {R.~R.}\ \bibnamefont
  {{Rafikov}}},\ }\bibfield  {title} {\bibinfo {title} {{Compact Object Binary
  Mergers Driven By Cluster Tides: A New Channel for LIGO/Virgo
  Gravitational-wave Events}},\ }\href
  {https://doi.org/10.3847/2041-8213/ab3468} {\bibfield  {journal} {\bibinfo
  {journal} {APJl}\ }\textbf {\bibinfo {volume} {881}},\ \bibinfo {eid} {L13}
  (\bibinfo {year} {2019})},\ \Eprint {https://arxiv.org/abs/1907.00994}
  {arXiv:1907.00994 [astro-ph.GA]} \BibitemShut {NoStop}%
\bibitem [{\citenamefont {{Bub}}\ and\ \citenamefont
  {{Petrovich}}(2020)}]{2020ApJ...894...15B}%
  \BibitemOpen
  \bibfield  {author} {\bibinfo {author} {\bibfnamefont {M.~W.}\ \bibnamefont
  {{Bub}}}\ and\ \bibinfo {author} {\bibfnamefont {C.}~\bibnamefont
  {{Petrovich}}},\ }\bibfield  {title} {\bibinfo {title} {{Compact-object
  Mergers in the Galactic Center: Evolution in Triaxial Clusters}},\ }\href
  {https://doi.org/10.3847/1538-4357/ab8461} {\bibfield  {journal} {\bibinfo
  {journal} {APJ}\ }\textbf {\bibinfo {volume} {894}},\ \bibinfo {eid} {15}
  (\bibinfo {year} {2020})},\ \Eprint {https://arxiv.org/abs/1910.02079}
  {arXiv:1910.02079 [astro-ph.HE]} \BibitemShut {NoStop}%
\bibitem [{\citenamefont {{Wang}}\ \emph {et~al.}(2021)\citenamefont {{Wang}},
  \citenamefont {{Stephan}}, \citenamefont {{Naoz}}, \citenamefont {{Hoang}},\
  and\ \citenamefont {{Breivik}}}]{2021ApJ...917...76W}%
  \BibitemOpen
  \bibfield  {author} {\bibinfo {author} {\bibfnamefont {H.}~\bibnamefont
  {{Wang}}}, \bibinfo {author} {\bibfnamefont {A.~P.}\ \bibnamefont
  {{Stephan}}}, \bibinfo {author} {\bibfnamefont {S.}~\bibnamefont {{Naoz}}},
  \bibinfo {author} {\bibfnamefont {B.-M.}\ \bibnamefont {{Hoang}}},\ and\
  \bibinfo {author} {\bibfnamefont {K.}~\bibnamefont {{Breivik}}},\ }\bibfield
  {title} {\bibinfo {title} {{Gravitational-wave Signatures from Compact Object
  Binaries in the Galactic Center}},\ }\href
  {https://doi.org/10.3847/1538-4357/ac088d} {\bibfield  {journal} {\bibinfo
  {journal} {APJ}\ }\textbf {\bibinfo {volume} {917}},\ \bibinfo {eid} {76}
  (\bibinfo {year} {2021})},\ \Eprint {https://arxiv.org/abs/2010.15841}
  {arXiv:2010.15841 [astro-ph.HE]} \BibitemShut {NoStop}%
\bibitem [{\citenamefont {{Clesse}}\ and\ \citenamefont
  {{Garc{\'\i}a-Bellido}}(2018)}]{2018PDU....22..137C}%
  \BibitemOpen
  \bibfield  {author} {\bibinfo {author} {\bibfnamefont {S.}~\bibnamefont
  {{Clesse}}}\ and\ \bibinfo {author} {\bibfnamefont {J.}~\bibnamefont
  {{Garc{\'\i}a-Bellido}}},\ }\bibfield  {title} {\bibinfo {title} {{Seven
  hints for primordial black hole dark matter}},\ }\href
  {https://doi.org/10.1016/j.dark.2018.08.004} {\bibfield  {journal} {\bibinfo
  {journal} {Physics of the Dark Universe}\ }\textbf {\bibinfo {volume} {22}},\
  \bibinfo {pages} {137} (\bibinfo {year} {2018})},\ \Eprint
  {https://arxiv.org/abs/1711.10458} {arXiv:1711.10458 [astro-ph.CO]}
  \BibitemShut {NoStop}%
\bibitem [{\citenamefont {{Clesse}}\ and\ \citenamefont
  {{Garcia-Bellido}}(2020)}]{2020arXiv200706481C}%
  \BibitemOpen
  \bibfield  {author} {\bibinfo {author} {\bibfnamefont {S.}~\bibnamefont
  {{Clesse}}}\ and\ \bibinfo {author} {\bibfnamefont {J.}~\bibnamefont
  {{Garcia-Bellido}}},\ }\bibfield  {title} {\bibinfo {title} {{GW190425,
  GW190521 and GW190814: Three candidate mergers of primordial black holes from
  the QCD epoch}},\ }\href@noop {} {\bibfield  {journal} {\bibinfo  {journal}
  {arXiv e-prints}\ ,\ \bibinfo {eid} {arXiv:2007.06481}} (\bibinfo {year}
  {2020})},\ \Eprint {https://arxiv.org/abs/2007.06481} {arXiv:2007.06481
  [astro-ph.CO]} \BibitemShut {NoStop}%
\bibitem [{\citenamefont {{De Luca}}\ \emph {et~al.}(2020)\citenamefont {{De
  Luca}}, \citenamefont {{Franciolini}}, \citenamefont {{Pani}},\ and\
  \citenamefont {{Riotto}}}]{2020JCAP...06..044D}%
  \BibitemOpen
  \bibfield  {author} {\bibinfo {author} {\bibfnamefont {V.}~\bibnamefont {{De
  Luca}}}, \bibinfo {author} {\bibfnamefont {G.}~\bibnamefont {{Franciolini}}},
  \bibinfo {author} {\bibfnamefont {P.}~\bibnamefont {{Pani}}},\ and\ \bibinfo
  {author} {\bibfnamefont {A.}~\bibnamefont {{Riotto}}},\ }\bibfield  {title}
  {\bibinfo {title} {{Primordial black holes confront LIGO/Virgo data: current
  situation}},\ }\href {https://doi.org/10.1088/1475-7516/2020/06/044}
  {\bibfield  {journal} {\bibinfo  {journal} {JCAP}\ }\textbf {\bibinfo
  {volume} {2020}}\bibfield  {number} {\bibinfo  {number} { (6)},\ \bibinfo
  {eid} {044}},\ }\Eprint {https://arxiv.org/abs/2005.05641} {arXiv:2005.05641
  [astro-ph.CO]} \BibitemShut {NoStop}%
\bibitem [{\citenamefont {{B{\oe}hm}}\ \emph {et~al.}(2021)\citenamefont
  {{B{\oe}hm}}, \citenamefont {{Kobakhidze}}, \citenamefont {{O'Hare}},
  \citenamefont {{Picker}},\ and\ \citenamefont
  {{Sakellariadou}}}]{2021JCAP...03..078B}%
  \BibitemOpen
  \bibfield  {author} {\bibinfo {author} {\bibfnamefont {C.}~\bibnamefont
  {{B{\oe}hm}}}, \bibinfo {author} {\bibfnamefont {A.}~\bibnamefont
  {{Kobakhidze}}}, \bibinfo {author} {\bibfnamefont {C.~A.~J.}\ \bibnamefont
  {{O'Hare}}}, \bibinfo {author} {\bibfnamefont {Z.~S.~C.}\ \bibnamefont
  {{Picker}}},\ and\ \bibinfo {author} {\bibfnamefont {M.}~\bibnamefont
  {{Sakellariadou}}},\ }\bibfield  {title} {\bibinfo {title} {{Eliminating the
  LIGO bounds on primordial black hole dark matter}},\ }\href
  {https://doi.org/10.1088/1475-7516/2021/03/078} {\bibfield  {journal}
  {\bibinfo  {journal} {JCAP}\ }\textbf {\bibinfo {volume} {2021}}\bibfield
  {number} {\bibinfo  {number} { (3)},\ \bibinfo {eid} {078}},\ }\Eprint
  {https://arxiv.org/abs/2008.10743} {arXiv:2008.10743 [astro-ph.CO]}
  \BibitemShut {NoStop}%
\bibitem [{\citenamefont {{Dominik}}\ \emph {et~al.}(2012)\citenamefont
  {{Dominik}}, \citenamefont {{Belczynski}}, \citenamefont {{Fryer}},
  \citenamefont {{Holz}}, \citenamefont {{Berti}}, \citenamefont {{Bulik}},
  \citenamefont {{Mandel}},\ and\ \citenamefont
  {{O'Shaughnessy}}}]{2012ApJ...759...52D}%
  \BibitemOpen
  \bibfield  {author} {\bibinfo {author} {\bibfnamefont {M.}~\bibnamefont
  {{Dominik}}}, \bibinfo {author} {\bibfnamefont {K.}~\bibnamefont
  {{Belczynski}}}, \bibinfo {author} {\bibfnamefont {C.}~\bibnamefont
  {{Fryer}}}, \bibinfo {author} {\bibfnamefont {D.~E.}\ \bibnamefont {{Holz}}},
  \bibinfo {author} {\bibfnamefont {E.}~\bibnamefont {{Berti}}}, \bibinfo
  {author} {\bibfnamefont {T.}~\bibnamefont {{Bulik}}}, \bibinfo {author}
  {\bibfnamefont {I.}~\bibnamefont {{Mandel}}},\ and\ \bibinfo {author}
  {\bibfnamefont {R.}~\bibnamefont {{O'Shaughnessy}}},\ }\bibfield  {title}
  {\bibinfo {title} {{Double Compact Objects. I. The Significance of the Common
  Envelope on Merger Rates}},\ }\href
  {https://doi.org/10.1088/0004-637X/759/1/52} {\bibfield  {journal} {\bibinfo
  {journal} {APJ}\ }\textbf {\bibinfo {volume} {759}},\ \bibinfo {eid} {52}
  (\bibinfo {year} {2012})},\ \Eprint {https://arxiv.org/abs/1202.4901}
  {arXiv:1202.4901 [astro-ph.HE]} \BibitemShut {NoStop}%
\bibitem [{\citenamefont {{Belczynski}}\ \emph {et~al.}(2016)\citenamefont
  {{Belczynski}}, \citenamefont {{Holz}}, \citenamefont {{Bulik}},\ and\
  \citenamefont {{O'Shaughnessy}}}]{2016Natur.534..512B}%
  \BibitemOpen
  \bibfield  {author} {\bibinfo {author} {\bibfnamefont {K.}~\bibnamefont
  {{Belczynski}}}, \bibinfo {author} {\bibfnamefont {D.~E.}\ \bibnamefont
  {{Holz}}}, \bibinfo {author} {\bibfnamefont {T.}~\bibnamefont {{Bulik}}},\
  and\ \bibinfo {author} {\bibfnamefont {R.}~\bibnamefont {{O'Shaughnessy}}},\
  }\bibfield  {title} {\bibinfo {title} {{The first gravitational-wave source
  from the isolated evolution of two stars in the 40-100 solar mass range}},\
  }\href {https://doi.org/10.1038/nature18322} {\bibfield  {journal} {\bibinfo
  {journal} {Nature}\ }\textbf {\bibinfo {volume} {534}},\ \bibinfo {pages}
  {512} (\bibinfo {year} {2016})},\ \Eprint {https://arxiv.org/abs/1602.04531}
  {arXiv:1602.04531 [astro-ph.HE]} \BibitemShut {NoStop}%
\bibitem [{\citenamefont {{Giacobbo}}\ and\ \citenamefont
  {{Mapelli}}(2018)}]{2018MNRAS.480.2011G}%
  \BibitemOpen
  \bibfield  {author} {\bibinfo {author} {\bibfnamefont {N.}~\bibnamefont
  {{Giacobbo}}}\ and\ \bibinfo {author} {\bibfnamefont {M.}~\bibnamefont
  {{Mapelli}}},\ }\bibfield  {title} {\bibinfo {title} {{The progenitors of
  compact-object binaries: impact of metallicity, common envelope and natal
  kicks}},\ }\href {https://doi.org/10.1093/mnras/sty1999} {\bibfield
  {journal} {\bibinfo  {journal} {MNRAS}\ }\textbf {\bibinfo {volume} {480}},\
  \bibinfo {pages} {2011} (\bibinfo {year} {2018})},\ \Eprint
  {https://arxiv.org/abs/1806.00001} {arXiv:1806.00001 [astro-ph.HE]}
  \BibitemShut {NoStop}%
\bibitem [{\citenamefont {{Olejak}}\ \emph {et~al.}(2021)\citenamefont
  {{Olejak}}, \citenamefont {{Belczynski}},\ and\ \citenamefont
  {{Ivanova}}}]{2021A&A...651A.100O}%
  \BibitemOpen
  \bibfield  {author} {\bibinfo {author} {\bibfnamefont {A.}~\bibnamefont
  {{Olejak}}}, \bibinfo {author} {\bibfnamefont {K.}~\bibnamefont
  {{Belczynski}}},\ and\ \bibinfo {author} {\bibfnamefont {N.}~\bibnamefont
  {{Ivanova}}},\ }\bibfield  {title} {\bibinfo {title} {{Impact of common
  envelope development criteria on the formation of LIGO/Virgo sources}},\
  }\href {https://doi.org/10.1051/0004-6361/202140520} {\bibfield  {journal}
  {\bibinfo  {journal} {AAP}\ }\textbf {\bibinfo {volume} {651}},\ \bibinfo
  {eid} {A100} (\bibinfo {year} {2021})},\ \Eprint
  {https://arxiv.org/abs/2102.05649} {arXiv:2102.05649 [astro-ph.HE]}
  \BibitemShut {NoStop}%
\bibitem [{\citenamefont {{Broekgaarden}}\ \emph
  {et~al.}(2021{\natexlab{a}})\citenamefont {{Broekgaarden}}, \citenamefont
  {{Berger}}, \citenamefont {{Neijssel}}, \citenamefont {{Vigna-G{\'o}mez}},
  \citenamefont {{Chattopadhyay}}, \citenamefont {{Stevenson}}, \citenamefont
  {{Chruslinska}}, \citenamefont {{Justham}}, \citenamefont {{de Mink}},\ and\
  \citenamefont {{Mandel}}}]{2021MNRAS.508.5028B}%
  \BibitemOpen
  \bibfield  {author} {\bibinfo {author} {\bibfnamefont {F.~S.}\ \bibnamefont
  {{Broekgaarden}}}, \bibinfo {author} {\bibfnamefont {E.}~\bibnamefont
  {{Berger}}}, \bibinfo {author} {\bibfnamefont {C.~J.}\ \bibnamefont
  {{Neijssel}}}, \bibinfo {author} {\bibfnamefont {A.}~\bibnamefont
  {{Vigna-G{\'o}mez}}}, \bibinfo {author} {\bibfnamefont {D.}~\bibnamefont
  {{Chattopadhyay}}}, \bibinfo {author} {\bibfnamefont {S.}~\bibnamefont
  {{Stevenson}}}, \bibinfo {author} {\bibfnamefont {M.}~\bibnamefont
  {{Chruslinska}}}, \bibinfo {author} {\bibfnamefont {S.}~\bibnamefont
  {{Justham}}}, \bibinfo {author} {\bibfnamefont {S.~E.}\ \bibnamefont {{de
  Mink}}},\ and\ \bibinfo {author} {\bibfnamefont {I.}~\bibnamefont
  {{Mandel}}},\ }\bibfield  {title} {\bibinfo {title} {{Impact of massive
  binary star and cosmic evolution on gravitational wave observations I: black
  hole-neutron star mergers}},\ }\href {https://doi.org/10.1093/mnras/stab2716}
  {\bibfield  {journal} {\bibinfo  {journal} {MNRAS}\ }\textbf {\bibinfo
  {volume} {508}},\ \bibinfo {pages} {5028} (\bibinfo {year}
  {2021}{\natexlab{a}})},\ \Eprint {https://arxiv.org/abs/2103.02608}
  {arXiv:2103.02608 [astro-ph.HE]} \BibitemShut {NoStop}%
\bibitem [{\citenamefont {{Broekgaarden}}\ \emph
  {et~al.}(2021{\natexlab{b}})\citenamefont {{Broekgaarden}}, \citenamefont
  {{Berger}}, \citenamefont {{Stevenson}}, \citenamefont {{Justham}},
  \citenamefont {{Mandel}}, \citenamefont {{Chru{\'s}li{\'n}ska}},
  \citenamefont {{van Son}}, \citenamefont {{Wagg}}, \citenamefont
  {{Vigna-G{\'o}mez}}, \citenamefont {{de Mink}}, \citenamefont
  {{Chattopadhyay}},\ and\ \citenamefont {{Neijssel}}}]{2021arXiv211205763B}%
  \BibitemOpen
  \bibfield  {author} {\bibinfo {author} {\bibfnamefont {F.~S.}\ \bibnamefont
  {{Broekgaarden}}}, \bibinfo {author} {\bibfnamefont {E.}~\bibnamefont
  {{Berger}}}, \bibinfo {author} {\bibfnamefont {S.}~\bibnamefont
  {{Stevenson}}}, \bibinfo {author} {\bibfnamefont {S.}~\bibnamefont
  {{Justham}}}, \bibinfo {author} {\bibfnamefont {I.}~\bibnamefont {{Mandel}}},
  \bibinfo {author} {\bibfnamefont {M.}~\bibnamefont {{Chru{\'s}li{\'n}ska}}},
  \bibinfo {author} {\bibfnamefont {L.~A.~C.}\ \bibnamefont {{van Son}}},
  \bibinfo {author} {\bibfnamefont {T.}~\bibnamefont {{Wagg}}}, \bibinfo
  {author} {\bibfnamefont {A.}~\bibnamefont {{Vigna-G{\'o}mez}}}, \bibinfo
  {author} {\bibfnamefont {S.~E.}\ \bibnamefont {{de Mink}}}, \bibinfo {author}
  {\bibfnamefont {D.}~\bibnamefont {{Chattopadhyay}}},\ and\ \bibinfo {author}
  {\bibfnamefont {C.~J.}\ \bibnamefont {{Neijssel}}},\ }\bibfield  {title}
  {\bibinfo {title} {{Impact of Massive Binary Star and Cosmic Evolution on
  Gravitational Wave Observations II: Double Compact Object Rates and
  Properties}},\ }\href@noop {} {\bibfield  {journal} {\bibinfo  {journal}
  {arXiv e-prints}\ ,\ \bibinfo {eid} {arXiv:2112.05763}} (\bibinfo {year}
  {2021}{\natexlab{b}})},\ \Eprint {https://arxiv.org/abs/2112.05763}
  {arXiv:2112.05763 [astro-ph.HE]} \BibitemShut {NoStop}%
\bibitem [{\citenamefont {{Sana}}\ \emph {et~al.}(2012)\citenamefont {{Sana}},
  \citenamefont {{de Mink}}, \citenamefont {{de Koter}}, \citenamefont
  {{Langer}}, \citenamefont {{Evans}}, \citenamefont {{Gieles}}, \citenamefont
  {{Gosset}}, \citenamefont {{Izzard}}, \citenamefont {{Le Bouquin}},\ and\
  \citenamefont {{Schneider}}}]{2012Sci...337..444S}%
  \BibitemOpen
  \bibfield  {author} {\bibinfo {author} {\bibfnamefont {H.}~\bibnamefont
  {{Sana}}}, \bibinfo {author} {\bibfnamefont {S.~E.}\ \bibnamefont {{de
  Mink}}}, \bibinfo {author} {\bibfnamefont {A.}~\bibnamefont {{de Koter}}},
  \bibinfo {author} {\bibfnamefont {N.}~\bibnamefont {{Langer}}}, \bibinfo
  {author} {\bibfnamefont {C.~J.}\ \bibnamefont {{Evans}}}, \bibinfo {author}
  {\bibfnamefont {M.}~\bibnamefont {{Gieles}}}, \bibinfo {author}
  {\bibfnamefont {E.}~\bibnamefont {{Gosset}}}, \bibinfo {author}
  {\bibfnamefont {R.~G.}\ \bibnamefont {{Izzard}}}, \bibinfo {author}
  {\bibfnamefont {J.~B.}\ \bibnamefont {{Le Bouquin}}},\ and\ \bibinfo {author}
  {\bibfnamefont {F.~R.~N.}\ \bibnamefont {{Schneider}}},\ }\bibfield  {title}
  {\bibinfo {title} {{Binary Interaction Dominates the Evolution of Massive
  Stars}},\ }\href {https://doi.org/10.1126/science.1223344} {\bibfield
  {journal} {\bibinfo  {journal} {Science}\ }\textbf {\bibinfo {volume}
  {337}},\ \bibinfo {pages} {444} (\bibinfo {year} {2012})},\ \Eprint
  {https://arxiv.org/abs/1207.6397} {arXiv:1207.6397 [astro-ph.SR]}
  \BibitemShut {NoStop}%
\bibitem [{\citenamefont {{Paczy{\'n}ski}}(1967)}]{1967AcA....17..355P}%
  \BibitemOpen
  \bibfield  {author} {\bibinfo {author} {\bibfnamefont {B.}~\bibnamefont
  {{Paczy{\'n}ski}}},\ }\bibfield  {title} {\bibinfo {title} {{Evolution of
  Close Binaries. V. The Evolution of Massive Binaries and the Formation of the
  Wolf-Rayet Stars}},\ }\href@noop {} {\bibfield  {journal} {\bibinfo
  {journal} {actaa}\ }\textbf {\bibinfo {volume} {17}},\ \bibinfo {pages} {355}
  (\bibinfo {year} {1967})}\BibitemShut {NoStop}%
\bibitem [{\citenamefont {{Podsiadlowski}}\ \emph {et~al.}(1992)\citenamefont
  {{Podsiadlowski}}, \citenamefont {{Joss}},\ and\ \citenamefont
  {{Hsu}}}]{1992ApJ...391..246P}%
  \BibitemOpen
  \bibfield  {author} {\bibinfo {author} {\bibfnamefont {P.}~\bibnamefont
  {{Podsiadlowski}}}, \bibinfo {author} {\bibfnamefont {P.~C.}\ \bibnamefont
  {{Joss}}},\ and\ \bibinfo {author} {\bibfnamefont {J.~J.~L.}\ \bibnamefont
  {{Hsu}}},\ }\bibfield  {title} {\bibinfo {title} {{Presupernova Evolution in
  Massive Interacting Binaries}},\ }\href {https://doi.org/10.1086/171341}
  {\bibfield  {journal} {\bibinfo  {journal} {APJ}\ }\textbf {\bibinfo {volume}
  {391}},\ \bibinfo {pages} {246} (\bibinfo {year} {1992})}\BibitemShut
  {NoStop}%
\bibitem [{\citenamefont {{de Mink}}\ \emph {et~al.}(2013)\citenamefont {{de
  Mink}}, \citenamefont {{Langer}}, \citenamefont {{Izzard}}, \citenamefont
  {{Sana}},\ and\ \citenamefont {{de Koter}}}]{2013ApJ...764..166D}%
  \BibitemOpen
  \bibfield  {author} {\bibinfo {author} {\bibfnamefont {S.~E.}\ \bibnamefont
  {{de Mink}}}, \bibinfo {author} {\bibfnamefont {N.}~\bibnamefont {{Langer}}},
  \bibinfo {author} {\bibfnamefont {R.~G.}\ \bibnamefont {{Izzard}}}, \bibinfo
  {author} {\bibfnamefont {H.}~\bibnamefont {{Sana}}},\ and\ \bibinfo {author}
  {\bibfnamefont {A.}~\bibnamefont {{de Koter}}},\ }\bibfield  {title}
  {\bibinfo {title} {{The Rotation Rates of Massive Stars: The Role of Binary
  Interaction through Tides, Mass Transfer, and Mergers}},\ }\href
  {https://doi.org/10.1088/0004-637X/764/2/166} {\bibfield  {journal} {\bibinfo
   {journal} {APJ}\ }\textbf {\bibinfo {volume} {764}},\ \bibinfo {eid} {166}
  (\bibinfo {year} {2013})},\ \Eprint {https://arxiv.org/abs/1211.3742}
  {arXiv:1211.3742 [astro-ph.SR]} \BibitemShut {NoStop}%
\bibitem [{\citenamefont {{Raucq}}\ \emph {et~al.}(2016)\citenamefont
  {{Raucq}}, \citenamefont {{Rauw}}, \citenamefont {{Gosset}}, \citenamefont
  {{Naz{\'e}}}, \citenamefont {{Mahy}}, \citenamefont {{Herv{\'e}}},\ and\
  \citenamefont {{Martins}}}]{2016A&A...588A..10R}%
  \BibitemOpen
  \bibfield  {author} {\bibinfo {author} {\bibfnamefont {F.}~\bibnamefont
  {{Raucq}}}, \bibinfo {author} {\bibfnamefont {G.}~\bibnamefont {{Rauw}}},
  \bibinfo {author} {\bibfnamefont {E.}~\bibnamefont {{Gosset}}}, \bibinfo
  {author} {\bibfnamefont {Y.}~\bibnamefont {{Naz{\'e}}}}, \bibinfo {author}
  {\bibfnamefont {L.}~\bibnamefont {{Mahy}}}, \bibinfo {author} {\bibfnamefont
  {A.}~\bibnamefont {{Herv{\'e}}}},\ and\ \bibinfo {author} {\bibfnamefont
  {F.}~\bibnamefont {{Martins}}},\ }\bibfield  {title} {\bibinfo {title}
  {{Observational signatures of past mass-exchange episodes in massive
  binaries: The case of HD 149 404}},\ }\href
  {https://doi.org/10.1051/0004-6361/201527543} {\bibfield  {journal} {\bibinfo
   {journal} {AAP}\ }\textbf {\bibinfo {volume} {588}},\ \bibinfo {eid} {A10}
  (\bibinfo {year} {2016})},\ \Eprint {https://arxiv.org/abs/1601.08083}
  {arXiv:1601.08083 [astro-ph.SR]} \BibitemShut {NoStop}%
\bibitem [{\citenamefont {{Stegmann}}\ and\ \citenamefont
  {{Antonini}}(2021)}]{2021PhRvD.103f3007S}%
  \BibitemOpen
  \bibfield  {author} {\bibinfo {author} {\bibfnamefont {J.}~\bibnamefont
  {{Stegmann}}}\ and\ \bibinfo {author} {\bibfnamefont {F.}~\bibnamefont
  {{Antonini}}},\ }\bibfield  {title} {\bibinfo {title} {{Flipping spins in
  mass transferring binaries and origin of spin-orbit misalignment in binary
  black holes}},\ }\href {https://doi.org/10.1103/PhysRevD.103.063007}
  {\bibfield  {journal} {\bibinfo  {journal} {PRD}\ }\textbf {\bibinfo {volume}
  {103}},\ \bibinfo {eid} {063007} (\bibinfo {year} {2021})},\ \Eprint
  {https://arxiv.org/abs/2012.06329} {arXiv:2012.06329 [astro-ph.HE]}
  \BibitemShut {NoStop}%
\bibitem [{\citenamefont {{Menon}}\ \emph {et~al.}(2021)\citenamefont
  {{Menon}}, \citenamefont {{Langer}}, \citenamefont {{de Mink}}, \citenamefont
  {{Justham}}, \citenamefont {{Sen}}, \citenamefont {{Sz{\'e}csi}},
  \citenamefont {{de Koter}}, \citenamefont {{Abdul-Masih}}, \citenamefont
  {{Sana}}, \citenamefont {{Mahy}},\ and\ \citenamefont
  {{Marchant}}}]{2021MNRAS.507.5013M}%
  \BibitemOpen
  \bibfield  {author} {\bibinfo {author} {\bibfnamefont {A.}~\bibnamefont
  {{Menon}}}, \bibinfo {author} {\bibfnamefont {N.}~\bibnamefont {{Langer}}},
  \bibinfo {author} {\bibfnamefont {S.~E.}\ \bibnamefont {{de Mink}}}, \bibinfo
  {author} {\bibfnamefont {S.}~\bibnamefont {{Justham}}}, \bibinfo {author}
  {\bibfnamefont {K.}~\bibnamefont {{Sen}}}, \bibinfo {author} {\bibfnamefont
  {D.}~\bibnamefont {{Sz{\'e}csi}}}, \bibinfo {author} {\bibfnamefont
  {A.}~\bibnamefont {{de Koter}}}, \bibinfo {author} {\bibfnamefont
  {M.}~\bibnamefont {{Abdul-Masih}}}, \bibinfo {author} {\bibfnamefont
  {H.}~\bibnamefont {{Sana}}}, \bibinfo {author} {\bibfnamefont
  {L.}~\bibnamefont {{Mahy}}},\ and\ \bibinfo {author} {\bibfnamefont
  {P.}~\bibnamefont {{Marchant}}},\ }\bibfield  {title} {\bibinfo {title}
  {{Detailed evolutionary models of massive contact binaries - I. Model grids
  and synthetic populations for the Magellanic Clouds}},\ }\href
  {https://doi.org/10.1093/mnras/stab2276} {\bibfield  {journal} {\bibinfo
  {journal} {MNRAS}\ }\textbf {\bibinfo {volume} {507}},\ \bibinfo {pages}
  {5013} (\bibinfo {year} {2021})},\ \Eprint {https://arxiv.org/abs/2011.13459}
  {arXiv:2011.13459 [astro-ph.SR]} \BibitemShut {NoStop}%
\bibitem [{\citenamefont {{Moe}}\ and\ \citenamefont {{Di
  Stefano}}(2017)}]{2017ApJS..230...15M}%
  \BibitemOpen
  \bibfield  {author} {\bibinfo {author} {\bibfnamefont {M.}~\bibnamefont
  {{Moe}}}\ and\ \bibinfo {author} {\bibfnamefont {R.}~\bibnamefont {{Di
  Stefano}}},\ }\bibfield  {title} {\bibinfo {title} {{Mind Your Ps and Qs: The
  Interrelation between Period (P) and Mass-ratio (Q) Distributions of Binary
  Stars}},\ }\href {https://doi.org/10.3847/1538-4365/aa6fb6} {\bibfield
  {journal} {\bibinfo  {journal} {ApJs}\ }\textbf {\bibinfo {volume} {230}},\
  \bibinfo {eid} {15} (\bibinfo {year} {2017})},\ \Eprint
  {https://arxiv.org/abs/1606.05347} {arXiv:1606.05347 [astro-ph.SR]}
  \BibitemShut {NoStop}%
\bibitem [{\citenamefont {{Bordier}}\ \emph {et~al.}(2022)\citenamefont
  {{Bordier}}, \citenamefont {{Frost}}, \citenamefont {{Sana}}, \citenamefont
  {{Reggiani}}, \citenamefont {{M{\'e}rand}}, \citenamefont {{Rainot}},
  \citenamefont {{Ram{\'\i}rez-Tannus}},\ and\ \citenamefont {{de
  Wit}}}]{2022arXiv220305036B}%
  \BibitemOpen
  \bibfield  {author} {\bibinfo {author} {\bibfnamefont {E.}~\bibnamefont
  {{Bordier}}}, \bibinfo {author} {\bibfnamefont {A.~J.}\ \bibnamefont
  {{Frost}}}, \bibinfo {author} {\bibfnamefont {H.}~\bibnamefont {{Sana}}},
  \bibinfo {author} {\bibfnamefont {M.}~\bibnamefont {{Reggiani}}}, \bibinfo
  {author} {\bibfnamefont {A.}~\bibnamefont {{M{\'e}rand}}}, \bibinfo {author}
  {\bibfnamefont {A.}~\bibnamefont {{Rainot}}}, \bibinfo {author}
  {\bibfnamefont {M.~C.}\ \bibnamefont {{Ram{\'\i}rez-Tannus}}},\ and\ \bibinfo
  {author} {\bibfnamefont {W.~J.}\ \bibnamefont {{de Wit}}},\ }\bibfield
  {title} {\bibinfo {title} {{On the origin of close massive binaries in the
  M17 star-forming region}},\ }\href@noop {} {\bibfield  {journal} {\bibinfo
  {journal} {arXiv e-prints}\ ,\ \bibinfo {eid} {arXiv:2203.05036}} (\bibinfo
  {year} {2022})},\ \Eprint {https://arxiv.org/abs/2203.05036}
  {arXiv:2203.05036 [astro-ph.SR]} \BibitemShut {NoStop}%
\bibitem [{\citenamefont {{Stegmann}}\ \emph {et~al.}(2021)\citenamefont
  {{Stegmann}}, \citenamefont {{Antonini}},\ and\ \citenamefont
  {{Moe}}}]{2021arXiv211210786S}%
  \BibitemOpen
  \bibfield  {author} {\bibinfo {author} {\bibfnamefont {J.}~\bibnamefont
  {{Stegmann}}}, \bibinfo {author} {\bibfnamefont {F.}~\bibnamefont
  {{Antonini}}},\ and\ \bibinfo {author} {\bibfnamefont {M.}~\bibnamefont
  {{Moe}}},\ }\bibfield  {title} {\bibinfo {title} {{Evolution of massive
  stellar triples and implications for compact object binary formation}},\
  }\href@noop {} {\bibfield  {journal} {\bibinfo  {journal} {arXiv e-prints}\
  ,\ \bibinfo {eid} {arXiv:2112.10786}} (\bibinfo {year} {2021})},\ \Eprint
  {https://arxiv.org/abs/2112.10786} {arXiv:2112.10786 [astro-ph.SR]}
  \BibitemShut {NoStop}%
\bibitem [{\citenamefont {{Toonen}}\ \emph {et~al.}(2020)\citenamefont
  {{Toonen}}, \citenamefont {{Portegies Zwart}}, \citenamefont {{Hamers}},\
  and\ \citenamefont {{Bandopadhyay}}}]{2020A&A...640A..16T}%
  \BibitemOpen
  \bibfield  {author} {\bibinfo {author} {\bibfnamefont {S.}~\bibnamefont
  {{Toonen}}}, \bibinfo {author} {\bibfnamefont {S.}~\bibnamefont {{Portegies
  Zwart}}}, \bibinfo {author} {\bibfnamefont {A.~S.}\ \bibnamefont
  {{Hamers}}},\ and\ \bibinfo {author} {\bibfnamefont {D.}~\bibnamefont
  {{Bandopadhyay}}},\ }\bibfield  {title} {\bibinfo {title} {{The evolution of
  stellar triples. The most common evolutionary pathways}},\ }\href
  {https://doi.org/10.1051/0004-6361/201936835} {\bibfield  {journal} {\bibinfo
   {journal} {AAP}\ }\textbf {\bibinfo {volume} {640}},\ \bibinfo {eid} {A16}
  (\bibinfo {year} {2020})},\ \Eprint {https://arxiv.org/abs/2004.07848}
  {arXiv:2004.07848 [astro-ph.SR]} \BibitemShut {NoStop}%
\bibitem [{\citenamefont {{Hamers}}\ and\ \citenamefont
  {{Dosopoulou}}(2019)}]{2019ApJ...872..119H}%
  \BibitemOpen
  \bibfield  {author} {\bibinfo {author} {\bibfnamefont {A.~S.}\ \bibnamefont
  {{Hamers}}}\ and\ \bibinfo {author} {\bibfnamefont {F.}~\bibnamefont
  {{Dosopoulou}}},\ }\bibfield  {title} {\bibinfo {title} {{An Analytic Model
  for Mass Transfer in Binaries with Arbitrary Eccentricity, with Applications
  to Triple-star Systems}},\ }\href {https://doi.org/10.3847/1538-4357/ab001d}
  {\bibfield  {journal} {\bibinfo  {journal} {APJ}\ }\textbf {\bibinfo {volume}
  {872}},\ \bibinfo {eid} {119} (\bibinfo {year} {2019})},\ \Eprint
  {https://arxiv.org/abs/1812.05624} {arXiv:1812.05624 [astro-ph.SR]}
  \BibitemShut {NoStop}%
\bibitem [{\citenamefont {{Silsbee}}\ and\ \citenamefont
  {{Tremaine}}(2017)}]{2017ApJ...836...39S}%
  \BibitemOpen
  \bibfield  {author} {\bibinfo {author} {\bibfnamefont {K.}~\bibnamefont
  {{Silsbee}}}\ and\ \bibinfo {author} {\bibfnamefont {S.}~\bibnamefont
  {{Tremaine}}},\ }\bibfield  {title} {\bibinfo {title} {{Lidov-Kozai Cycles
  with Gravitational Radiation: Merging Black Holes in Isolated Triple
  Systems}},\ }\href {https://doi.org/10.3847/1538-4357/aa5729} {\bibfield
  {journal} {\bibinfo  {journal} {APJ}\ }\textbf {\bibinfo {volume} {836}},\
  \bibinfo {eid} {39} (\bibinfo {year} {2017})},\ \Eprint
  {https://arxiv.org/abs/1608.07642} {arXiv:1608.07642 [astro-ph.HE]}
  \BibitemShut {NoStop}%
\bibitem [{\citenamefont {{Antonini}}\ \emph {et~al.}(2017)\citenamefont
  {{Antonini}}, \citenamefont {{Toonen}},\ and\ \citenamefont
  {{Hamers}}}]{2017ApJ...841...77A}%
  \BibitemOpen
  \bibfield  {author} {\bibinfo {author} {\bibfnamefont {F.}~\bibnamefont
  {{Antonini}}}, \bibinfo {author} {\bibfnamefont {S.}~\bibnamefont
  {{Toonen}}},\ and\ \bibinfo {author} {\bibfnamefont {A.~S.}\ \bibnamefont
  {{Hamers}}},\ }\bibfield  {title} {\bibinfo {title} {{Binary Black Hole
  Mergers from Field Triples: Properties, Rates, and the Impact of Stellar
  Evolution}},\ }\href {https://doi.org/10.3847/1538-4357/aa6f5e} {\bibfield
  {journal} {\bibinfo  {journal} {APJ}\ }\textbf {\bibinfo {volume} {841}},\
  \bibinfo {eid} {77} (\bibinfo {year} {2017})},\ \Eprint
  {https://arxiv.org/abs/1703.06614} {arXiv:1703.06614 [astro-ph.GA]}
  \BibitemShut {NoStop}%
\bibitem [{\citenamefont {{Antonini}}\ \emph {et~al.}(2018)\citenamefont
  {{Antonini}}, \citenamefont {{Rodriguez}}, \citenamefont {{Petrovich}},\ and\
  \citenamefont {{Fischer}}}]{2018MNRAS.480L..58A}%
  \BibitemOpen
  \bibfield  {author} {\bibinfo {author} {\bibfnamefont {F.}~\bibnamefont
  {{Antonini}}}, \bibinfo {author} {\bibfnamefont {C.~L.}\ \bibnamefont
  {{Rodriguez}}}, \bibinfo {author} {\bibfnamefont {C.}~\bibnamefont
  {{Petrovich}}},\ and\ \bibinfo {author} {\bibfnamefont {C.~L.}\ \bibnamefont
  {{Fischer}}},\ }\bibfield  {title} {\bibinfo {title} {{Precessional dynamics
  of black hole triples: binary mergers with near-zero effective spin}},\
  }\href {https://doi.org/10.1093/mnrasl/sly126} {\bibfield  {journal}
  {\bibinfo  {journal} {MNRAS}\ }\textbf {\bibinfo {volume} {480}},\ \bibinfo
  {pages} {L58} (\bibinfo {year} {2018})},\ \Eprint
  {https://arxiv.org/abs/1711.07142} {arXiv:1711.07142 [astro-ph.HE]}
  \BibitemShut {NoStop}%
\bibitem [{\citenamefont {{Liu}}\ and\ \citenamefont
  {{Lai}}(2018)}]{2018ApJ...863...68L}%
  \BibitemOpen
  \bibfield  {author} {\bibinfo {author} {\bibfnamefont {B.}~\bibnamefont
  {{Liu}}}\ and\ \bibinfo {author} {\bibfnamefont {D.}~\bibnamefont {{Lai}}},\
  }\bibfield  {title} {\bibinfo {title} {{Black Hole and Neutron Star Binary
  Mergers in Triple Systems: Merger Fraction and Spin-Orbit Misalignment}},\
  }\href {https://doi.org/10.3847/1538-4357/aad09f} {\bibfield  {journal}
  {\bibinfo  {journal} {APJ}\ }\textbf {\bibinfo {volume} {863}},\ \bibinfo
  {eid} {68} (\bibinfo {year} {2018})},\ \Eprint
  {https://arxiv.org/abs/1805.03202} {arXiv:1805.03202 [astro-ph.HE]}
  \BibitemShut {NoStop}%
\bibitem [{\citenamefont {{Rodriguez}}\ and\ \citenamefont
  {{Antonini}}(2018)}]{2018ApJ...863....7R}%
  \BibitemOpen
  \bibfield  {author} {\bibinfo {author} {\bibfnamefont {C.~L.}\ \bibnamefont
  {{Rodriguez}}}\ and\ \bibinfo {author} {\bibfnamefont {F.}~\bibnamefont
  {{Antonini}}},\ }\bibfield  {title} {\bibinfo {title} {{A Triple Origin for
  the Heavy and Low-spin Binary Black Holes Detected by LIGO/VIRGO}},\ }\href
  {https://doi.org/10.3847/1538-4357/aacea4} {\bibfield  {journal} {\bibinfo
  {journal} {APJ}\ }\textbf {\bibinfo {volume} {863}},\ \bibinfo {eid} {7}
  (\bibinfo {year} {2018})},\ \Eprint {https://arxiv.org/abs/1805.08212}
  {arXiv:1805.08212 [astro-ph.HE]} \BibitemShut {NoStop}%
\bibitem [{\citenamefont {{Fragione}}\ and\ \citenamefont
  {{Kocsis}}(2020)}]{2020MNRAS.493.3920F}%
  \BibitemOpen
  \bibfield  {author} {\bibinfo {author} {\bibfnamefont {G.}~\bibnamefont
  {{Fragione}}}\ and\ \bibinfo {author} {\bibfnamefont {B.}~\bibnamefont
  {{Kocsis}}},\ }\bibfield  {title} {\bibinfo {title} {{Effective spin
  distribution of black hole mergers in triples}},\ }\href
  {https://doi.org/10.1093/mnras/staa443} {\bibfield  {journal} {\bibinfo
  {journal} {MNRAS}\ }\textbf {\bibinfo {volume} {493}},\ \bibinfo {pages}
  {3920} (\bibinfo {year} {2020})},\ \Eprint {https://arxiv.org/abs/1910.00407}
  {arXiv:1910.00407 [astro-ph.GA]} \BibitemShut {NoStop}%
\bibitem [{\citenamefont {{Martinez}}\ \emph {et~al.}(2021)\citenamefont
  {{Martinez}}, \citenamefont {{Rodriguez}},\ and\ \citenamefont
  {{Fragione}}}]{2021arXiv210501671M}%
  \BibitemOpen
  \bibfield  {author} {\bibinfo {author} {\bibfnamefont {M.~A.~S.}\
  \bibnamefont {{Martinez}}}, \bibinfo {author} {\bibfnamefont {C.~L.}\
  \bibnamefont {{Rodriguez}}},\ and\ \bibinfo {author} {\bibfnamefont
  {G.}~\bibnamefont {{Fragione}}},\ }\bibfield  {title} {\bibinfo {title} {{On
  the Mass Ratio Distribution of Black Hole Mergers in Triple Systems}},\
  }\href@noop {} {\bibfield  {journal} {\bibinfo  {journal} {arXiv e-prints}\
  ,\ \bibinfo {eid} {arXiv:2105.01671}} (\bibinfo {year} {2021})},\ \Eprint
  {https://arxiv.org/abs/2105.01671} {arXiv:2105.01671 [astro-ph.SR]}
  \BibitemShut {NoStop}%
\bibitem [{\citenamefont {{Fishbach}}\ \emph {et~al.}(2017)\citenamefont
  {{Fishbach}}, \citenamefont {{Holz}},\ and\ \citenamefont
  {{Farr}}}]{2017ApJ...840L..24F}%
  \BibitemOpen
  \bibfield  {author} {\bibinfo {author} {\bibfnamefont {M.}~\bibnamefont
  {{Fishbach}}}, \bibinfo {author} {\bibfnamefont {D.~E.}\ \bibnamefont
  {{Holz}}},\ and\ \bibinfo {author} {\bibfnamefont {B.}~\bibnamefont
  {{Farr}}},\ }\bibfield  {title} {\bibinfo {title} {{Are LIGO's Black Holes
  Made from Smaller Black Holes?}},\ }\href
  {https://doi.org/10.3847/2041-8213/aa7045} {\bibfield  {journal} {\bibinfo
  {journal} {APJL}\ }\textbf {\bibinfo {volume} {840}},\ \bibinfo {eid} {L24}
  (\bibinfo {year} {2017})},\ \Eprint {https://arxiv.org/abs/1703.06869}
  {arXiv:1703.06869 [astro-ph.HE]} \BibitemShut {NoStop}%
\bibitem [{\citenamefont {{Vigna-G{\'o}mez}}\ \emph {et~al.}(2021)\citenamefont
  {{Vigna-G{\'o}mez}}, \citenamefont {{Toonen}}, \citenamefont
  {{Ramirez-Ruiz}}, \citenamefont {{Leigh}}, \citenamefont {{Riley}},\ and\
  \citenamefont {{Haster}}}]{2021ApJ...907L..19V}%
  \BibitemOpen
  \bibfield  {author} {\bibinfo {author} {\bibfnamefont {A.}~\bibnamefont
  {{Vigna-G{\'o}mez}}}, \bibinfo {author} {\bibfnamefont {S.}~\bibnamefont
  {{Toonen}}}, \bibinfo {author} {\bibfnamefont {E.}~\bibnamefont
  {{Ramirez-Ruiz}}}, \bibinfo {author} {\bibfnamefont {N.~W.~C.}\ \bibnamefont
  {{Leigh}}}, \bibinfo {author} {\bibfnamefont {J.}~\bibnamefont {{Riley}}},\
  and\ \bibinfo {author} {\bibfnamefont {C.-J.}\ \bibnamefont {{Haster}}},\
  }\bibfield  {title} {\bibinfo {title} {{Massive Stellar Triples Leading to
  Sequential Binary Black Hole Mergers in the Field}},\ }\href
  {https://doi.org/10.3847/2041-8213/abd5b7} {\bibfield  {journal} {\bibinfo
  {journal} {APJL}\ }\textbf {\bibinfo {volume} {907}},\ \bibinfo {eid} {L19}
  (\bibinfo {year} {2021})},\ \Eprint {https://arxiv.org/abs/2010.13669}
  {arXiv:2010.13669 [astro-ph.HE]} \BibitemShut {NoStop}%
\bibitem [{\citenamefont {{Gerosa}}\ and\ \citenamefont
  {{Fishbach}}(2021)}]{2021NatAs...5..749G}%
  \BibitemOpen
  \bibfield  {author} {\bibinfo {author} {\bibfnamefont {D.}~\bibnamefont
  {{Gerosa}}}\ and\ \bibinfo {author} {\bibfnamefont {M.}~\bibnamefont
  {{Fishbach}}},\ }\bibfield  {title} {\bibinfo {title} {{Hierarchical mergers
  of stellar-mass black holes and their gravitational-wave signatures}},\
  }\href {https://doi.org/10.1038/s41550-021-01398-w} {\bibfield  {journal}
  {\bibinfo  {journal} {Nature Astronomy}\ }\textbf {\bibinfo {volume} {5}},\
  \bibinfo {pages} {749} (\bibinfo {year} {2021})},\ \Eprint
  {https://arxiv.org/abs/2105.03439} {arXiv:2105.03439 [astro-ph.HE]}
  \BibitemShut {NoStop}%
\bibitem [{\citenamefont {{Giacobbo}}\ \emph {et~al.}(2018)\citenamefont
  {{Giacobbo}}, \citenamefont {{Mapelli}},\ and\ \citenamefont
  {{Spera}}}]{2018MNRAS.474.2959G}%
  \BibitemOpen
  \bibfield  {author} {\bibinfo {author} {\bibfnamefont {N.}~\bibnamefont
  {{Giacobbo}}}, \bibinfo {author} {\bibfnamefont {M.}~\bibnamefont
  {{Mapelli}}},\ and\ \bibinfo {author} {\bibfnamefont {M.}~\bibnamefont
  {{Spera}}},\ }\bibfield  {title} {\bibinfo {title} {{Merging black hole
  binaries: the effects of progenitor's metallicity, mass-loss rate and
  Eddington factor}},\ }\href {https://doi.org/10.1093/mnras/stx2933}
  {\bibfield  {journal} {\bibinfo  {journal} {MNRAS}\ }\textbf {\bibinfo
  {volume} {474}},\ \bibinfo {pages} {2959} (\bibinfo {year} {2018})},\ \Eprint
  {https://arxiv.org/abs/1711.03556} {arXiv:1711.03556 [astro-ph.SR]}
  \BibitemShut {NoStop}%
\bibitem [{\citenamefont {{Giacobbo}}\ and\ \citenamefont
  {{Mapelli}}(2019)}]{2019MNRAS.482.2234G}%
  \BibitemOpen
  \bibfield  {author} {\bibinfo {author} {\bibfnamefont {N.}~\bibnamefont
  {{Giacobbo}}}\ and\ \bibinfo {author} {\bibfnamefont {M.}~\bibnamefont
  {{Mapelli}}},\ }\bibfield  {title} {\bibinfo {title} {{The impact of
  electron-capture supernovae on merging double neutron stars}},\ }\href
  {https://doi.org/10.1093/mnras/sty2848} {\bibfield  {journal} {\bibinfo
  {journal} {MNRAS}\ }\textbf {\bibinfo {volume} {482}},\ \bibinfo {pages}
  {2234} (\bibinfo {year} {2019})},\ \Eprint {https://arxiv.org/abs/1805.11100}
  {arXiv:1805.11100 [astro-ph.SR]} \BibitemShut {NoStop}%
\bibitem [{\citenamefont {{Giacobbo}}\ and\ \citenamefont
  {{Mapelli}}(2020)}]{2020ApJ...891..141G}%
  \BibitemOpen
  \bibfield  {author} {\bibinfo {author} {\bibfnamefont {N.}~\bibnamefont
  {{Giacobbo}}}\ and\ \bibinfo {author} {\bibfnamefont {M.}~\bibnamefont
  {{Mapelli}}},\ }\bibfield  {title} {\bibinfo {title} {{Revising Natal Kick
  Prescriptions in Population Synthesis Simulations}},\ }\href
  {https://doi.org/10.3847/1538-4357/ab7335} {\bibfield  {journal} {\bibinfo
  {journal} {APJ}\ }\textbf {\bibinfo {volume} {891}},\ \bibinfo {eid} {141}
  (\bibinfo {year} {2020})},\ \Eprint {https://arxiv.org/abs/1909.06385}
  {arXiv:1909.06385 [astro-ph.HE]} \BibitemShut {NoStop}%
\bibitem [{\citenamefont {{Hurley}}\ \emph {et~al.}(2002)\citenamefont
  {{Hurley}}, \citenamefont {{Tout}},\ and\ \citenamefont
  {{Pols}}}]{2002MNRAS.329..897H}%
  \BibitemOpen
  \bibfield  {author} {\bibinfo {author} {\bibfnamefont {J.~R.}\ \bibnamefont
  {{Hurley}}}, \bibinfo {author} {\bibfnamefont {C.~A.}\ \bibnamefont
  {{Tout}}},\ and\ \bibinfo {author} {\bibfnamefont {O.~R.}\ \bibnamefont
  {{Pols}}},\ }\bibfield  {title} {\bibinfo {title} {{Evolution of binary stars
  and the effect of tides on binary populations}},\ }\href
  {https://doi.org/10.1046/j.1365-8711.2002.05038.x} {\bibfield  {journal}
  {\bibinfo  {journal} {MNRAS}\ }\textbf {\bibinfo {volume} {329}},\ \bibinfo
  {pages} {897} (\bibinfo {year} {2002})},\ \Eprint
  {https://arxiv.org/abs/astro-ph/0201220} {arXiv:astro-ph/0201220 [astro-ph]}
  \BibitemShut {NoStop}%
\bibitem [{\citenamefont {{Lidov}}(1962)}]{1962P&SS....9..719L}%
  \BibitemOpen
  \bibfield  {author} {\bibinfo {author} {\bibfnamefont {M.~L.}\ \bibnamefont
  {{Lidov}}},\ }\bibfield  {title} {\bibinfo {title} {{The evolution of orbits
  of artificial satellites of planets under the action of gravitational
  perturbations of external bodies}},\ }\href
  {https://doi.org/10.1016/0032-0633(62)90129-0} {\bibfield  {journal}
  {\bibinfo  {journal} {Planetary and Space Science}\ }\textbf {\bibinfo
  {volume} {9}},\ \bibinfo {pages} {719} (\bibinfo {year} {1962})}\BibitemShut
  {NoStop}%
\bibitem [{\citenamefont {{Kozai}}(1962)}]{1962AJ.....67..591K}%
  \BibitemOpen
  \bibfield  {author} {\bibinfo {author} {\bibfnamefont {Y.}~\bibnamefont
  {{Kozai}}},\ }\bibfield  {title} {\bibinfo {title} {{Secular perturbations of
  asteroids with high inclination and eccentricity}},\ }\href
  {https://doi.org/10.1086/108790} {\bibfield  {journal} {\bibinfo  {journal}
  {AJ}\ }\textbf {\bibinfo {volume} {67}},\ \bibinfo {pages} {591} (\bibinfo
  {year} {1962})}\BibitemShut {NoStop}%
\bibitem [{\citenamefont {{Liu}}\ \emph {et~al.}(2019)\citenamefont {{Liu}},
  \citenamefont {{Lai}},\ and\ \citenamefont {{Wang}}}]{2019ApJ...881...41L}%
  \BibitemOpen
  \bibfield  {author} {\bibinfo {author} {\bibfnamefont {B.}~\bibnamefont
  {{Liu}}}, \bibinfo {author} {\bibfnamefont {D.}~\bibnamefont {{Lai}}},\ and\
  \bibinfo {author} {\bibfnamefont {Y.-H.}\ \bibnamefont {{Wang}}},\ }\bibfield
   {title} {\bibinfo {title} {{Black Hole and Neutron Star Binary Mergers in
  Triple Systems. II. Merger Eccentricity and Spin-Orbit Misalignment}},\
  }\href {https://doi.org/10.3847/1538-4357/ab2dfb} {\bibfield  {journal}
  {\bibinfo  {journal} {APJ}\ }\textbf {\bibinfo {volume} {881}},\ \bibinfo
  {eid} {41} (\bibinfo {year} {2019})},\ \Eprint
  {https://arxiv.org/abs/1905.00427} {arXiv:1905.00427 [astro-ph.HE]}
  \BibitemShut {NoStop}%
\bibitem [{\citenamefont {{Toonen}}\ \emph {et~al.}(2022)\citenamefont
  {{Toonen}}, \citenamefont {{Boekholt}},\ and\ \citenamefont {{Portegies
  Zwart}}}]{2022A&A...661A..61T}%
  \BibitemOpen
  \bibfield  {author} {\bibinfo {author} {\bibfnamefont {S.}~\bibnamefont
  {{Toonen}}}, \bibinfo {author} {\bibfnamefont {T.~C.~N.}\ \bibnamefont
  {{Boekholt}}},\ and\ \bibinfo {author} {\bibfnamefont {S.}~\bibnamefont
  {{Portegies Zwart}}},\ }\bibfield  {title} {\bibinfo {title} {{Stellar
  triples on the edge. Comprehensive overview of the evolution of destabilised
  triples leading to stellar and binary exotica}},\ }\href
  {https://doi.org/10.1051/0004-6361/202141991} {\bibfield  {journal} {\bibinfo
   {journal} {AAP}\ }\textbf {\bibinfo {volume} {661}},\ \bibinfo {eid} {A61}
  (\bibinfo {year} {2022})},\ \Eprint {https://arxiv.org/abs/2108.04272}
  {arXiv:2108.04272 [astro-ph.SR]} \BibitemShut {NoStop}%
\bibitem [{\citenamefont {{Mardling}}\ and\ \citenamefont
  {{Aarseth}}(2001)}]{2001MNRAS.321..398M}%
  \BibitemOpen
  \bibfield  {author} {\bibinfo {author} {\bibfnamefont {R.~A.}\ \bibnamefont
  {{Mardling}}}\ and\ \bibinfo {author} {\bibfnamefont {S.~J.}\ \bibnamefont
  {{Aarseth}}},\ }\bibfield  {title} {\bibinfo {title} {{Tidal interactions in
  star cluster simulations}},\ }\href
  {https://doi.org/10.1046/j.1365-8711.2001.03974.x} {\bibfield  {journal}
  {\bibinfo  {journal} {MNRAS}\ }\textbf {\bibinfo {volume} {321}},\ \bibinfo
  {pages} {398} (\bibinfo {year} {2001})}\BibitemShut {NoStop}%
\bibitem [{\citenamefont {{Eggleton}}(1983)}]{1983ApJ...268..368E}%
  \BibitemOpen
  \bibfield  {author} {\bibinfo {author} {\bibfnamefont {P.~P.}\ \bibnamefont
  {{Eggleton}}},\ }\bibfield  {title} {\bibinfo {title} {{Aproximations to the
  radii of Roche lobes.}},\ }\href {https://doi.org/10.1086/160960} {\bibfield
  {journal} {\bibinfo  {journal} {APJ}\ }\textbf {\bibinfo {volume} {268}},\
  \bibinfo {pages} {368} (\bibinfo {year} {1983})}\BibitemShut {NoStop}%
\bibitem [{\citenamefont {{Glanz}}\ and\ \citenamefont
  {{Perets}}(2021)}]{2021MNRAS.500.1921G}%
  \BibitemOpen
  \bibfield  {author} {\bibinfo {author} {\bibfnamefont {H.}~\bibnamefont
  {{Glanz}}}\ and\ \bibinfo {author} {\bibfnamefont {H.~B.}\ \bibnamefont
  {{Perets}}},\ }\bibfield  {title} {\bibinfo {title} {{Simulations of common
  envelope evolution in triple systems: circumstellar case}},\ }\href
  {https://doi.org/10.1093/mnras/staa3242} {\bibfield  {journal} {\bibinfo
  {journal} {MNRAS}\ }\textbf {\bibinfo {volume} {500}},\ \bibinfo {pages}
  {1921} (\bibinfo {year} {2021})},\ \Eprint {https://arxiv.org/abs/2004.00020}
  {arXiv:2004.00020 [astro-ph.SR]} \BibitemShut {NoStop}%
\bibitem [{\citenamefont {{Leigh}}(2022)}]{2022arXiv220305357L}%
  \BibitemOpen
  \bibfield  {author} {\bibinfo {author} {\bibfnamefont {Y.~G. S. T.~N.}\
  \bibnamefont {{Leigh}}},\ }\bibfield  {title} {\bibinfo {title} {{Stellar
  Triples as a Source for Ba Stars}},\ }\href@noop {} {\bibfield  {journal}
  {\bibinfo  {journal} {arXiv e-prints}\ ,\ \bibinfo {eid} {arXiv:2203.05357}}
  (\bibinfo {year} {2022})},\ \Eprint {https://arxiv.org/abs/2203.05357}
  {arXiv:2203.05357 [astro-ph.SR]} \BibitemShut {NoStop}%
\bibitem [{\citenamefont {{Glebbeek}}\ and\ \citenamefont
  {{Pols}}(2008)}]{Glebbeek}%
  \BibitemOpen
  \bibfield  {author} {\bibinfo {author} {\bibfnamefont {E.}~\bibnamefont
  {{Glebbeek}}}\ and\ \bibinfo {author} {\bibfnamefont {O.~R.}\ \bibnamefont
  {{Pols}}},\ }\bibfield  {title} {\bibinfo {title} {{Evolution of stellar
  collision products in open clusters. II. A grid of low-mass collisions}},\
  }\href {https://doi.org/10.1051/0004-6361:200809931} {\bibfield  {journal}
  {\bibinfo  {journal} {AAP}\ }\textbf {\bibinfo {volume} {488}},\ \bibinfo
  {pages} {1017} (\bibinfo {year} {2008})},\ \Eprint
  {https://arxiv.org/abs/0806.0865} {arXiv:0806.0865 [astro-ph]} \BibitemShut
  {NoStop}%
\bibitem [{\citenamefont {{Hurley}}\ \emph {et~al.}(2000)\citenamefont
  {{Hurley}}, \citenamefont {{Pols}},\ and\ \citenamefont
  {{Tout}}}]{2000MNRAS.315..543H}%
  \BibitemOpen
  \bibfield  {author} {\bibinfo {author} {\bibfnamefont {J.~R.}\ \bibnamefont
  {{Hurley}}}, \bibinfo {author} {\bibfnamefont {O.~R.}\ \bibnamefont
  {{Pols}}},\ and\ \bibinfo {author} {\bibfnamefont {C.~A.}\ \bibnamefont
  {{Tout}}},\ }\bibfield  {title} {\bibinfo {title} {{Comprehensive analytic
  formulae for stellar evolution as a function of mass and metallicity}},\
  }\href {https://doi.org/10.1046/j.1365-8711.2000.03426.x} {\bibfield
  {journal} {\bibinfo  {journal} {MNRAS}\ }\textbf {\bibinfo {volume} {315}},\
  \bibinfo {pages} {543} (\bibinfo {year} {2000})},\ \Eprint
  {https://arxiv.org/abs/astro-ph/0001295} {arXiv:astro-ph/0001295 [astro-ph]}
  \BibitemShut {NoStop}%
\bibitem [{\citenamefont {{Schneider}}\ \emph {et~al.}(2016)\citenamefont
  {{Schneider}}, \citenamefont {{Podsiadlowski}}, \citenamefont {{Langer}},
  \citenamefont {{Castro}},\ and\ \citenamefont
  {{Fossati}}}]{2016MNRAS.457.2355S}%
  \BibitemOpen
  \bibfield  {author} {\bibinfo {author} {\bibfnamefont {F.~R.~N.}\
  \bibnamefont {{Schneider}}}, \bibinfo {author} {\bibfnamefont
  {P.}~\bibnamefont {{Podsiadlowski}}}, \bibinfo {author} {\bibfnamefont
  {N.}~\bibnamefont {{Langer}}}, \bibinfo {author} {\bibfnamefont
  {N.}~\bibnamefont {{Castro}}},\ and\ \bibinfo {author} {\bibfnamefont
  {L.}~\bibnamefont {{Fossati}}},\ }\bibfield  {title} {\bibinfo {title}
  {{Rejuvenation of stellar mergers and the origin of magnetic fields in
  massive stars}},\ }\href {https://doi.org/10.1093/mnras/stw148} {\bibfield
  {journal} {\bibinfo  {journal} {MNRAS}\ }\textbf {\bibinfo {volume} {457}},\
  \bibinfo {pages} {2355} (\bibinfo {year} {2016})},\ \Eprint
  {https://arxiv.org/abs/1601.05084} {arXiv:1601.05084 [astro-ph.SR]}
  \BibitemShut {NoStop}%
\bibitem [{\citenamefont {{Tout}}\ \emph {et~al.}(1997)\citenamefont {{Tout}},
  \citenamefont {{Aarseth}}, \citenamefont {{Pols}},\ and\ \citenamefont
  {{Eggleton}}}]{1997MNRAS.291..732T}%
  \BibitemOpen
  \bibfield  {author} {\bibinfo {author} {\bibfnamefont {C.~A.}\ \bibnamefont
  {{Tout}}}, \bibinfo {author} {\bibfnamefont {S.~J.}\ \bibnamefont
  {{Aarseth}}}, \bibinfo {author} {\bibfnamefont {O.~R.}\ \bibnamefont
  {{Pols}}},\ and\ \bibinfo {author} {\bibfnamefont {P.~P.}\ \bibnamefont
  {{Eggleton}}},\ }\bibfield  {title} {\bibinfo {title} {{Rapid binary star
  evolution for N-body simulations and population synthesis}},\ }\href
  {https://doi.org/10.1093/mnras/291.4.732} {\bibfield  {journal} {\bibinfo
  {journal} {MNRAS}\ }\textbf {\bibinfo {volume} {291}},\ \bibinfo {pages}
  {732} (\bibinfo {year} {1997})}\BibitemShut {NoStop}%
\bibitem [{\citenamefont {Glebbeek}\ \emph {et~al.}(2013)\citenamefont
  {Glebbeek}, \citenamefont {Gaburov}, \citenamefont {Portegies~Zwart},\ and\
  \citenamefont {Pols}}]{10.1093/mnras/stt1268}%
  \BibitemOpen
  \bibfield  {author} {\bibinfo {author} {\bibfnamefont {E.}~\bibnamefont
  {Glebbeek}}, \bibinfo {author} {\bibfnamefont {E.}~\bibnamefont {Gaburov}},
  \bibinfo {author} {\bibfnamefont {S.}~\bibnamefont {Portegies~Zwart}},\ and\
  \bibinfo {author} {\bibfnamefont {O.~R.}\ \bibnamefont {Pols}},\ }\bibfield
  {title} {\bibinfo {title} {{Structure and evolution of high-mass stellar
  mergers}},\ }\href {https://doi.org/10.1093/mnras/stt1268} {\bibfield
  {journal} {\bibinfo  {journal} {MNRAS}\ }\textbf {\bibinfo {volume} {434}},\
  \bibinfo {pages} {3497} (\bibinfo {year} {2013})},\ \Eprint
  {https://arxiv.org/abs/https://academic.oup.com/mnras/article-pdf/434/4/3497/18501344/stt1268.pdf}
  {https://academic.oup.com/mnras/article-pdf/434/4/3497/18501344/stt1268.pdf}
  \BibitemShut {NoStop}%
\bibitem [{\citenamefont {{Suzuki}}\ \emph {et~al.}(2007)\citenamefont
  {{Suzuki}}, \citenamefont {{Nakasato}}, \citenamefont {{Baumgardt}},
  \citenamefont {{Ibukiyama}}, \citenamefont {{Makino}},\ and\ \citenamefont
  {{Ebisuzaki}}}]{2007ApJ...668..435S}%
  \BibitemOpen
  \bibfield  {author} {\bibinfo {author} {\bibfnamefont {T.~K.}\ \bibnamefont
  {{Suzuki}}}, \bibinfo {author} {\bibfnamefont {N.}~\bibnamefont
  {{Nakasato}}}, \bibinfo {author} {\bibfnamefont {H.}~\bibnamefont
  {{Baumgardt}}}, \bibinfo {author} {\bibfnamefont {A.}~\bibnamefont
  {{Ibukiyama}}}, \bibinfo {author} {\bibfnamefont {J.}~\bibnamefont
  {{Makino}}},\ and\ \bibinfo {author} {\bibfnamefont {T.}~\bibnamefont
  {{Ebisuzaki}}},\ }\bibfield  {title} {\bibinfo {title} {{Evolution of
  Collisionally Merged Massive Stars}},\ }\href
  {https://doi.org/10.1086/521214} {\bibfield  {journal} {\bibinfo  {journal}
  {APJ}\ }\textbf {\bibinfo {volume} {668}},\ \bibinfo {pages} {435} (\bibinfo
  {year} {2007})},\ \Eprint {https://arxiv.org/abs/astro-ph/0703290}
  {arXiv:astro-ph/0703290 [astro-ph]} \BibitemShut {NoStop}%
\bibitem [{\citenamefont {{Hirai}}\ \emph {et~al.}(2021)\citenamefont
  {{Hirai}}, \citenamefont {{Podsiadlowski}}, \citenamefont {{Owocki}},
  \citenamefont {{Schneider}},\ and\ \citenamefont
  {{Smith}}}]{2021MNRAS.503.4276H}%
  \BibitemOpen
  \bibfield  {author} {\bibinfo {author} {\bibfnamefont {R.}~\bibnamefont
  {{Hirai}}}, \bibinfo {author} {\bibfnamefont {P.}~\bibnamefont
  {{Podsiadlowski}}}, \bibinfo {author} {\bibfnamefont {S.~P.}\ \bibnamefont
  {{Owocki}}}, \bibinfo {author} {\bibfnamefont {F.~R.~N.}\ \bibnamefont
  {{Schneider}}},\ and\ \bibinfo {author} {\bibfnamefont {N.}~\bibnamefont
  {{Smith}}},\ }\bibfield  {title} {\bibinfo {title} {{Simulating the formation
  of {\ensuremath{\eta}} Carinae's surrounding nebula through unstable triple
  evolution and stellar merger-induced eruption}},\ }\href
  {https://doi.org/10.1093/mnras/stab571} {\bibfield  {journal} {\bibinfo
  {journal} {MNRAS}\ }\textbf {\bibinfo {volume} {503}},\ \bibinfo {pages}
  {4276} (\bibinfo {year} {2021})},\ \Eprint {https://arxiv.org/abs/2011.12434}
  {arXiv:2011.12434 [astro-ph.SR]} \BibitemShut {NoStop}%
\bibitem [{\citenamefont {{Portegies Zwart}}\ and\ \citenamefont {{van den
  Heuvel}}(2016)}]{2016MNRAS.456.3401P}%
  \BibitemOpen
  \bibfield  {author} {\bibinfo {author} {\bibfnamefont {S.~F.}\ \bibnamefont
  {{Portegies Zwart}}}\ and\ \bibinfo {author} {\bibfnamefont {E.~P.~J.}\
  \bibnamefont {{van den Heuvel}}},\ }\bibfield  {title} {\bibinfo {title}
  {{Was the nineteenth century giant eruption of Eta Carinae a merger event in
  a triple system?}},\ }\href {https://doi.org/10.1093/mnras/stv2787}
  {\bibfield  {journal} {\bibinfo  {journal} {MNRAS}\ }\textbf {\bibinfo
  {volume} {456}},\ \bibinfo {pages} {3401} (\bibinfo {year} {2016})},\ \Eprint
  {https://arxiv.org/abs/1511.06889} {arXiv:1511.06889 [astro-ph.SR]}
  \BibitemShut {NoStop}%
\bibitem [{\citenamefont {{Schneider}}\ \emph {et~al.}(2020)\citenamefont
  {{Schneider}}, \citenamefont {{Ohlmann}}, \citenamefont {{Podsiadlowski}},
  \citenamefont {{R{\"o}pke}}, \citenamefont {{Balbus}},\ and\ \citenamefont
  {{Pakmor}}}]{2020MNRAS.495.2796S}%
  \BibitemOpen
  \bibfield  {author} {\bibinfo {author} {\bibfnamefont {F.~R.~N.}\
  \bibnamefont {{Schneider}}}, \bibinfo {author} {\bibfnamefont {S.~T.}\
  \bibnamefont {{Ohlmann}}}, \bibinfo {author} {\bibfnamefont {P.}~\bibnamefont
  {{Podsiadlowski}}}, \bibinfo {author} {\bibfnamefont {F.~K.}\ \bibnamefont
  {{R{\"o}pke}}}, \bibinfo {author} {\bibfnamefont {S.~A.}\ \bibnamefont
  {{Balbus}}},\ and\ \bibinfo {author} {\bibfnamefont {R.}~\bibnamefont
  {{Pakmor}}},\ }\bibfield  {title} {\bibinfo {title} {{Long-term evolution of
  a magnetic massive merger product}},\ }\href
  {https://doi.org/10.1093/mnras/staa1326} {\bibfield  {journal} {\bibinfo
  {journal} {MNRAS}\ }\textbf {\bibinfo {volume} {495}},\ \bibinfo {pages}
  {2796} (\bibinfo {year} {2020})},\ \Eprint {https://arxiv.org/abs/2005.05335}
  {arXiv:2005.05335 [astro-ph.SR]} \BibitemShut {NoStop}%
\bibitem [{\citenamefont {{Peters}}(1964)}]{1964PhRv..136.1224P}%
  \BibitemOpen
  \bibfield  {author} {\bibinfo {author} {\bibfnamefont {P.~C.}\ \bibnamefont
  {{Peters}}},\ }\bibfield  {title} {\bibinfo {title} {{Gravitational Radiation
  and the Motion of Two Point Masses}},\ }\href
  {https://doi.org/10.1103/PhysRev.136.B1224} {\bibfield  {journal} {\bibinfo
  {journal} {Physical Review}\ }\textbf {\bibinfo {volume} {136}},\ \bibinfo
  {pages} {1224} (\bibinfo {year} {1964})}\BibitemShut {NoStop}%
\bibitem [{\citenamefont {{Ivanova}}\ \emph {et~al.}(2013)\citenamefont
  {{Ivanova}}, \citenamefont {{Justham}}, \citenamefont {{Chen}}, \citenamefont
  {{De Marco}}, \citenamefont {{Fryer}}, \citenamefont {{Gaburov}},
  \citenamefont {{Ge}}, \citenamefont {{Glebbeek}}, \citenamefont {{Han}},
  \citenamefont {{Li}}, \citenamefont {{Lu}}, \citenamefont {{Marsh}},
  \citenamefont {{Podsiadlowski}}, \citenamefont {{Potter}}, \citenamefont
  {{Soker}}, \citenamefont {{Taam}}, \citenamefont {{Tauris}}, \citenamefont
  {{van den Heuvel}},\ and\ \citenamefont {{Webbink}}}]{2013A&ARv..21...59I}%
  \BibitemOpen
  \bibfield  {author} {\bibinfo {author} {\bibfnamefont {N.}~\bibnamefont
  {{Ivanova}}}, \bibinfo {author} {\bibfnamefont {S.}~\bibnamefont
  {{Justham}}}, \bibinfo {author} {\bibfnamefont {X.}~\bibnamefont {{Chen}}},
  \bibinfo {author} {\bibfnamefont {O.}~\bibnamefont {{De Marco}}}, \bibinfo
  {author} {\bibfnamefont {C.~L.}\ \bibnamefont {{Fryer}}}, \bibinfo {author}
  {\bibfnamefont {E.}~\bibnamefont {{Gaburov}}}, \bibinfo {author}
  {\bibfnamefont {H.}~\bibnamefont {{Ge}}}, \bibinfo {author} {\bibfnamefont
  {E.}~\bibnamefont {{Glebbeek}}}, \bibinfo {author} {\bibfnamefont
  {Z.}~\bibnamefont {{Han}}}, \bibinfo {author} {\bibfnamefont {X.~D.}\
  \bibnamefont {{Li}}}, \bibinfo {author} {\bibfnamefont {G.}~\bibnamefont
  {{Lu}}}, \bibinfo {author} {\bibfnamefont {T.}~\bibnamefont {{Marsh}}},
  \bibinfo {author} {\bibfnamefont {P.}~\bibnamefont {{Podsiadlowski}}},
  \bibinfo {author} {\bibfnamefont {A.}~\bibnamefont {{Potter}}}, \bibinfo
  {author} {\bibfnamefont {N.}~\bibnamefont {{Soker}}}, \bibinfo {author}
  {\bibfnamefont {R.}~\bibnamefont {{Taam}}}, \bibinfo {author} {\bibfnamefont
  {T.~M.}\ \bibnamefont {{Tauris}}}, \bibinfo {author} {\bibfnamefont
  {E.~P.~J.}\ \bibnamefont {{van den Heuvel}}},\ and\ \bibinfo {author}
  {\bibfnamefont {R.~F.}\ \bibnamefont {{Webbink}}},\ }\bibfield  {title}
  {\bibinfo {title} {{Common envelope evolution: where we stand and how we can
  move forward}},\ }\href {https://doi.org/10.1007/s00159-013-0059-2}
  {\bibfield  {journal} {\bibinfo  {journal} {AAPR}\ }\textbf {\bibinfo
  {volume} {21}},\ \bibinfo {eid} {59} (\bibinfo {year} {2013})},\ \Eprint
  {https://arxiv.org/abs/1209.4302} {arXiv:1209.4302 [astro-ph.HE]}
  \BibitemShut {NoStop}%
\bibitem [{\citenamefont {Spera}\ \emph {et~al.}(2015)\citenamefont {Spera},
  \citenamefont {Mapelli},\ and\ \citenamefont
  {Bressan}}]{10.1093/mnras/stv1161}%
  \BibitemOpen
  \bibfield  {author} {\bibinfo {author} {\bibfnamefont {M.}~\bibnamefont
  {Spera}}, \bibinfo {author} {\bibfnamefont {M.}~\bibnamefont {Mapelli}},\
  and\ \bibinfo {author} {\bibfnamefont {A.}~\bibnamefont {Bressan}},\
  }\bibfield  {title} {\bibinfo {title} {{The mass spectrum of compact remnants
  from the parsec stellar evolution tracks}},\ }\href
  {https://doi.org/10.1093/mnras/stv1161} {\bibfield  {journal} {\bibinfo
  {journal} {Monthly Notices of the Royal Astronomical Society}\ }\textbf
  {\bibinfo {volume} {451}},\ \bibinfo {pages} {4086} (\bibinfo {year}
  {2015})},\ \Eprint
  {https://arxiv.org/abs/https://academic.oup.com/mnras/article-pdf/451/4/4086/3864018/stv1161.pdf}
  {https://academic.oup.com/mnras/article-pdf/451/4/4086/3864018/stv1161.pdf}
  \BibitemShut {NoStop}%
\bibitem [{\citenamefont {{Kroupa}}(2001)}]{2001MNRAS.322..231K}%
  \BibitemOpen
  \bibfield  {author} {\bibinfo {author} {\bibfnamefont {P.}~\bibnamefont
  {{Kroupa}}},\ }\bibfield  {title} {\bibinfo {title} {{On the variation of the
  initial mass function}},\ }\href
  {https://doi.org/10.1046/j.1365-8711.2001.04022.x} {\bibfield  {journal}
  {\bibinfo  {journal} {MNRAS}\ }\textbf {\bibinfo {volume} {322}},\ \bibinfo
  {pages} {231} (\bibinfo {year} {2001})},\ \Eprint
  {https://arxiv.org/abs/astro-ph/0009005} {arXiv:astro-ph/0009005 [astro-ph]}
  \BibitemShut {NoStop}%
\bibitem [{\citenamefont {{Ballone}}\ \emph {et~al.}(2022)\citenamefont
  {{Ballone}}, \citenamefont {{Costa}}, \citenamefont {{Mapelli}},\ and\
  \citenamefont {{MacLeod}}}]{2022arXiv220403493B}%
  \BibitemOpen
  \bibfield  {author} {\bibinfo {author} {\bibfnamefont {A.}~\bibnamefont
  {{Ballone}}}, \bibinfo {author} {\bibfnamefont {G.}~\bibnamefont {{Costa}}},
  \bibinfo {author} {\bibfnamefont {M.}~\bibnamefont {{Mapelli}}},\ and\
  \bibinfo {author} {\bibfnamefont {M.}~\bibnamefont {{MacLeod}}},\ }\bibfield
  {title} {\bibinfo {title} {{Formation of black holes in the pair-instability
  mass gap: Hydrodynamical simulation of a massive star collision}},\
  }\href@noop {} {\bibfield  {journal} {\bibinfo  {journal} {arXiv e-prints}\
  ,\ \bibinfo {eid} {arXiv:2204.03493}} (\bibinfo {year} {2022})},\ \Eprint
  {https://arxiv.org/abs/2204.03493} {arXiv:2204.03493 [astro-ph.SR]}
  \BibitemShut {NoStop}%
\bibitem [{\citenamefont {{Abbott, B.~P. et
  al.}}(2020{\natexlab{a}})}]{2020PhRvL.125j1102A}%
  \BibitemOpen
  \bibfield  {author} {\bibinfo {author} {\bibnamefont {{Abbott, B.~P. et
  al.}}},\ }\bibfield  {title} {\bibinfo {title} {{GW190521: A Binary Black
  Hole Merger with a Total Mass of 150 M$_{{\ensuremath{\odot}}}$}},\ }\href
  {https://doi.org/10.1103/PhysRevLett.125.101102} {\bibfield  {journal}
  {\bibinfo  {journal} {PRL}\ }\textbf {\bibinfo {volume} {125}},\ \bibinfo
  {eid} {101102} (\bibinfo {year} {2020}{\natexlab{a}})},\ \Eprint
  {https://arxiv.org/abs/2009.01075} {arXiv:2009.01075 [gr-qc]} \BibitemShut
  {NoStop}%
\bibitem [{\citenamefont {{Di Carlo}}\ \emph
  {et~al.}(2020{\natexlab{a}})\citenamefont {{Di Carlo}}, \citenamefont
  {{Mapelli}}, \citenamefont {{Bouffanais}}, \citenamefont {{Giacobbo}},
  \citenamefont {{Santoliquido}}, \citenamefont {{Bressan}}, \citenamefont
  {{Spera}},\ and\ \citenamefont {{Haardt}}}]{2020MNRAS.497.1043D}%
  \BibitemOpen
  \bibfield  {author} {\bibinfo {author} {\bibfnamefont {U.~N.}\ \bibnamefont
  {{Di Carlo}}}, \bibinfo {author} {\bibfnamefont {M.}~\bibnamefont
  {{Mapelli}}}, \bibinfo {author} {\bibfnamefont {Y.}~\bibnamefont
  {{Bouffanais}}}, \bibinfo {author} {\bibfnamefont {N.}~\bibnamefont
  {{Giacobbo}}}, \bibinfo {author} {\bibfnamefont {F.}~\bibnamefont
  {{Santoliquido}}}, \bibinfo {author} {\bibfnamefont {A.}~\bibnamefont
  {{Bressan}}}, \bibinfo {author} {\bibfnamefont {M.}~\bibnamefont {{Spera}}},\
  and\ \bibinfo {author} {\bibfnamefont {F.}~\bibnamefont {{Haardt}}},\
  }\bibfield  {title} {\bibinfo {title} {{Binary black holes in the pair
  instability mass gap}},\ }\href {https://doi.org/10.1093/mnras/staa1997}
  {\bibfield  {journal} {\bibinfo  {journal} {MNRAS}\ }\textbf {\bibinfo
  {volume} {497}},\ \bibinfo {pages} {1043} (\bibinfo {year}
  {2020}{\natexlab{a}})},\ \Eprint {https://arxiv.org/abs/1911.01434}
  {arXiv:1911.01434 [astro-ph.HE]} \BibitemShut {NoStop}%
\bibitem [{\citenamefont {{Di Carlo}}\ \emph
  {et~al.}(2020{\natexlab{b}})\citenamefont {{Di Carlo}}, \citenamefont
  {{Mapelli}}, \citenamefont {{Giacobbo}}, \citenamefont {{Spera}},
  \citenamefont {{Bouffanais}}, \citenamefont {{Rastello}}, \citenamefont
  {{Santoliquido}}, \citenamefont {{Pasquato}}, \citenamefont {{Ballone}},
  \citenamefont {{Trani}}, \citenamefont {{Torniamenti}},\ and\ \citenamefont
  {{Haardt}}}]{2020MNRAS.498..495D}%
  \BibitemOpen
  \bibfield  {author} {\bibinfo {author} {\bibfnamefont {U.~N.}\ \bibnamefont
  {{Di Carlo}}}, \bibinfo {author} {\bibfnamefont {M.}~\bibnamefont
  {{Mapelli}}}, \bibinfo {author} {\bibfnamefont {N.}~\bibnamefont
  {{Giacobbo}}}, \bibinfo {author} {\bibfnamefont {M.}~\bibnamefont {{Spera}}},
  \bibinfo {author} {\bibfnamefont {Y.}~\bibnamefont {{Bouffanais}}}, \bibinfo
  {author} {\bibfnamefont {S.}~\bibnamefont {{Rastello}}}, \bibinfo {author}
  {\bibfnamefont {F.}~\bibnamefont {{Santoliquido}}}, \bibinfo {author}
  {\bibfnamefont {M.}~\bibnamefont {{Pasquato}}}, \bibinfo {author}
  {\bibfnamefont {A.}~\bibnamefont {{Ballone}}}, \bibinfo {author}
  {\bibfnamefont {A.~A.}\ \bibnamefont {{Trani}}}, \bibinfo {author}
  {\bibfnamefont {S.}~\bibnamefont {{Torniamenti}}},\ and\ \bibinfo {author}
  {\bibfnamefont {F.}~\bibnamefont {{Haardt}}},\ }\bibfield  {title} {\bibinfo
  {title} {{Binary black holes in young star clusters: the impact of
  metallicity}},\ }\href {https://doi.org/10.1093/mnras/staa2286} {\bibfield
  {journal} {\bibinfo  {journal} {MNRAS}\ }\textbf {\bibinfo {volume} {498}},\
  \bibinfo {pages} {495} (\bibinfo {year} {2020}{\natexlab{b}})},\ \Eprint
  {https://arxiv.org/abs/2004.09525} {arXiv:2004.09525 [astro-ph.HE]}
  \BibitemShut {NoStop}%
\bibitem [{\citenamefont {{Kremer}}\ \emph {et~al.}(2020)\citenamefont
  {{Kremer}}, \citenamefont {{Spera}}, \citenamefont {{Becker}}, \citenamefont
  {{Chatterjee}}, \citenamefont {{Di Carlo}}, \citenamefont {{Fragione}},
  \citenamefont {{Rodriguez}}, \citenamefont {{Ye}},\ and\ \citenamefont
  {{Rasio}}}]{2020ApJ...903...45K}%
  \BibitemOpen
  \bibfield  {author} {\bibinfo {author} {\bibfnamefont {K.}~\bibnamefont
  {{Kremer}}}, \bibinfo {author} {\bibfnamefont {M.}~\bibnamefont {{Spera}}},
  \bibinfo {author} {\bibfnamefont {D.}~\bibnamefont {{Becker}}}, \bibinfo
  {author} {\bibfnamefont {S.}~\bibnamefont {{Chatterjee}}}, \bibinfo {author}
  {\bibfnamefont {U.~N.}\ \bibnamefont {{Di Carlo}}}, \bibinfo {author}
  {\bibfnamefont {G.}~\bibnamefont {{Fragione}}}, \bibinfo {author}
  {\bibfnamefont {C.~L.}\ \bibnamefont {{Rodriguez}}}, \bibinfo {author}
  {\bibfnamefont {C.~S.}\ \bibnamefont {{Ye}}},\ and\ \bibinfo {author}
  {\bibfnamefont {F.~A.}\ \bibnamefont {{Rasio}}},\ }\bibfield  {title}
  {\bibinfo {title} {{Populating the Upper Black Hole Mass Gap through Stellar
  Collisions in Young Star Clusters}},\ }\href
  {https://doi.org/10.3847/1538-4357/abb945} {\bibfield  {journal} {\bibinfo
  {journal} {APJ}\ }\textbf {\bibinfo {volume} {903}},\ \bibinfo {eid} {45}
  (\bibinfo {year} {2020})},\ \Eprint {https://arxiv.org/abs/2006.10771}
  {arXiv:2006.10771 [astro-ph.HE]} \BibitemShut {NoStop}%
\bibitem [{\citenamefont {{Fryer}}\ \emph {et~al.}(2012)\citenamefont
  {{Fryer}}, \citenamefont {{Belczynski}}, \citenamefont {{Wiktorowicz}},
  \citenamefont {{Dominik}}, \citenamefont {{Kalogera}},\ and\ \citenamefont
  {{Holz}}}]{2012ApJ...749...91F}%
  \BibitemOpen
  \bibfield  {author} {\bibinfo {author} {\bibfnamefont {C.~L.}\ \bibnamefont
  {{Fryer}}}, \bibinfo {author} {\bibfnamefont {K.}~\bibnamefont
  {{Belczynski}}}, \bibinfo {author} {\bibfnamefont {G.}~\bibnamefont
  {{Wiktorowicz}}}, \bibinfo {author} {\bibfnamefont {M.}~\bibnamefont
  {{Dominik}}}, \bibinfo {author} {\bibfnamefont {V.}~\bibnamefont
  {{Kalogera}}},\ and\ \bibinfo {author} {\bibfnamefont {D.~E.}\ \bibnamefont
  {{Holz}}},\ }\bibfield  {title} {\bibinfo {title} {{Compact Remnant Mass
  Function: Dependence on the Explosion Mechanism and Metallicity}},\ }\href
  {https://doi.org/10.1088/0004-637X/749/1/91} {\bibfield  {journal} {\bibinfo
  {journal} {APJ}\ }\textbf {\bibinfo {volume} {749}},\ \bibinfo {eid} {91}
  (\bibinfo {year} {2012})},\ \Eprint {https://arxiv.org/abs/1110.1726}
  {arXiv:1110.1726 [astro-ph.SR]} \BibitemShut {NoStop}%
\bibitem [{\citenamefont {{Stevenson}}\ \emph {et~al.}(2017)\citenamefont
  {{Stevenson}}, \citenamefont {{Vigna-G{\'o}mez}}, \citenamefont {{Mandel}},
  \citenamefont {{Barrett}}, \citenamefont {{Neijssel}}, \citenamefont
  {{Perkins}},\ and\ \citenamefont {{de Mink}}}]{2017NatCo...814906S}%
  \BibitemOpen
  \bibfield  {author} {\bibinfo {author} {\bibfnamefont {S.}~\bibnamefont
  {{Stevenson}}}, \bibinfo {author} {\bibfnamefont {A.}~\bibnamefont
  {{Vigna-G{\'o}mez}}}, \bibinfo {author} {\bibfnamefont {I.}~\bibnamefont
  {{Mandel}}}, \bibinfo {author} {\bibfnamefont {J.~W.}\ \bibnamefont
  {{Barrett}}}, \bibinfo {author} {\bibfnamefont {C.~J.}\ \bibnamefont
  {{Neijssel}}}, \bibinfo {author} {\bibfnamefont {D.}~\bibnamefont
  {{Perkins}}},\ and\ \bibinfo {author} {\bibfnamefont {S.~E.}\ \bibnamefont
  {{de Mink}}},\ }\bibfield  {title} {\bibinfo {title} {{Formation of the first
  three gravitational-wave observations through isolated binary evolution}},\
  }\href {https://doi.org/10.1038/ncomms14906} {\bibfield  {journal} {\bibinfo
  {journal} {Nature Communications}\ }\textbf {\bibinfo {volume} {8}},\
  \bibinfo {eid} {14906} (\bibinfo {year} {2017})},\ \Eprint
  {https://arxiv.org/abs/1704.01352} {arXiv:1704.01352 [astro-ph.HE]}
  \BibitemShut {NoStop}%
\bibitem [{\citenamefont {{Olejak}}\ \emph {et~al.}(2020)\citenamefont
  {{Olejak}}, \citenamefont {{Fishbach}}, \citenamefont {{Belczynski}},
  \citenamefont {{Holz}}, \citenamefont {{Lasota}}, \citenamefont {{Miller}},\
  and\ \citenamefont {{Bulik}}}]{2020ApJ...901L..39O}%
  \BibitemOpen
  \bibfield  {author} {\bibinfo {author} {\bibfnamefont {A.}~\bibnamefont
  {{Olejak}}}, \bibinfo {author} {\bibfnamefont {M.}~\bibnamefont
  {{Fishbach}}}, \bibinfo {author} {\bibfnamefont {K.}~\bibnamefont
  {{Belczynski}}}, \bibinfo {author} {\bibfnamefont {D.~E.}\ \bibnamefont
  {{Holz}}}, \bibinfo {author} {\bibfnamefont {J.~P.}\ \bibnamefont
  {{Lasota}}}, \bibinfo {author} {\bibfnamefont {M.~C.}\ \bibnamefont
  {{Miller}}},\ and\ \bibinfo {author} {\bibfnamefont {T.}~\bibnamefont
  {{Bulik}}},\ }\bibfield  {title} {\bibinfo {title} {{The Origin of
  Inequality: Isolated Formation of a 30+10 M$_{{\ensuremath{\odot}}}$ Binary
  Black Hole Merger}},\ }\href {https://doi.org/10.3847/2041-8213/abb5b5}
  {\bibfield  {journal} {\bibinfo  {journal} {APJL}\ }\textbf {\bibinfo
  {volume} {901}},\ \bibinfo {eid} {L39} (\bibinfo {year} {2020})},\ \Eprint
  {https://arxiv.org/abs/2004.11866} {arXiv:2004.11866 [astro-ph.HE]}
  \BibitemShut {NoStop}%
\bibitem [{\citenamefont {{Belczynski}}\ \emph {et~al.}(2020)\citenamefont
  {{Belczynski}}, \citenamefont {{Klencki}}, \citenamefont {{Fields}},
  \citenamefont {{Olejak}}, \citenamefont {{Berti}}, \citenamefont {{Meynet}},
  \citenamefont {{Fryer}}, \citenamefont {{Holz}}, \citenamefont
  {{O'Shaughnessy}}, \citenamefont {{Brown}}, \citenamefont {{Bulik}},
  \citenamefont {{Leung}}, \citenamefont {{Nomoto}}, \citenamefont {{Madau}},
  \citenamefont {{Hirschi}}, \citenamefont {{Kaiser}}, \citenamefont {{Jones}},
  \citenamefont {{Mondal}}, \citenamefont {{Chruslinska}}, \citenamefont
  {{Drozda}}, \citenamefont {{Gerosa}}, \citenamefont {{Doctor}}, \citenamefont
  {{Giersz}}, \citenamefont {{Ekstrom}}, \citenamefont {{Georgy}},
  \citenamefont {{Askar}}, \citenamefont {{Baibhav}}, \citenamefont
  {{Wysocki}}, \citenamefont {{Natan}}, \citenamefont {{Farr}}, \citenamefont
  {{Wiktorowicz}}, \citenamefont {{Coleman Miller}}, \citenamefont {{Farr}},\
  and\ \citenamefont {{Lasota}}}]{2020A&A...636A.104B}%
  \BibitemOpen
  \bibfield  {author} {\bibinfo {author} {\bibfnamefont {K.}~\bibnamefont
  {{Belczynski}}}, \bibinfo {author} {\bibfnamefont {J.}~\bibnamefont
  {{Klencki}}}, \bibinfo {author} {\bibfnamefont {C.~E.}\ \bibnamefont
  {{Fields}}}, \bibinfo {author} {\bibfnamefont {A.}~\bibnamefont {{Olejak}}},
  \bibinfo {author} {\bibfnamefont {E.}~\bibnamefont {{Berti}}}, \bibinfo
  {author} {\bibfnamefont {G.}~\bibnamefont {{Meynet}}}, \bibinfo {author}
  {\bibfnamefont {C.~L.}\ \bibnamefont {{Fryer}}}, \bibinfo {author}
  {\bibfnamefont {D.~E.}\ \bibnamefont {{Holz}}}, \bibinfo {author}
  {\bibfnamefont {R.}~\bibnamefont {{O'Shaughnessy}}}, \bibinfo {author}
  {\bibfnamefont {D.~A.}\ \bibnamefont {{Brown}}}, \bibinfo {author}
  {\bibfnamefont {T.}~\bibnamefont {{Bulik}}}, \bibinfo {author} {\bibfnamefont
  {S.~C.}\ \bibnamefont {{Leung}}}, \bibinfo {author} {\bibfnamefont
  {K.}~\bibnamefont {{Nomoto}}}, \bibinfo {author} {\bibfnamefont
  {P.}~\bibnamefont {{Madau}}}, \bibinfo {author} {\bibfnamefont
  {R.}~\bibnamefont {{Hirschi}}}, \bibinfo {author} {\bibfnamefont
  {E.}~\bibnamefont {{Kaiser}}}, \bibinfo {author} {\bibfnamefont
  {S.}~\bibnamefont {{Jones}}}, \bibinfo {author} {\bibfnamefont
  {S.}~\bibnamefont {{Mondal}}}, \bibinfo {author} {\bibfnamefont
  {M.}~\bibnamefont {{Chruslinska}}}, \bibinfo {author} {\bibfnamefont
  {P.}~\bibnamefont {{Drozda}}}, \bibinfo {author} {\bibfnamefont
  {D.}~\bibnamefont {{Gerosa}}}, \bibinfo {author} {\bibfnamefont
  {Z.}~\bibnamefont {{Doctor}}}, \bibinfo {author} {\bibfnamefont
  {M.}~\bibnamefont {{Giersz}}}, \bibinfo {author} {\bibfnamefont
  {S.}~\bibnamefont {{Ekstrom}}}, \bibinfo {author} {\bibfnamefont
  {C.}~\bibnamefont {{Georgy}}}, \bibinfo {author} {\bibfnamefont
  {A.}~\bibnamefont {{Askar}}}, \bibinfo {author} {\bibfnamefont
  {V.}~\bibnamefont {{Baibhav}}}, \bibinfo {author} {\bibfnamefont
  {D.}~\bibnamefont {{Wysocki}}}, \bibinfo {author} {\bibfnamefont
  {T.}~\bibnamefont {{Natan}}}, \bibinfo {author} {\bibfnamefont {W.~M.}\
  \bibnamefont {{Farr}}}, \bibinfo {author} {\bibfnamefont {G.}~\bibnamefont
  {{Wiktorowicz}}}, \bibinfo {author} {\bibfnamefont {M.}~\bibnamefont
  {{Coleman Miller}}}, \bibinfo {author} {\bibfnamefont {B.}~\bibnamefont
  {{Farr}}},\ and\ \bibinfo {author} {\bibfnamefont {J.~P.}\ \bibnamefont
  {{Lasota}}},\ }\bibfield  {title} {\bibinfo {title} {{Evolutionary roads
  leading to low effective spins, high black hole masses, and O1/O2 rates for
  LIGO/Virgo binary black holes}},\ }\href
  {https://doi.org/10.1051/0004-6361/201936528} {\bibfield  {journal} {\bibinfo
   {journal} {AAP}\ }\textbf {\bibinfo {volume} {636}},\ \bibinfo {eid} {A104}
  (\bibinfo {year} {2020})},\ \Eprint {https://arxiv.org/abs/1706.07053}
  {arXiv:1706.07053 [astro-ph.HE]} \BibitemShut {NoStop}%
\bibitem [{\citenamefont {{Zevin}}\ and\ \citenamefont
  {{Bavera}}(2022)}]{2022arXiv220302515Z}%
  \BibitemOpen
  \bibfield  {author} {\bibinfo {author} {\bibfnamefont {M.}~\bibnamefont
  {{Zevin}}}\ and\ \bibinfo {author} {\bibfnamefont {S.~S.}\ \bibnamefont
  {{Bavera}}},\ }\bibfield  {title} {\bibinfo {title} {{Suspicious Siblings:
  The Distribution of Mass and Spin Across Component Black Holes in Isolated
  Binary Evolution}},\ }\href@noop {} {\bibfield  {journal} {\bibinfo
  {journal} {arXiv e-prints}\ ,\ \bibinfo {eid} {arXiv:2203.02515}} (\bibinfo
  {year} {2022})},\ \Eprint {https://arxiv.org/abs/2203.02515}
  {arXiv:2203.02515 [astro-ph.HE]} \BibitemShut {NoStop}%
\bibitem [{\citenamefont {{Wellstein}}\ \emph {et~al.}(2001)\citenamefont
  {{Wellstein}}, \citenamefont {{Langer}},\ and\ \citenamefont
  {{Braun}}}]{2001A&A...369..939W}%
  \BibitemOpen
  \bibfield  {author} {\bibinfo {author} {\bibfnamefont {S.}~\bibnamefont
  {{Wellstein}}}, \bibinfo {author} {\bibfnamefont {N.}~\bibnamefont
  {{Langer}}},\ and\ \bibinfo {author} {\bibfnamefont {H.}~\bibnamefont
  {{Braun}}},\ }\bibfield  {title} {\bibinfo {title} {{Formation of contact in
  massive close binaries}},\ }\href
  {https://doi.org/10.1051/0004-6361:20010151} {\bibfield  {journal} {\bibinfo
  {journal} {AAP}\ }\textbf {\bibinfo {volume} {369}},\ \bibinfo {pages} {939}
  (\bibinfo {year} {2001})},\ \Eprint {https://arxiv.org/abs/astro-ph/0102244}
  {arXiv:astro-ph/0102244 [astro-ph]} \BibitemShut {NoStop}%
\bibitem [{\citenamefont {{de Mink}}\ \emph {et~al.}(2007)\citenamefont {{de
  Mink}}, \citenamefont {{Pols}},\ and\ \citenamefont
  {{Hilditch}}}]{2007A&A...467.1181D}%
  \BibitemOpen
  \bibfield  {author} {\bibinfo {author} {\bibfnamefont {S.~E.}\ \bibnamefont
  {{de Mink}}}, \bibinfo {author} {\bibfnamefont {O.~R.}\ \bibnamefont
  {{Pols}}},\ and\ \bibinfo {author} {\bibfnamefont {R.~W.}\ \bibnamefont
  {{Hilditch}}},\ }\bibfield  {title} {\bibinfo {title} {{Efficiency of mass
  transfer in massive close binaries. Tests from double-lined eclipsing
  binaries in the SMC}},\ }\href {https://doi.org/10.1051/0004-6361:20067007}
  {\bibfield  {journal} {\bibinfo  {journal} {AAP}\ }\textbf {\bibinfo {volume}
  {467}},\ \bibinfo {pages} {1181} (\bibinfo {year} {2007})},\ \Eprint
  {https://arxiv.org/abs/astro-ph/0703480} {arXiv:astro-ph/0703480 [astro-ph]}
  \BibitemShut {NoStop}%
\bibitem [{\citenamefont {{Erdem}}\ and\ \citenamefont
  {{{\"O}zt{\"u}rk}}(2014)}]{2014MNRAS.441.1166E}%
  \BibitemOpen
  \bibfield  {author} {\bibinfo {author} {\bibfnamefont {A.}~\bibnamefont
  {{Erdem}}}\ and\ \bibinfo {author} {\bibfnamefont {O.}~\bibnamefont
  {{{\"O}zt{\"u}rk}}},\ }\bibfield  {title} {\bibinfo {title}
  {{Non-conservative mass transfers in Algols}},\ }\href
  {https://doi.org/10.1093/mnras/stu630} {\bibfield  {journal} {\bibinfo
  {journal} {MNRAS}\ }\textbf {\bibinfo {volume} {441}},\ \bibinfo {pages}
  {1166} (\bibinfo {year} {2014})}\BibitemShut {NoStop}%
\bibitem [{\citenamefont {{Schneider}}\ \emph {et~al.}(2015)\citenamefont
  {{Schneider}}, \citenamefont {{Izzard}}, \citenamefont {{Langer}},\ and\
  \citenamefont {{de Mink}}}]{2015ApJ...805...20S}%
  \BibitemOpen
  \bibfield  {author} {\bibinfo {author} {\bibfnamefont {F.~R.~N.}\
  \bibnamefont {{Schneider}}}, \bibinfo {author} {\bibfnamefont {R.~G.}\
  \bibnamefont {{Izzard}}}, \bibinfo {author} {\bibfnamefont {N.}~\bibnamefont
  {{Langer}}},\ and\ \bibinfo {author} {\bibfnamefont {S.~E.}\ \bibnamefont
  {{de Mink}}},\ }\bibfield  {title} {\bibinfo {title} {{Evolution of Mass
  Functions of Coeval Stars through Wind Mass Loss and Binary Interactions}},\
  }\href {https://doi.org/10.1088/0004-637X/805/1/20} {\bibfield  {journal}
  {\bibinfo  {journal} {APJ}\ }\textbf {\bibinfo {volume} {805}},\ \bibinfo
  {eid} {20} (\bibinfo {year} {2015})},\ \Eprint
  {https://arxiv.org/abs/1504.01735} {arXiv:1504.01735 [astro-ph.SR]}
  \BibitemShut {NoStop}%
\bibitem [{\citenamefont {{Shao}}\ and\ \citenamefont
  {{Li}}(2016)}]{2016ApJ...833..108S}%
  \BibitemOpen
  \bibfield  {author} {\bibinfo {author} {\bibfnamefont {Y.}~\bibnamefont
  {{Shao}}}\ and\ \bibinfo {author} {\bibfnamefont {X.-D.}\ \bibnamefont
  {{Li}}},\ }\bibfield  {title} {\bibinfo {title} {{Nonconservative Mass
  Transfer in Massive Binaries and the Formation of Wolf-Rayet+O Binaries}},\
  }\href {https://doi.org/10.3847/1538-4357/833/1/108} {\bibfield  {journal}
  {\bibinfo  {journal} {APJ}\ }\textbf {\bibinfo {volume} {833}},\ \bibinfo
  {eid} {108} (\bibinfo {year} {2016})},\ \Eprint
  {https://arxiv.org/abs/1610.04307} {arXiv:1610.04307 [astro-ph.SR]}
  \BibitemShut {NoStop}%
\bibitem [{\citenamefont {{Mennekens}}\ and\ \citenamefont
  {{Vanbeveren}}(2017)}]{2017A&A...599A..84M}%
  \BibitemOpen
  \bibfield  {author} {\bibinfo {author} {\bibfnamefont {N.}~\bibnamefont
  {{Mennekens}}}\ and\ \bibinfo {author} {\bibfnamefont {D.}~\bibnamefont
  {{Vanbeveren}}},\ }\bibfield  {title} {\bibinfo {title} {{A comparison
  between observed Algol-type double stars in the solar neighborhood and
  evolutionary computations of galactic case A binaries with a B-type primary
  at birth}},\ }\href {https://doi.org/10.1051/0004-6361/201630131} {\bibfield
  {journal} {\bibinfo  {journal} {AAP}\ }\textbf {\bibinfo {volume} {599}},\
  \bibinfo {eid} {A84} (\bibinfo {year} {2017})},\ \Eprint
  {https://arxiv.org/abs/1611.08398} {arXiv:1611.08398 [astro-ph.SR]}
  \BibitemShut {NoStop}%
\bibitem [{\citenamefont {{Dervi{\c{s}}o{\v{g}}lu}}\ \emph
  {et~al.}(2018)\citenamefont {{Dervi{\c{s}}o{\v{g}}lu}}, \citenamefont
  {{Pavlovski}}, \citenamefont {{Lehmann}}, \citenamefont {{Southworth}},\ and\
  \citenamefont {{Bewsher}}}]{2018MNRAS.481.5660D}%
  \BibitemOpen
  \bibfield  {author} {\bibinfo {author} {\bibfnamefont {A.}~\bibnamefont
  {{Dervi{\c{s}}o{\v{g}}lu}}}, \bibinfo {author} {\bibfnamefont
  {K.}~\bibnamefont {{Pavlovski}}}, \bibinfo {author} {\bibfnamefont
  {H.}~\bibnamefont {{Lehmann}}}, \bibinfo {author} {\bibfnamefont
  {J.}~\bibnamefont {{Southworth}}},\ and\ \bibinfo {author} {\bibfnamefont
  {D.}~\bibnamefont {{Bewsher}}},\ }\bibfield  {title} {\bibinfo {title}
  {{Evidence for conservative mass transfer in the classical Algol system
  {\ensuremath{\delta}} Librae from its surface carbon-to-nitrogen abundance
  ratio}},\ }\href {https://doi.org/10.1093/mnras/sty2684} {\bibfield
  {journal} {\bibinfo  {journal} {MNRAS}\ }\textbf {\bibinfo {volume} {481}},\
  \bibinfo {pages} {5660} (\bibinfo {year} {2018})},\ \Eprint
  {https://arxiv.org/abs/1810.01465} {arXiv:1810.01465 [astro-ph.SR]}
  \BibitemShut {NoStop}%
\bibitem [{\citenamefont {{Tiwari}}(2021{\natexlab{a}})}]{2021CQGra..38o5007T}%
  \BibitemOpen
  \bibfield  {author} {\bibinfo {author} {\bibfnamefont {V.}~\bibnamefont
  {{Tiwari}}},\ }\bibfield  {title} {\bibinfo {title} {{VAMANA: modeling binary
  black hole population with minimal assumptions}},\ }\href
  {https://doi.org/10.1088/1361-6382/ac0b54} {\bibfield  {journal} {\bibinfo
  {journal} {Classical and Quantum Gravity}\ }\textbf {\bibinfo {volume}
  {38}},\ \bibinfo {eid} {155007} (\bibinfo {year} {2021}{\natexlab{a}})},\
  \Eprint {https://arxiv.org/abs/2006.15047} {arXiv:2006.15047 [astro-ph.HE]}
  \BibitemShut {NoStop}%
\bibitem [{\citenamefont {{Tiwari}}\ and\ \citenamefont
  {{Fairhurst}}(2021)}]{2021ApJ...913L..19T}%
  \BibitemOpen
  \bibfield  {author} {\bibinfo {author} {\bibfnamefont {V.}~\bibnamefont
  {{Tiwari}}}\ and\ \bibinfo {author} {\bibfnamefont {S.}~\bibnamefont
  {{Fairhurst}}},\ }\bibfield  {title} {\bibinfo {title} {{The Emergence of
  Structure in the Binary Black Hole Mass Distribution}},\ }\href
  {https://doi.org/10.3847/2041-8213/abfbe7} {\bibfield  {journal} {\bibinfo
  {journal} {APJL}\ }\textbf {\bibinfo {volume} {913}},\ \bibinfo {eid} {L19}
  (\bibinfo {year} {2021})},\ \Eprint {https://arxiv.org/abs/2011.04502}
  {arXiv:2011.04502 [astro-ph.HE]} \BibitemShut {NoStop}%
\bibitem [{\citenamefont {{Tiwari}}(2021{\natexlab{b}})}]{2021arXiv211113991T}%
  \BibitemOpen
  \bibfield  {author} {\bibinfo {author} {\bibfnamefont {V.}~\bibnamefont
  {{Tiwari}}},\ }\bibfield  {title} {\bibinfo {title} {{Exploring Features in
  the Binary Black Hole Population}},\ }\href@noop {} {\bibfield  {journal}
  {\bibinfo  {journal} {arXiv e-prints}\ ,\ \bibinfo {eid} {arXiv:2111.13991}}
  (\bibinfo {year} {2021}{\natexlab{b}})},\ \Eprint
  {https://arxiv.org/abs/2111.13991} {arXiv:2111.13991 [astro-ph.HE]}
  \BibitemShut {NoStop}%
\bibitem [{\citenamefont {{Madau}}\ and\ \citenamefont
  {{Fragos}}(2017)}]{2017ApJ...840...39M}%
  \BibitemOpen
  \bibfield  {author} {\bibinfo {author} {\bibfnamefont {P.}~\bibnamefont
  {{Madau}}}\ and\ \bibinfo {author} {\bibfnamefont {T.}~\bibnamefont
  {{Fragos}}},\ }\bibfield  {title} {\bibinfo {title} {{Radiation Backgrounds
  at Cosmic Dawn: X-Rays from Compact Binaries}},\ }\href
  {https://doi.org/10.3847/1538-4357/aa6af9} {\bibfield  {journal} {\bibinfo
  {journal} {APJ}\ }\textbf {\bibinfo {volume} {840}},\ \bibinfo {eid} {39}
  (\bibinfo {year} {2017})},\ \Eprint {https://arxiv.org/abs/1606.07887}
  {arXiv:1606.07887 [astro-ph.GA]} \BibitemShut {NoStop}%
\bibitem [{\citenamefont {{Mandel}}\ and\ \citenamefont
  {{Broekgaarden}}(2021)}]{2021arXiv210714239M}%
  \BibitemOpen
  \bibfield  {author} {\bibinfo {author} {\bibfnamefont {I.}~\bibnamefont
  {{Mandel}}}\ and\ \bibinfo {author} {\bibfnamefont {F.~S.}\ \bibnamefont
  {{Broekgaarden}}},\ }\bibfield  {title} {\bibinfo {title} {{Rates of Compact
  Object Coalescences}},\ }\href@noop {} {\bibfield  {journal} {\bibinfo
  {journal} {arXiv e-prints}\ ,\ \bibinfo {eid} {arXiv:2107.14239}} (\bibinfo
  {year} {2021})},\ \Eprint {https://arxiv.org/abs/2107.14239}
  {arXiv:2107.14239 [astro-ph.HE]} \BibitemShut {NoStop}%
\bibitem [{\citenamefont {{Abbott, B.~P. et
  al.}}(2020{\natexlab{b}})}]{2020PhRvD.102d3015A}%
  \BibitemOpen
  \bibfield  {author} {\bibinfo {author} {\bibnamefont {{Abbott, B.~P. et
  al.}}},\ }\bibfield  {title} {\bibinfo {title} {{GW190412: Observation of a
  binary-black-hole coalescence with asymmetric masses}},\ }\href
  {https://doi.org/10.1103/PhysRevD.102.043015} {\bibfield  {journal} {\bibinfo
   {journal} {PRD}\ }\textbf {\bibinfo {volume} {102}},\ \bibinfo {eid}
  {043015} (\bibinfo {year} {2020}{\natexlab{b}})},\ \Eprint
  {https://arxiv.org/abs/2004.08342} {arXiv:2004.08342 [astro-ph.HE]}
  \BibitemShut {NoStop}%
\bibitem [{\citenamefont {{Packet}}(1981)}]{1981A&A...102...17P}%
  \BibitemOpen
  \bibfield  {author} {\bibinfo {author} {\bibfnamefont {W.}~\bibnamefont
  {{Packet}}},\ }\bibfield  {title} {\bibinfo {title} {{On the spin-up of the
  mass accreting component in a close binary system}},\ }\href@noop {}
  {\bibfield  {journal} {\bibinfo  {journal} {AAP}\ }\textbf {\bibinfo {volume}
  {102}},\ \bibinfo {pages} {17} (\bibinfo {year} {1981})}\BibitemShut
  {NoStop}%
\bibitem [{\citenamefont {{Renzo}}\ and\ \citenamefont
  {{G{\"o}tberg}}(2021)}]{2021ApJ...923..277R}%
  \BibitemOpen
  \bibfield  {author} {\bibinfo {author} {\bibfnamefont {M.}~\bibnamefont
  {{Renzo}}}\ and\ \bibinfo {author} {\bibfnamefont {Y.}~\bibnamefont
  {{G{\"o}tberg}}},\ }\bibfield  {title} {\bibinfo {title} {{Evolution of
  Accretor Stars in Massive Binaries: Broader Implications from Modeling
  {\ensuremath{\zeta}} Ophiuchi}},\ }\href
  {https://doi.org/10.3847/1538-4357/ac29c5} {\bibfield  {journal} {\bibinfo
  {journal} {APJ}\ }\textbf {\bibinfo {volume} {923}},\ \bibinfo {eid} {277}
  (\bibinfo {year} {2021})},\ \Eprint {https://arxiv.org/abs/2107.10933}
  {arXiv:2107.10933 [astro-ph.SR]} \BibitemShut {NoStop}%
\bibitem [{\citenamefont {{Schneider}}\ \emph {et~al.}(2019)\citenamefont
  {{Schneider}}, \citenamefont {{Ohlmann}}, \citenamefont {{Podsiadlowski}},
  \citenamefont {{R{\"o}pke}}, \citenamefont {{Balbus}}, \citenamefont
  {{Pakmor}},\ and\ \citenamefont {{Springel}}}]{2019Natur.574..211S}%
  \BibitemOpen
  \bibfield  {author} {\bibinfo {author} {\bibfnamefont {F.~R.~N.}\
  \bibnamefont {{Schneider}}}, \bibinfo {author} {\bibfnamefont {S.~T.}\
  \bibnamefont {{Ohlmann}}}, \bibinfo {author} {\bibfnamefont {P.}~\bibnamefont
  {{Podsiadlowski}}}, \bibinfo {author} {\bibfnamefont {F.~K.}\ \bibnamefont
  {{R{\"o}pke}}}, \bibinfo {author} {\bibfnamefont {S.~A.}\ \bibnamefont
  {{Balbus}}}, \bibinfo {author} {\bibfnamefont {R.}~\bibnamefont {{Pakmor}}},\
  and\ \bibinfo {author} {\bibfnamefont {V.}~\bibnamefont {{Springel}}},\
  }\bibfield  {title} {\bibinfo {title} {{Stellar mergers as the origin of
  magnetic massive stars}},\ }\href {https://doi.org/10.1038/s41586-019-1621-5}
  {\bibfield  {journal} {\bibinfo  {journal} {Nature}\ }\textbf {\bibinfo
  {volume} {574}},\ \bibinfo {pages} {211} (\bibinfo {year} {2019})},\ \Eprint
  {https://arxiv.org/abs/1910.14058} {arXiv:1910.14058 [astro-ph.SR]}
  \BibitemShut {NoStop}%
\bibitem [{\citenamefont {{Fuller}}\ and\ \citenamefont
  {{Ma}}(2019)}]{2019ApJ...881L...1F}%
  \BibitemOpen
  \bibfield  {author} {\bibinfo {author} {\bibfnamefont {J.}~\bibnamefont
  {{Fuller}}}\ and\ \bibinfo {author} {\bibfnamefont {L.}~\bibnamefont
  {{Ma}}},\ }\bibfield  {title} {\bibinfo {title} {{Most Black Holes Are Born
  Very Slowly Rotating}},\ }\href {https://doi.org/10.3847/2041-8213/ab339b}
  {\bibfield  {journal} {\bibinfo  {journal} {APJL}\ }\textbf {\bibinfo
  {volume} {881}},\ \bibinfo {eid} {L1} (\bibinfo {year} {2019})},\ \Eprint
  {https://arxiv.org/abs/1907.03714} {arXiv:1907.03714 [astro-ph.SR]}
  \BibitemShut {NoStop}%
\bibitem [{\citenamefont {{Safarzadeh}}\ \emph {et~al.}(2020)\citenamefont
  {{Safarzadeh}}, \citenamefont {{Biscoveanu}},\ and\ \citenamefont
  {{Loeb}}}]{2020ApJ...901..137S}%
  \BibitemOpen
  \bibfield  {author} {\bibinfo {author} {\bibfnamefont {M.}~\bibnamefont
  {{Safarzadeh}}}, \bibinfo {author} {\bibfnamefont {S.}~\bibnamefont
  {{Biscoveanu}}},\ and\ \bibinfo {author} {\bibfnamefont {A.}~\bibnamefont
  {{Loeb}}},\ }\bibfield  {title} {\bibinfo {title} {{Constraining the Delay
  Time Distribution of Compact Binary Objects from the Stochastic
  Gravitational-wave Background Searches}},\ }\href
  {https://doi.org/10.3847/1538-4357/abb1af} {\bibfield  {journal} {\bibinfo
  {journal} {APJ}\ }\textbf {\bibinfo {volume} {901}},\ \bibinfo {eid} {137}
  (\bibinfo {year} {2020})},\ \Eprint {https://arxiv.org/abs/2004.12999}
  {arXiv:2004.12999 [astro-ph.HE]} \BibitemShut {NoStop}%
\bibitem [{\citenamefont {{Planck Collaboration}}(2016)}]{2016A&A...594A..13P}%
  \BibitemOpen
  \bibfield  {author} {\bibinfo {author} {\bibnamefont {{Planck
  Collaboration}}},\ }\bibfield  {title} {\bibinfo {title} {{Planck 2015
  results. XIII. Cosmological parameters}},\ }\href
  {https://doi.org/10.1051/0004-6361/201525830} {\bibfield  {journal} {\bibinfo
   {journal} {AAP}\ }\textbf {\bibinfo {volume} {594}},\ \bibinfo {eid} {A13}
  (\bibinfo {year} {2016})},\ \Eprint {https://arxiv.org/abs/1502.01589}
  {arXiv:1502.01589 [astro-ph.CO]} \BibitemShut {NoStop}%
\bibitem [{\citenamefont {{Dvorkin}}\ \emph {et~al.}(2015)\citenamefont
  {{Dvorkin}}, \citenamefont {{Silk}}, \citenamefont {{Vangioni}},
  \citenamefont {{Petitjean}},\ and\ \citenamefont
  {{Olive}}}]{2015MNRAS.452L..36D}%
  \BibitemOpen
  \bibfield  {author} {\bibinfo {author} {\bibfnamefont {I.}~\bibnamefont
  {{Dvorkin}}}, \bibinfo {author} {\bibfnamefont {J.}~\bibnamefont {{Silk}}},
  \bibinfo {author} {\bibfnamefont {E.}~\bibnamefont {{Vangioni}}}, \bibinfo
  {author} {\bibfnamefont {P.}~\bibnamefont {{Petitjean}}},\ and\ \bibinfo
  {author} {\bibfnamefont {K.~A.}\ \bibnamefont {{Olive}}},\ }\bibfield
  {title} {\bibinfo {title} {{The origin of dispersion in DLA
  metallicities.}},\ }\href {https://doi.org/10.1093/mnrasl/slv085} {\bibfield
  {journal} {\bibinfo  {journal} {MNRAS}\ }\textbf {\bibinfo {volume} {452}},\
  \bibinfo {pages} {L36} (\bibinfo {year} {2015})},\ \Eprint
  {https://arxiv.org/abs/1506.06761} {arXiv:1506.06761 [astro-ph.CO]}
  \BibitemShut {NoStop}%
\end{thebibliography}%

\appendix

\section{BBH merger rate density}\label{sec:Merger rate density}
At cosmological redshift $z$, the merger rate of BBHs from a given stellar population can be calculated by a convolution of the delay time distribution and the cosmic star formation history \citep[][]{2020ApJ...901..137S} 
\begin{align}\label{eq:merger rate}
\mathcal{R}_{\rm BBH}(z)=&\int_{z_b=10}^{z_b=z}{\rm d}z_b\,\frac{{\rm d}t}{{\rm d}z}(z_b)\nonumber\\
&\times\int_{Z=10^{-4}}^{Z=0.03}{\rm d}Z\,\mathcal{F}_{\rm Z}(t-t_b)\mathcal{R}_{\rm SFR}(Z,z_b).
\end{align}
We adopt standard $\Lambda$CDM cosmology in which ${{\rm d}t}/{{\rm d}z}=-\left[(1+z)E(z)H_0\right]^{-1}$ and $E(z)=\left[\Omega_{M,0}(1+z)^3+\Omega_{K,0}(1+z)^2+\Omega_{\Lambda,0}\right]^{1/2}$ with $1/H_0=14\,\rm Gyr$, $\Omega_{M,0}=0.3$, $\Omega_{K,0}=0$, and $\Omega_{\Lambda,0}=0.7$ \citep[][]{2016A&A...594A..13P}.

%In Eq.~\eqref{eq:merger rate}, the inner integration w.r.t. $Z$ yields the volumetric rate of BBHs whose progenitor systems are born at redshift $z_b$ and merge after a delay time $t-t_b$. For a given merger channel, the function $\mathcal{F}_{\rm Z}(t-t_b)$ is the fraction of systems in a metallicity bin $\left[Z,Z+{\rm d}Z\right]$ which lead to BBH mergers with a delay time $t-t_b$ per ${\rm d}t$, i.e., the integral $\int_0^\infty{\rm d}t\,\mathcal{F}_{\rm Z}(t-t_b)$ yields the total fraction of BBH mergers in the metallicity bin. Thus, $\mathcal{F}_{\rm Z}$ entirely encapsulates the results of our population synthesis. 
The function $\mathcal{F}_{\rm Z}(t-t_b)$ results from the simulation of the stellar populations and describes the fraction of systems in a metallicity bin $\left[Z,Z+{\rm d}Z\right]$ which lead to BBH mergers with a delay time $t-t_b$ per ${\rm d}t$. The second term in the integration, $\mathcal{R}_{\rm SFR}(Z,z_b)$, is the cosmic star formation rate which we write as
\begin{equation}\label{eq:SFR}
    \mathcal{R}_{\rm SFR}(Z,z_b){\rm d}Z=\kappa\psi(z_b)p(Z,z_b){\rm d}Z,
\end{equation}
using the data fit from \citet[][]{2017ApJ...840...39M}
\begin{equation}\label{eq:SFR2}
    \psi(z_b)=0.01\frac{(1+z_b)^{2.6}}{1+\left[(1+z_b)/3.2\right]^{6.2}}\,\rm M_\odot\,Mpc^{-3}\,yr^{-1},
\end{equation}
and the chemical enrichment model where $Z$ at redshift $z_b$ follows a log-normal distribution
\begin{equation}\label{eq:log-normal}
    p(Z,z_b)=\frac{\log(e)}{\sqrt{2\pi\sigma_Z^2}Z}\exp\Big\{-\frac{[\log(Z/{\rm Z_\odot})-\mu(z_b)]^2}{2\sigma_Z^2}\Big\},
\end{equation}
with mean metallicity $\mu(z_b)=0.153-0.074z_b^{1.34}$ \citep[][]{2017ApJ...840...39M}. and standard deviation $\sigma_Z=0.25$ \citep[][]{2015MNRAS.452L..36D}. Different choices for the value $\sigma_Z$  For $\psi$ we assume a log-normal distribution with a standard deviation of $\sigma_\psi=0.5$.

Eq.~\eqref{eq:SFR2} describes the rate at which mass of all stars with mass $0.1$~--~$100\,\rm M_\odot$ is formed, regardless whether they are multiplicity systems or not. Since we are only simulating massive triples and binaries with $m_{1(2)}\geq5\,\rm M_\odot$ and $m_{3}\geq8\,\rm M_\odot$ (see Section~\ref{sec:Initial conditions}) the constant $\kappa$ is introduced to convert $\psi$ into a formation rate of progenitor systems with the assumed properties. To calculate $\kappa$ we set up an entire stellar population of any mass assuming the same parameter distribution as in Section~\ref{sec:Initial conditions} and multiplicity fractions as a function of the primary spectral type as reported by \citet{2017ApJS..230...15M}. Here, we only consider singles, binaries, and triples (neglecting the effect of quadruples and higher-order systems) and assume that any primary star with $m_1\geq20\,\rm M_\odot$ is in a triple. We then calculate $\kappa$ and the resulting BBH merger rate for each spectral type individually.

In practice, the uncertainty of $\mathcal{R}_{\rm BBH}(z)$ is estimated by Monte-Carlo sampling of Eq.~\eqref{eq:merger rate}.

\section{Chirp mass distribution}\label{sec:chirp}
\begin{figure*}
 \includegraphics[width=2\columnwidth]{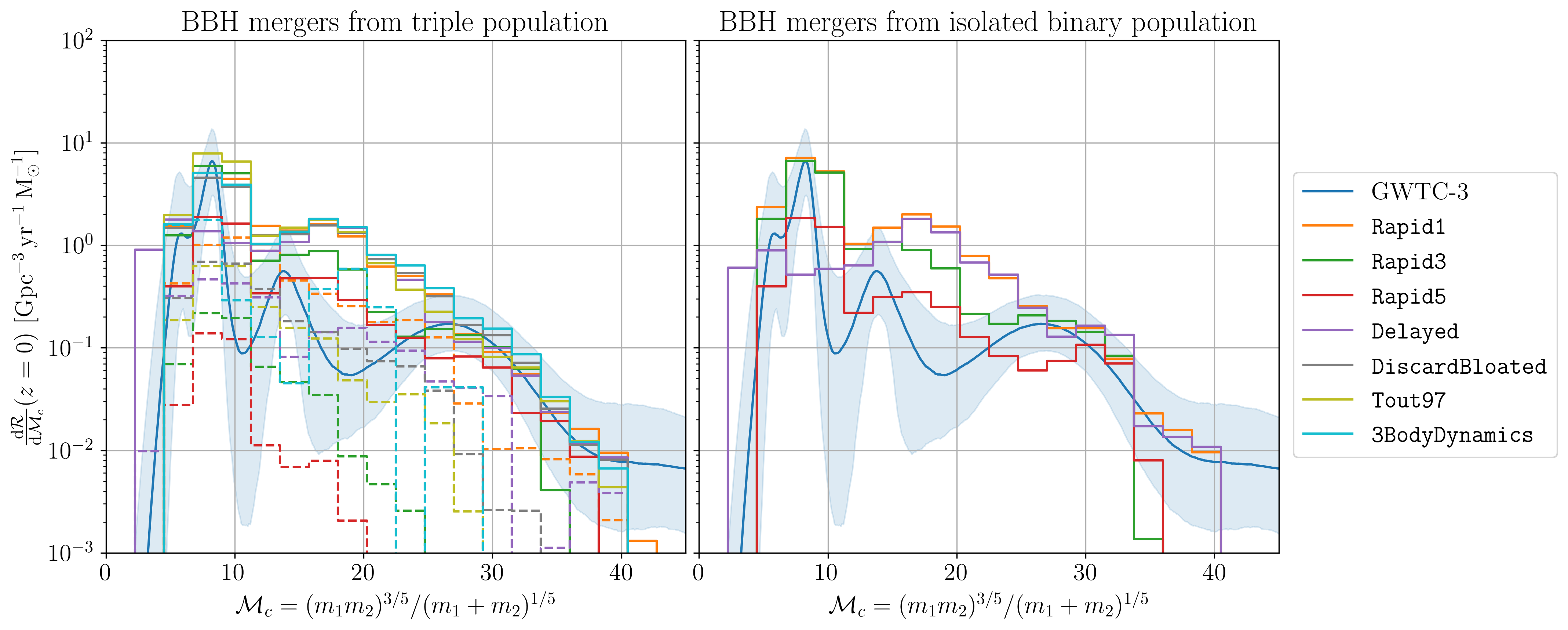}
 \caption{Chirp mass distribution of BBH mergers in the local Universe ($z=0$). As in Figure~\ref{fig:q-rate} dashed lines highlight the contribution from the outer binary channel.}
 \label{fig:chirp}
\end{figure*}
In Figure~\ref{fig:chirp}, we show the differential merger rate per chirp mass $\mathcal{M}_c=(m_1m_2)^{3/5}/(m_1+m_2)^{1/5}$ which determines the gravitational waveform at leading order. On coarse-grained scales, we find agreement of our models with the ``flexible mixture model" (GWTC-3) up to $\mathcal{M}_c\lesssim40\,\rm M_\odot$, but note that neither the isolated nor the triple population could reproduce substructure of the mass distribution that were discovered in the third observing run of the \citet[][]{2021arXiv211103634T}.

\end{document}